%% file: Main.tex
\documentclass[11 pt]{article}

\usepackage[margin=1 in]{geometry}
\usepackage[inline]{enumitem}
\usepackage{comment, graphicx, cancel, dsfont, bbm, float, verbatim, multirow, pdflscape, natbib, setspace, booktabs}
\usepackage{amsmath, amssymb, mathtools, mathrsfs, amsthm, thmtools, thm-restate}
\usepackage[hypertexnames=false]{hyperref}

\graphicspath{{./Figs/}}

\declaretheorem[numberwithin=section, style=remark]{remark}
\declaretheorem[numberwithin=section]{example}

\newcommand*{\thisdraft}{This draft: 26 December, 2025}
\newcommand*{\firstdraft}{First draft: 12 January, 2023} 
\date{\thisdraft \\ \firstdraft}

\hypersetup{
	colorlinks=true, 
	linkcolor=black, 
	urlcolor=blue, 
	citecolor=blue, 
	linktoc=all, 
}

\usepackage[framemethod=TikZ]{mdframed}
\mdfdefinestyle{MyFrame}{%
	linecolor=black,
	outerlinewidth=2pt,
	roundcorner=20pt,
	innertopmargin=\baselineskip,
	innerbottommargin=\baselineskip,
	innerrightmargin=20pt,
	innerleftmargin=20pt,
	backgroundcolor=gray!30!white}

\DeclareMathOperator*{\argmax}{arg\,max}
\DeclareMathOperator*{\argmin}{arg\,min}

\title{Robustness to Missing Data: \\ Breakdown Point Analysis}

\author{Daniel Ober-Reynolds\thanks{Email: doberreynolds@gmail.com. This paper was presented at the 2023 North American Summer Meeting of the Econometric Society. I want to thank Andres Santos, Denis Chetverikov, Jinyong Hahn, Zhipeng Liao, Rosa Matzkin, Shuyang Sheng, Manu Navjeevan, Lucas Zhang, two anonymous referees, and seminar participants for many helpful comments and suggestions.}}

\begin{document}

	\maketitle
	
	\input{./MissingDataBDP_Abstract.tex}

	\pagenumbering{gobble} 
	
	\clearpage 
	
	\pagenumbering{arabic} 
	
	\input{./Section_Introduction.tex}

	\input{./Section_Setting.tex}
	
	\input{./Section_Duality.tex}

	\input{./Section_Asymptotics.tex}

	\input{./Section_Simulations.tex}

	\input{./Section_Application.tex}

	\input{./Section_Conclusion.tex}
	
	\nocite{bandiera2020womendata} 
	\nocite{barham2024experimentaldata}
	\nocite{giacobino2024schoolgirlsdata}
	\nocite{molina2025attritiondata} 
	
	\bibliography{./MissingDataBDP_bibliography.bib}
	
	\clearpage
	
	\singlespacing
	
	\appendix
	
	\input{./Appendix_Notation.tex}
	
	\input{./Appendix_BDP_Partial_ID.tex}

	\input{./Appendix_MeasuringSelectionBDP.tex}

	\input{./Appendix_Duality.tex}
	
	\input{./Appendix_Estimation_TechnicalLemmas.tex}
	
	\input{./Appendix_Estimation_Consistency.tex}
	
	\input{./Appendix_Estimation_Inference.tex}
	
	\input{./Appendix_Examples.tex}
	
\end{document}

%% file: MissingDataBDP_Abstract.tex
\begin{abstract}
	Missing data is pervasive in econometric applications, and rarely is it plausible that the data are missing (completely) at random. This paper proposes a methodology for studying the robustness of results drawn from incomplete datasets. Selection is measured as the divergence from the distribution of complete observations to the distribution of incomplete observations. The \textit{breakdown point} is defined as the minimal amount of selection needed to overturn a given result. Reporting point estimates and lower confidence intervals of the breakdown point is a simple, concise way to communicate the robustness of a result. An estimator of the breakdown point is proposed and shown $\sqrt{n}$-consistent and asymptotically normal. This estimator can be applied directly to conclusions drawn from any model identified with the generalized method of moments (GMM) that satisfies mild assumptions. Simulations demonstrate the finite sample performance of the breakdown point estimator on averages, linear regression, and logistic regression. The methodology is illustrated by estimating the breakdown point of conclusions drawn from several randomized controlled trails suffering from missing data due to attrition.
\end{abstract}

\bigskip



%% file: Section_Introduction.tex
\section{Introduction}


Virtually every economic dataset is plagued by missing and incomplete records. 
Survey nonresponse is the most visible cause, and appears to be worsening over time. 
\cite{bollinger2019trouble} report that the Current Population Survey's Annual Social and Economics Supplement item and whole nonresponse has been increasing, reaching 43 percent in 2015. 
By linking these data with the Social Security Administration Detailed Earnings Record, the authors show that the distribution of nonresponders differs from that of responders even after conditioning on a large set of covariates. 


Samples with missing or incomplete observations fail to identify the population distribution \citep{manski2005partial}. 
To make progress, researchers commonly apply standard procedures to the complete observations. 
This practice is typically justified by assuming the data are ``missing completely at random'' (MCAR): the assumption that incomplete observations follow the same distribution as that of the complete observations. 
In many settings such an assumption is implausible. 
Without it, the conclusions drawn are uncomfortably qualified as being about the distribution of the complete observations, rather than the actual distribution of interest.


This paper proposes a method to investigate the robustness of a conclusion regarding the whole population. 
Results are more robust when overturning them would require more selection. 
To make this intuition precise, selection is measured as the divergence from the distribution of complete observations to the distribution of incomplete observations. 
Many statistical divergences can be used to measure selection.
Squared Hellinger is an attractive choice for this purpose, as it can be interpreted as a measure of how well the variables under study would predict an observation being complete. 
This gives the values of the selection measure context, allowing researchers to gauge how much selection can be expected in a given setting. 
The \textit{breakdown point} is the minimum amount of selection needed to overturn a conclusion. 
Readers who doubt the setting exhibits that much selection will find the conclusion compelling. 

The estimator of the breakdown point proposed below can be applied to conclusions drawn from a model identified with the generalized method of moments (GMM) \citep{hansen1982large}. 
This includes most models used in applied econometrics, including linear regression, instrumental variable models, binary choice models such as the logit model, and many more.
In a model identified with GMM, the breakdown point is the constrained minimum of the value function of a convex optimization problem. 
Estimators of the breakdown point are constructed from the dual of this convex inner problem, and shown to be $\sqrt{n}$-consistent and asymptotically normal. 
Lower confidence intervals are simple to construct. 
Reporting the point estimates and lower confidence intervals of the breakdown point is a simple, concise way to communicate a result's robustness. 

As a demonstration of breakdown point analysis, the paper concludes with an investigation of the robustness of results from three randomized controlled trails  \citep{barham2024experimental, bandiera2020women, giacobino2024schoolgirls}. 
Many randomized controlled trials conducted in developing countries suffer from missing data that results from attrition, often due to study subject migration. 
Results can change meaningfully when migrants are carefully tracked and included in the sample, suggesting that the data are not missing at random \citep{molina2025attrition}.
The three randomized controlled trials studied here are evaluated through intent-to-treat estimates implemented through linear regression. 
The breakdown point analyses thus show a range of breakdown point estimates resulting from variation in real world data, rather than variation in methodology. 
Some claims are notably more robust than others, even when a similar amount of data is missing and the estimates found by dropping incomplete observations are similar.


The breakdown point analysis proposed here has a number of advantages over existing methods for incomplete datasets. 
Sample selection models consider regressions with samples where the dependent variable is sometimes missing, and obtain point identification by modeling the selection process \citep{heckman1979sample, das2003nonparametric}. 
These models often require the data include a variable changing the probability of observation but not the dependent variable. 
This ``exclusion restriction'' is difficult to satisfy in many applications. 
Sample selection models can be identified without such an exclusion restriction provided the researcher makes additional functional form restrictions, as in \cite{escanciano2016identification}.
The breakdown point approach proposed here can be used on most common models identified with GMM, including but not restricted to regressions with missing outcomes. 
It requires neither additional data nor additional modeling assumptions.
The breakdown point can be estimated even if the incomplete observations are in fact completely missing, a distinct possibility when using survey data.

The econometric literature on missing data has also explored bounding the parameter of interest based on the support \citep{manski2005partial, horowitz2006identification}. 
If all parameter values within these ``worst-case'' bounds satisfy the researcher's conclusion, the conclusion is undoubtedly robust. 
Unfortunately, the bounds may be uninformative in practice. 
Proponents of this approach are well aware these bounds are conservative, and propose this exercise as a place to begin an investigation rather than end one. 
Additional identifying assumptions should then be considered, in order to make plain to readers what needs to be assumed to reach a given conclusion \citep{manski2013response}. 
The breakdown approach proposed here is a simple version of this exercise, as the assumption that selection is less than the breakdown point leads one to conclude the hypothesis under investigation. 


A growing literature advocates for breakdown analysis as a general, tractable method to assess the sensitivity of a result to relaxations of identifying assumptions. 
The term ``identification breakdown point'' can be found as early as \cite{horowitz1995identification} in the context of corrupted data.
One well-known example of breakdown analysis is \cite{altonji2005selection}, which considers linear regressions suffering from omitted variable bias and proposes measuring how strong selection on unobservables would need to be (relative to measured selection on observables) to attribute the entire estimated effect to selection. 
This idea was developed further in \cite{oster2019unobservable}, is widely used in empirical economics, and is an area of active research; see, e.g., \cite{masten2025effect}, \cite{diegert2025assessing}, and the references therein.
Breakdown analyses are also commonly used to evaluate the sensitivity of results to identifying assumptions made for causal inference, with recent examples including \cite{masten2020inference}, \cite{bonvini2022sensitivity}, \cite{rosenbaum2023sensitivity}, and \cite{spini2024robustness}.

This paper is not the first to notice the appeal of breakdown point analysis in the context of missing data. 
\cite{kline2013sensitivity} considers a setting with a missing scalar, propose measuring selection with the maximal Kolmogorov-Smirnov (KS) distance between the conditional distributions of complete and incomplete observations across all values of covariates, and advocates for ``reporting the minimal level of selection necessary to undermine a hypothesis,'' (p. 233). 
In their setting, one minus the KS distance can be interpreted as the proportion of the missing population assumed to be missing at random. 
The methodology proposed here has some notable advantages.
First, measuring selection with the maximal KS distance limits researchers to the case where only a scalar is missing, while measuring selection as proposed here allows any number of variables to be missing. 
Second, in a given setting it is easier to gauge whether the variables under study are likely to be good predictors of missingness than what share of the missing data is missing at random. 
This makes squared Hellinger a more natural measure of selection than KS distance. 
Which approach is more tractable will depend on the parameter of interest. 
\cite{kline2013sensitivity} derives sharp, closed form bounds to the conditional quantiles of the missing variable, and then frames the conclusion to be investigated in terms of those quantiles. 
This paper assumes the parameter of interest is identified with GMM and uses the model directly, but gives up closed form solutions.
In theory this could lead to computational difficulties, but the simulations in section \ref{Section: simulations} and application in section \ref{Section: application} present no issue. 


Estimation of the breakdown point remains tractable due to the use of $f$-divergences to measure selection. 
These divergences, defined and discussed in section \ref{Section: measuring selection and breakdown analysis, divergences}, have seen widespread use in the sensitivity analysis literature. 
\cite{christensen2023counterfactual} estimate the identified set of counterfactual predictions from structural models when the distribution of latent variables is allowed to vary within an $f$-divergence neighborhood of a given parametric specification. 
\cite{jin2022sensitivity} use $f$-divergences to characterize deviations from unconfoundedness in causal inference.
The most famous $f$-divergence may be Kullback-Leibler, which is used to study local model misspecification in \cite{bonhomme2022minimizing}, sensitivity to the choice of a Bayesian prior in \cite{ho2023global}, and breakdown of causal inference results in \cite{spini2024robustness}.


The remainder of this paper is structured as follows. 
Section \ref{Section: measuring selection and breakdown analysis} formalizes the setting, the proposed measure of selection, and the breakdown point. 
The dual problem is presented and discussed in section \ref{Section: duality}. 
Section \ref{Section: estimators and asymptotics} defines the estimator and states the main results on estimation and inference, which are proven in the Supplementary Material. 
Section \ref{Section: simulations} presents a simulation study investigating the finite sample performance of the estimators. 
The methodology is illustrated in section \ref{Section: application}, which contains estimates of the breakdown point of conclusions drawn from several randomized controlled trails suffering from attrition.
Section \ref{Section: conclusion} concludes.

%% file: Section_Setting.tex
\section{Measuring selection and breakdown analysis}
\label{Section: measuring selection and breakdown analysis}

Suppose the available data is the i.i.d. sample $\{(D_i, D_i Y_i, X_i)\}_{i=1}^n$, where $Z_i \equiv (Y_i, X_i)\in \mathbb{R}^{d_y} \times \mathbb{R}^{d_x}$ contains the variables of interest and $D_i \in \{0,1\}$ indicates whether $Y_i$ is observed. Note that $Y_i$ may be a vector, and $X_i$ may be empty. Variables are organized into the vectors $Y_i$ and $X_i$ based on whether they are occasionally missing; there is no need for $Y_i$ to be an outcome in the analysis. Let $p_D \equiv P(D = 1)$ denote the probability of observing $Y$, $P_1$ the distribution of $Z$ conditional on $D = 1$, and $P_0$ the distribution of $Z$ conditional on $D = 0$. $P_1$ and $P_0$ are called the \textit{complete case} and \textit{incomplete case} distributions respectively. The distribution of interest is the unconditional distribution of $Z$, given by $p_D P_1 + (1-p_D) P_0$. When $X$ is nonempty, the marginal distribution of $X$ conditional on $D = d$ is denoted $P_{dX}$. 

Two assumptions made below are worth highlighting when introducing the setting. First, it is assumed that $P_0$ is absolutely continuous with respect to $P_1$, meaning that for every set $A$ with $P((Y,X) \in A \mid D = 1) = 0$ one has $P((Y, X) \in A \mid D = 0) = 0$ as well. This assumption, denoted $P_0 \ll P_1$, facilitates measuring selection with a statistical divergence. It is natural in some settings and may be restrictive in others; see remark \ref{Remark: absolute continuity of P_0 wrt P_1} below for additional discussion. Second, $X$ is assumed to have the same finite support when $D = 0$ as when $D = 1$. This assumption can be relaxed, as discussed in remark \ref{Remark: always observed variables have finite support}.

To fix ideas, consider data collection via survey. $Y$ is a vector of data the survey hopes to collect, which is observed only if the recipient responds ($D = 1$). The survey's response rate, $p_D = P(D = 1)$, is essentially always less than one in practice. It is common for administrative data to provide basic information about a survey recipient (such as age, occupation, etc.), which is collected in $X$.

Analyses based on the complete observations may not convince researchers who worry that $P_0$ differs from $P_1$. Such concerns are common, as few settings plausibly satisfy the missing completely at random assumption. However, it is often similarly implausible that $P_0$ differs greatly from $P_1$. Researchers who convincingly argue that $P_0$ is not too different from $P_1$ can still convince their audience of conclusions drawn from an analysis of $P_1$.\footnote{In some cases, such as correctly specified regression models, it suffices that the conditional distributions $f_{Y \mid X = x, D = 0}(y \mid x)$ are the same as the identified $f_{Y \mid X = x, D = 1}(y \mid x)$. This weaker ``missing at random'' (MAR) assumption is also rarely plausible in practice, and analyses based on this assumption often rely heavily on the model being correctly specified.}

A quantitative measure of the difference between $P_1$ and $P_0$ is needed to make this argument formal and convincing. The statistics literature provides a natural solution in the form of \textit{divergences}: functions mapping two probability distributions to the nonnegative real line that take value zero if and only if the two distributions are the same. 
There are many such functions. To be useful as a measure of selection, a divergence should have a tractable interpretation, so that researchers can gauge whether a given amount of selection is reasonable for their setting.

\subsection{An interpretable measure of selection}
\label{Section: measuring selection and breakdown analysis, squared hellinger}

Missing data cause greater concern when researchers expect the variables of interest ($Z$) to be a good predictor of incompleteness ($D$). Consider again the example of data collection via survey. Researchers are rightfully more concerned about survey nonresponse when asking about the respondent's arrest record than when asking for opinions on recent television programming. People with criminal records may be less willing to answer questions about that record.\footnote{For example, \cite{brame2012cumulative} estimate the cumulative prevalence of arrest from ages 8 to 23 from a survey directly asking about prior arrests. The authors report upper and lower bounds derived by assuming the entire set of nonresponders had or had not been arrested, essentially the worst-case bounds advocated for by \cite{manski2005partial}.} This suggests that the distribution of responders may look quite different from the distribution of nonresponders, and that criminal records would be a good predictor of nonresponse.

To illustrate this more formally, let $f_1$ and $f_0$ be densities of $P_1$ and $P_0$ with respect to $p_D P_1 + (1-p_D)P_0$ respectively:
\begin{align*}
	&f_1(z) = \frac{P(D = 1 \mid Z = z)}{p_D}, &&f_0(z) = \frac{1-P(D=1 \mid Z = z)}{1-p_D}
\end{align*}
 An optimist may assume $D$ is independent of $Z$, implying that $P(D = 1 \mid Z = z) = P(D = 1) = p_D$ and $f_0 = f_1 = 1$. This would imply that $P_1$ and $P_0$ are the same distribution; i.e. the data are missing completely at random. In contrast, a pessimist may assume $D$ is close to a deterministic function of $Z$, allowing $Z$ to predict $D$ well. This would imply $P(D = 1 \mid Z = z)$ is close to $1$ or $0$ for many values of $z$, and that $f_1$ differs greatly from $f_0$.

As in the survey example, the setting often makes it clear whether $Z$ would be a good predictor of $D$. This heuristic is useful to identify and discuss selection concerns. The following lemma shows that measuring selection as the squared Hellinger distance between $P_0$ and $P_1$ captures this intuition, with larger values corresponding to $Z$ having greater capability of predicting $D$.\footnote{The Hellinger distance between probability measures $Q$ and $P$ is 
	\begin{equation*}
		H(Q, P) \equiv \left(\frac{1}{2} \int \left(\sqrt{\frac{dQ}{d\lambda}(z)} - \sqrt{\frac{dP}{d\lambda}(z)}\right)^2 d\lambda(z) \right)^{1/2}
	\end{equation*}
	where $\lambda$ is any measure dominating both $P$ and $Q$.}
\begin{restatable}{lemma}{lemmaSquaredHellingerInterpretation}
	\label{Lemma: squared hellinger interpretation}%
	\singlespacing
	
	Let $(Z, D) \in \mathbb{R}^{d_z} \times \{0,1\}$ be random variables with $p_D = P(D=1) \in (0,1)$. Let $Z \mid D = 1 \sim P_1$ and $Z \mid D = 0 \sim P_0$. Then
	\begin{equation}
		H^2(P_0, P_1) = 1 - \frac{E\left[\sqrt{\text{Var}(D \mid Z)}\right]}{\sqrt{\text{Var}(D)}} \label{Display: squared Hellinger interpretation}
	\end{equation}
	where the expectation is taken with respect to $p_D P_1 + (1-p_D) P_0$, the marginal distribution of $Z$.
\end{restatable}
All results are proven in the Supplementary Material. Equation \eqref{Display: squared Hellinger interpretation} states that the squared Hellinger distance between $P_0$ and $P_1$ is the expected percent of the standard deviation of $D$ reduced by conditioning on $Z$.  In the extreme case where $\text{Var}(D \mid Z) = \text{Var}(D)$, equation \eqref{Display: squared Hellinger interpretation} implies $H^2(P_0, P_1) = 0$ and the conditional distributions are the same. As the ability of $Z$ to predict $D$ grows, the variance of $D$ conditional on $Z$ decreases and $H^2(P_0, P_1)$ grows toward one. 

\begin{remark}
	\label{Remark: squared hellinger lower bound}
	It is shown in the Supplementary Material that
	\begin{equation*}
		H^2(P_0, P_1) = 1 - \frac{E\left[\sqrt{\text{Var}(D \mid Y, X)}\right]}{\sqrt{\text{Var}(D)}} \geq 1 - \frac{E[\sqrt{\text{Var}(D \mid X)}]}{\sqrt{\text{Var}(D)}} = H^2(P_{0X}, P_{1X})
	\end{equation*}
	where $P_{0X}$, $P_{1X}$ are the marginal distributions of $X$ conditional on $D = 0$ and $D = 1$ respectively. This lower bound on selection is identified from the sample, and motivates the common practice of comparing the distribution of $X$ conditional on $D=0$ with that of $X$ conditional on $D=1$; the distributions $P_0$ and $P_1$ can only be ``further'' apart.
\end{remark}

\subsection{Divergences}
\label{Section: measuring selection and breakdown analysis, divergences}

Squared Hellinger provides an intuitive measure of selection, but there are many other options. A function $d (\cdot \Vert \cdot)$ mapping two probability distributions $P$ and $Q$ to $\mathbb{R}$ is called a \textit{divergence} if $d(Q \Vert P) \geq 0$, with equality if and only if $P = Q$. Divergences need not be symmetric nor satisfy the triangle inequality. The set of \textit{$f$-divergences} are particularly well behaved. Given a convex function $f : \mathbb{R} \rightarrow [0,\infty]$ satisfying $f(t) = \infty$ for $t < 0$ and taking a unique minimum of $f(1) = 0$, the corresponding $f$-divergence is given by
\begin{align}
	d_f(Q \Vert P) \equiv \begin{cases}
		\int f\left(\frac{dQ}{dP}\right)dP & \text{ if } Q \ll P \\
		\infty & \text{ otherwise }
	\end{cases}, \label{Definition: f divergence}
\end{align}
where $Q \ll P$ denotes absolute continuity of $Q$ with respect to $P$. Many popular divergences are equal to $f$-divergences when $P$ dominates $Q$.\footnote{By convention, $f$ is specified on its \textit{effective domain}, the set $\text{dom}(f) \equiv \{t \in \mathbb{R} \; : \; f(t) < \infty\}$. The value of $f$ at $t = 0$ is set equal to $\lim_{t\rightarrow 0^+} f(t)$. Outside of $dom(f)$, $f$ takes the value $+\infty$.}

\begin{table}[H]
	\begin{small}
	\def\arraystretch{2.2}
	\caption{Common $f$-divergences}
	\label{Table: Common f divergences}
	\begin{center}
		\begin{tabular}{|l || l | l  | l |} \hline
			Name & Common formula & $f(t)$ & $\text{dom}(f)$ \\ \hline \hline
			Squared Hellinger & $H^2(Q, P) = \frac{1}{2}\int \left(\sqrt{\frac{dQ}{dP}(z)} - 1\right)^2 dP(z)$ & $f(t) = \frac{1}{2}(\sqrt{t} - 1)^2$ & $t \in [0, \infty)$ \\ \hline
			Kullback-Leibler (KL) & $KL(Q \Vert P) = \int \log\left(\frac{dQ}{dP}(z)\right)dQ(z)$ & $f(t) = t \log(t)-t+1$ & $t \in [0,\infty)$ \\ \hline
			``Reverse'' KL & $ KL(P \Vert Q) = \int \log\left(\frac{dP}{dQ}(z)\right)dP(z)$ & $f(t) = -\log(t)+t-1$ & $t \in (0, \infty)$ \\\hline
			Cressie-Read & -- & $f_\gamma(t) = \frac{t^\gamma - \gamma t + \gamma -1}{\gamma(\gamma - 1)}$ & -- \\ \hline
		\end{tabular}
	\end{center}
	\end{small}
\end{table}
Although squared Hellinger has intuitive appeal outlined in Section \ref{Section: measuring selection and breakdown analysis, squared hellinger}, the breakdown point analysis proposed in this paper remains tractable for any $f$-divergence listed in Table \ref{Table: Common f divergences}.\footnote{It is worth noting that the family of Cressie-Read divergences nests the other three as special cases. Squared Hellinger corresponds to $\frac{1}{2} f_{1/2}$. L'H\^opital's rule shows that Kullback-Leibler corresponds to $\lim_{\gamma \rightarrow 1} f_\gamma$ and Reverse Kullback-Leibler to $\lim_{\gamma \rightarrow 0} f_\gamma$. See \cite{broniatowski2012divergences} for additional discussion.} Precise assumptions regarding the $f$-divergence are collected in Assumption \ref{Assumption: setting} below.

\begin{remark}
	\label{Remark: absolute continuity of P_0 wrt P_1}
	Measuring selection with an $f$-divergence facilitates estimation and inference, as the space of distributions $Q$ with $d_f(Q \Vert P_1) < \infty$ corresponds to the set of densities with respect to $P_1$. In essence, measuring selection with an $f$-divergence assumes that $P_0$ is absolutely continuous with respect to $P_1$, denoted $P_0 \ll P_1$, as all distributions failing this requirement have infinite divergence from $P_1$.
	
	Absolute continuity is a natural assumption in some settings, but restrictive in others. For an example where $P_0 \ll P_1$ is natural, suppose $Y$ is a measure of time worked in a week measured in hours, obtained through a survey. Suppose the distribution $P_1$ displays positive mass at $Y = 0$ hours and $Y = 40$ hours, and is otherwise continuous. The assumption $P_0 \ll P_1$ here is natural, as it allows $P_0$ to have an atom of any size at $Y = 0$ and $Y = 40$ while ruling out distributions with positive mass at other points. For an example where absolute continuity fails, suppose $Y$ represents wage data where top coded observations are treated as missing. In this example, $P_1$ puts mass one below the top code while $P_0$ puts mass one above the top code, and $P_0 \not \ll P_1$.
\end{remark}

\subsection{Breakdown analysis in models identified with GMM}
\label{Section: measuring selection and breakdown analysis, breakdown analysis}

Suppose a preliminary analysis supports an alternative hypothesis $H_1$ over a null hypothesis $H_0$. For example, such an analysis may be based on the complete observations assuming MCAR, or using imputation and assuming $Y$ is MAR conditional on $X$. The breakdown point is the minimum amount of selection needed to overturn such a conclusion. When selection is measured in terms of the squared Hellinger distance, the breakdown point translates the claim that $H_0$ is true into a claim about the ability of $Z$ to predict $D$. Specifically, if $H_0$ were true then $1 - \frac{E[\sqrt{\text{Var}(D \mid Z)}]}{\sqrt{\text{Var}(D)}}$ would be weakly larger than the breakdown point. If this is implausible, then $H_0$ is similarly implausible.

This section formalizes this idea for models identified with the generalized method of moments (GMM). Suppose the parameter of interest $\beta \in \textbf{B} \subseteq \mathbb{R}^{d_b}$ is characterized as the unique solution to a finite set of moment conditions,
\begin{equation*}
	E[g(Z, \beta)] = 0 \in \mathbb{R}^{d_g}
\end{equation*}
where the expectation is taken with respect to the unconditional distribution, $p_D P_1 + (1-p_D) P_0$. The conclusion to be investigated is that $\beta$ falls outside a particular set $\textbf{B}_0 \subset \textbf{B}$, motivating the null and alternative hypotheses
\begin{align*}
	&H_0 \; : \; \beta \in \textbf{B}_0, &&H_1 \; : \; \beta \in \textbf{B} \setminus \textbf{B}_0
\end{align*}

Recall that the observed data is $\{(D_i, D_i Y_i, X_i)\}_{i=1}^n$, where $D_i = \mathbbm{1}\{Y_i \text{ is observed}\}$. The sample identifies $P_1$, $p_D$, and $P_{0X}$. A hypothetical distribution of the incomplete observations $Q$ \textit{rationalizes} the parameter $b$ if it has the identified marginal distribution of $X$, $Q_X = P_{0X}$, and the implied unconditional distribution $p_D P_1 + (1-p_D) Q$ solves the moment conditions for $b$. The set of such distributions implying finite selection is
\begin{equation}
	\textbf{P}^b \equiv \left\{Q \; : \; Q \ll P_1, \; Q_X = P_{0X}, \; p_D E_{P_1}[g(Z, b)] + (1-p_D) E_Q[g(Z, b)] = 0\right\}. \label{Definition: Set of distributions rationalizing parameter}
\end{equation}
The \textit{breakdown point} $\delta^{BP}$ is the minimum selection needed to rationalize the null hypothesis:
\begin{equation}
	\delta^{BP} \equiv \inf_{b \in \textbf{B}_0} \inf_{Q \in \textbf{P}^b} d_f(Q \Vert P_1), \label{Definition: breakdown point}
\end{equation}
where the infimum over the empty set is understood to be infinity. A simple example illustrates the idea.
\begin{example}
	\label{Example: Squared Hellinger Expectation}
	\singlespacing
	Let $Y \in \mathbb{R}$ and $\beta = E[Y] = p_D E_{P_1}[Y] + (1-p_D)E_{P_0}[Y]$. Let $p_D = 0.7$ and $P_1$ 
	be $\mathcal{U}[0,1]$. The claim to support is $H_1 \; : \; \beta > 0.4$, and selection is measured with squared Hellinger. $\textbf{P}^b$ is the set of continuous distributions on $[0,1]$ with expectation $\frac{b - p_D/2}{1-p_D}$, so that $Q \in \textbf{P}^b$ implies 
	\begin{equation*}
		p_D E_{P_1}[Y] + (1-p_D)E_Q[Y] = \frac{p_D}{2} + (1-p_D) \frac{b - p_D/2}{1-p_D} = b
	\end{equation*}
	The inner minimization in display \eqref{Definition: breakdown point} chooses the distribution that minimizes selection while rationalizing $b$. The outer minimization chooses the parameter that minimizes selection while rationalizing $H_0 \; : \; \beta \leq 0.4$. Unsurprisingly, the outer minimization is solved by $b = 0.4$. The breakdown point $\delta^{BP}$ is slightly above $0.2$. A researcher convinced $H^2(P_0, P_1)$ is less than $0.2$ should conclude $\beta > 0.4$. 
	\begin{figure}[H]
		\caption{$\inf_{Q \in \textbf{P}^b} d_f(Q \Vert P_1)$ and $p_D P_1 + (1-pD)Q^*$, where $Q^* \in \textbf{P}^{0.4}$ minimizes selection.}
		\label{Figure: numerical example}
		\vspace{0.5 cm}
		\begin{center}
			\includegraphics[scale=0.61, trim={0 0 0 83}, clip]{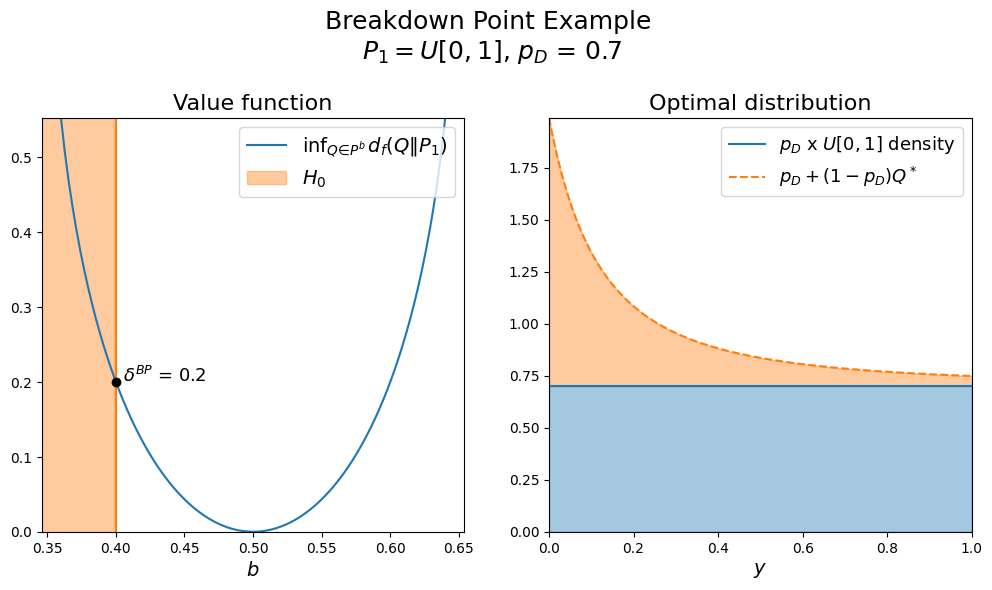}
		\end{center}
	\end{figure}
\end{example}

\begin{remark}
	\label{Remark: 0.2 is a big number}
	
	Example \ref{Example: Squared Hellinger Expectation}, while simple, can help anchor expectations for the values of $\delta^{BP}$ when squared Hellinger is used to measure selection. It is clear from observation of the left panel of Figure \ref{Figure: numerical example} that the value of $\delta^{BP}$ would rise quite quickly as the $\beta$ in $H_0 : \beta \leq 0.4$ shrinks toward the lower Manksi bound of $0.35$. $\beta$ would need to be quite close to the theoretical bound to see breakdown point values of $0.5$ or higher. The right-hand side panel of Figure \ref{Figure: numerical example} shows the distribution $Q^*$ closest to $P_1$ in terms of squared Hellinger that rationalizes an unconditional mean of $0.4$. $Q^*$ is quite different from the uniform distribution $P_1$, and in many settings where the complete data follows a uniform distribution it would be implausible that the missing observations follow such a distinct distribution. This suggests a breakdown point of $0.2$ should be treated as quite large for a squared Hellinger breakdown point. Depending on the context, even smaller values could provide reassurance to many researchers.
\end{remark}

Breakdown analysis can also be framed as an exercise in partial identification, as in \cite{kline2013sensitivity},  \cite{masten2020inference}, and \cite{diegert2025assessing}. In this framing, the researcher considers assumptions of the form $d_f(P_0, P_1) \leq \delta$ for some $\delta > 0$, which continuously relax the assumption $P_0 = P_1$. The identified set for $\beta$ grows with $\delta$. As long as the identified set is a subset of $\textbf{B} \setminus \textbf{B}_0$, it is clear the researcher's conclusion holds. The breakdown point $\delta^{BP}$ can then be defined as either the largest $\delta$ for which the identified set is contained in $\textbf{B} \setminus \textbf{B}_0$, or the smallest $\delta$ for which the identified set has nontrivial intersection with $\textbf{B}_0$ (the latter of which corresponds to the definition given in display \eqref{Definition: breakdown point}). For further discussion of this equivalent framing of the breakdown point, see the Supplementary Material. 

The remainder of this paper constructs a $\sqrt{n}$-consistent and asymptotically normal estimator of $\delta^{BP}$, and constructs a lower confidence interval for $\delta^{BP}$. Researchers working with partially complete datasets should discuss the plausible amount of selection in their setting, and report the point estimate and the lower confidence interval for $\delta^{BP}$ for each asserted conclusion. This will make plain to readers which conclusions are more sensitive to missing data concerns, and whether crucial results are sufficiently robust.

\subsection{Preview of results}
\label{Section: measuring selection and breakdown analysis, preview of results}

Estimation of $\delta^{BP}$ proceeds by separating the optimizations in \eqref{Definition: breakdown point}. Define the \textit{primal problem}
\begin{equation}
	\nu(b) \equiv \inf_{Q\in \textbf{P}^b} d_f(Q \Vert P_1), \label{Definition: primal problem}
\end{equation}
and notice that $\delta^{BP} = \inf_{b \in \textbf{B}_0} \nu(b)$. The first step is to estimate the value function $\nu$ over a set $B \subseteq \textbf{B}$ large enough that $\inf_{b \in \textbf{B}_0} \nu(b) = \inf_{b \in B \cap \textbf{B}_0} \nu(b)$, while the second step estimates $\delta^{BP}$ through a simple plug-in estimator.

The primal problem is an infinite dimensional convex optimization problem over the space of probability distributions, but one that is very well studied in convex analysis. In particular, when $\textbf{P}^b$ defined in display \eqref{Definition: Set of distributions rationalizing parameter} is characterized by a finite number of moment conditions, the primal problem has a well behaved, finite-dimensional  dual problem with the same value function \citep{borwein1991duality, borwein1993partially, csiszar1999mem, broniatowski2006minimization}. Section \ref{Section: duality} discusses this dual problem and the assumptions needed to make use of it. Under regularity conditions discussed in Section \ref{Section: estimators and asymptotics}, sample analogue estimators of $\nu$ based on this dual problem are uniformly consistent and asymptotically Gaussian on compact subsets of the parameter space. Differentiability of the infimum then implies convergence in distribution of the plug-in estimator.

To conclude this section, Assumption \ref{Assumption: setting} collects conditions on the setting, the GMM model, and the $f$-divergence used to measure selection.

\begin{restatable}[Setting]{assumption}{assumptionSetting}
	\label{Assumption: setting}%
	\singlespacing
	$\{(D_i, D_i Y_i, X_i)\}_{i=1}^n$ is an i.i.d. sample from a distribution satisfying 
	\begin{enumerate}[label=(\roman*)]
		\item $p_D = P(D=1) \in (0,1)$, \label{Assumption: setting, missing data}
		\item $X \mid D = 1$ and $X \mid D = 0$ have the same finite support $\{x_1, \ldots, x_K\}$, \label{Assumption: setting, support of always observed variables}
		\item $E\left[\sup_{b \in \boldsymbol{B}} \lVert g(Z, b) \rVert \mid D = 1\right] < \infty$, where $Z = (Y,X)$, \label{Assumption: setting, GMM}
		\item $P_0 \ll P_1$, and \label{Assumption: setting, absolute continuity}
		\item $f : \mathbb{R} \rightarrow [0,\infty]$ is closed, proper, strictly convex, essentially smooth, takes its unique minimum of $f(t) = 0$ at $t=1$, and satisfies $f(t) = \infty$ for all $t < 0$. The interior of $\text{dom}(f) \equiv \{t \in \mathbb{R} \; : \; f(t) < \infty\}$, denoted $(\ell, u)$, satisfies $\ell < 1 < u$, and $f$ is twice continuously differentiable on $(\ell, u)$. \label{Assumption: setting, divergence}
	\end{enumerate}
\end{restatable}
The finite support condition, assumption \ref{Assumption: setting} \ref{Assumption: setting, support of always observed variables}, simplifies estimation and inference by ensuring that $\textbf{P}^b$ is characterized by a finite number of moments. This assumption is not needed to define the breakdown point, but is used to ensure the duality results of section \ref{Section: duality} hold and can be used to define the estimator. This assumption can also be relaxed. In settings where assumption \ref{Assumption: setting} \ref{Assumption: setting, support of always observed variables} fails, researchers can still perform a breakdown point analysis through a conservative procedure described in remark \ref{Remark: always observed variables have finite support} below. 

Condition \ref{Assumption: setting, divergence} ensures the $f$-divergence used to measure selection is well behaved, and is satisfied by every divergence in Table \ref{Table: Common f divergences}. In particular, strict convexity of $f$ ensures the primal problem \eqref{Definition: primal problem} has a unique solution ($P_1$-almost surely). $f$ is required to be essentially smooth to ensure the dual problem has a unique solution. The requirements that $f(x)$ take a unique minimum of $0$ at $x=1$ and $f(x) = \infty$ for $x < 0$ ensures that $d_f(Q \Vert P)$ is a well-defined $f$-divergence.

%% file: Section_Duality.tex
\section{Duality}
\label{Section: duality}

As defined in display \eqref{Definition: primal problem}, $\nu(b)$ is the value function of an infinite dimensional convex optimization problem. Fortunately, when selection is measured with an $f$-divergence, this minimization becomes a well-studied problem known by various names: maximal entropy \citep{csiszar1999mem}, partially finite programming \citep{borwein1991duality}, or $f$-divergence projection \citep{broniatowski2006minimization}. The convex analysis results in these papers connect the primal problem in display \eqref{Definition: primal problem} to a finite dimensional dual problem that is much easier to study and estimate. Under mild conditions, the value function of this dual problem coincides with the value function of the primal.

To state the dual problem, first note that the primal can be viewed as a problem over the set of densities with respect to $P_1$:
\begin{align*}
	\nu(b) = &\inf_q E\left[f(q(Y, X)) \mid D = 1\right] \\
	&\text{s.t. } E[h(Y, X, b) q(Y,X) \mid D =1] = c(b) 
\end{align*}
where
\begin{align}
	&h(y, x, b) \equiv 
	\begin{pmatrix}
		g(y, x, b) \\
		\mathbbm{1}\{x = x_1\} \\
		\vdots \\
		\mathbbm{1}\{x = x_K\}
	\end{pmatrix},
	&&c(b) \equiv 
	\begin{pmatrix}
		\frac{-p_D}{1-p_D} E[g(Y, X, b) \mid D = 1] \\
		P(X = x_1 \mid D = 0) \\
		\vdots \\
		P(X = x_K \mid D = 0)
	\end{pmatrix}, \label{Definition: constraint notation, h(z,b) and c(b)}
\end{align}
As shown in \cite{borwein1991duality}, the dual problem corresponding to \eqref{Definition: primal problem} is given by
\begin{equation}
	V(b) \equiv \sup_{\lambda \in \mathbb{R}^{d_g + K}} \lambda^\intercal c(b) - E\left[f^*\left(\lambda^\intercal h(Y, X, b)\right) \mid D = 1\right] \label{Definition: dual problem}
\end{equation}
where $f^*$ is the convex conjugate of $f$, given by $f^*(r) \equiv \sup_{t \in\mathbb{R}}\{rt - f(t)\}$. For convenience, table \ref{Table: Common f divergence conjugates} summarizes the convex conjugate for several common divergences. 

\begin{table}[H]
	\def\arraystretch{1.6}
	\caption{Common $f$-divergence conjugates and effective domains}
	\label{Table: Common f divergence conjugates}
	\begin{center}
		\begin{tabular}{|l || l | l | l | l |} \hline
			Name & $f(t)$ & $\ell$, $u$ & $f^*(r)$ & $\ell^*$, $u^*$  \\ \hline \hline
			Squared Hellinger  & $\frac{1}{2}(\sqrt{t} - 1)^2$ & $\ell = 0$, $u = \infty$ & $\frac{1}{2}\left(\frac{1}{1-2r} - 1\right)$ & $\ell^* = -\infty$, $u^* = \frac{1}{2}$  \\ \hline
			Kullback-Leibler (KL) & $t\log(t) - t + 1$ & $\ell = 0$, $u = \infty$ & $\exp(r) - 1$ & $\ell^* = -\infty$, $u^* = \infty$\\ \hline
			``Reverse'' KL &  $-\log(t) + t - 1$ & $\ell = 0$, $u = \infty$ & $-\log(1-r)$ & $\ell^* = -\infty$, $u^* = 1$ \\\hline
			Cressie-Read & $f_\gamma(t) = \frac{t^\gamma - \gamma t + \gamma -1}{\gamma(\gamma - 1)}$ & --  & $\frac{1}{\gamma}(\gamma r - r + 1)^{\frac{\gamma}{\gamma - 1}} - \frac{1}{\gamma}$ & -- \\ \hline
		\end{tabular}
	\end{center}
\end{table}

\begin{remark}
	\label{Remark: ensuring a probability density}
	To ensure $q$ corresponds to a probability density, the constraints must enforce $\int q(z) dP(z) = 1$. This is implied by the constraints ensuring $Q_X = P_{0X}$ when $X$ is present. If $X$ is empty, set $h(z, b) = \begin{pmatrix} g(z, b)^\intercal & 1 \end{pmatrix}^\intercal \in \mathbb{R}^{d_g + 1}$ and $c(b) = \begin{pmatrix} \frac{-p_D}{(1-p_D)} E[g(Y, X, b) \mid D = 1]^\intercal & 1 \end{pmatrix}^\intercal \in \mathbb{R}^{d_g + 1}$ to ensure $q$ integrates to $1$.
\end{remark}

\subsection{Weak and strong duality}
\label{Subsection: weak and strong duality}

Assumption \ref{Assumption: setting} suffices to show $V(b) \leq \nu(b)$. This fact is known as \textit{weak duality}, and implies that 
\begin{equation}
	\inf_{b \in B \cap \textbf{B}_0} V(b) \leq \inf_{b \in B \cap  \textbf{B}_0} \nu(b) = \delta^{BP}
	\label{Display: weak duality}
\end{equation}
for any $B \subseteq \textbf{B}$. This inequality shows that using the dual problem for estimation of the breakdown point is at worst conservative. If $\inf_{b \in B \cap \textbf{B}_0} V(b)$ is large enough to assuage selection concerns, researchers are assured that the breakdown point can only be larger.

Assuming only slightly more ensures \textit{strong duality} holds, that is, $V(b) = \nu(b)$. Recall from Assumption \ref{Assumption: setting} \ref{Assumption: setting, divergence} that the interior of $\text{dom}(f) = \{t \in \mathbb{R} \; : \; f(t) < \infty\}$ is denoted $(\ell, u)$.  

\begin{restatable}[Strong duality]{assumption}{assumptionStrongDuality}
	\label{Assumption: strong duality}%
	\singlespacing
	
	$B \subseteq \textbf{B}$ is convex, compact, and satisfies $\inf_{b \in \textbf{B}_0} \nu(b) = \inf_{b \in B \cap \textbf{B}_0} \nu(b)$. Furthermore, for each $b \in B$,
	\begin{enumerate}[label=(\roman*)]
		\item there exists $Q^b \in \textbf{P}^b$ such that $\ell < \frac{\partial Q^b}{\partial P_1}(z) < u$, almost surely $P_1$, and \label{Assumption: strong duality, constraint qualification}
		\item $\lambda(b)$ solving \eqref{Definition: dual problem} is in the interior of $\{\lambda \; : \; E[\lvert f^*(\lambda^\intercal h(Z, b)) \rvert \mid D = 1] < \infty\}$. \label{Assumption: strong duality, interior dual solution}
	\end{enumerate}
\end{restatable}
That strong duality holds under these conditions is a well-known result.\footnote{To the authors knowledge, the first to show strong duality holds under similar conditions was \cite{borwein1991duality}. The proof of theorem \ref{Theorem: strong duality}, found in the Supplementary Material, uses a result due to \cite{csiszar1999mem}.}

\begin{restatable}[Strong duality]{theorem}{theoremStrongDuality}
	\label{Theorem: strong duality}
	\singlespacing
	
	Suppose assumptions \ref{Assumption: setting} and \ref{Assumption: strong duality} hold. Then for each $b \in B$, $\nu(b) = V(b)$, with dual attainment. 
\end{restatable}

The first order condition of the dual problem \eqref{Definition: dual problem} provides intuition. Exchanging expectation and differentiation, the first order condition is 
\begin{align*}
	\begin{pmatrix}
		\frac{-p_D}{1-p_D} E_{P_1}[g(Y, X, b)] \\
		P(X = x_1 \mid D = 0) \\
		\vdots \\
		P(X = x_K \mid D = 0)
	\end{pmatrix} = 
	E_{P_1}\left[(f^*)'\left(\lambda(b)^\intercal h(Y, X, b)\right)
	\begin{pmatrix}
		g(Y, X, b) \\
		\mathbbm{1}\{X = x_1\} \\
		\vdots \\
		\mathbbm{1}\{X = x_K\}
	\end{pmatrix}
	\right]
\end{align*}
where $\lambda(b) \in \mathbb{R}^{d_g + K}$ solves the dual problem. Consider $(f^*)'(\lambda(b)^\intercal h(y, x, b))$ as a density with respect to $P_1$. 
Notice that the first $d_g$ equations of the first order condition ensure $p_D E_{P_1}[g(Y, X, b)] + (1-p_D) E_{P_1}[(f^*)'\left(\lambda(b)^\intercal h(Y, X, b)\right)g(Y, X, b)] = 0$, while the remaining $K$ equalities ensure the marginal distribution of $X$ matches $P_{0X}$. In fact, the proof of theorem \ref{Theorem: strong duality} shows that under assumptions \ref{Assumption: setting} and \ref{Assumption: strong duality}, $(f^*)'\left(\lambda(b)^\intercal h(y, x, b)\right)$ is the $P_1$-density of the solution to the primal problem. 

Assumption \ref{Assumption: strong duality} ensures the set on which $\nu$ is estimated is large enough to estimate the breakdown point, but not so large as to contain parameter values that cannot be rationalized with a well behaved $P_1$-density. To illustrate, consider again example \ref{Example: Squared Hellinger Expectation}. $Y$ is a scalar, $\beta = E[Y] = p_D E_{P_1}[Y] + (1-p_D) E_{P_0}[Y]$, and $P_1$ is $\mathcal{U}[0,1]$. For tractability suppose that Kullback-Leibler is used to measure selection. Since $P_0$ takes values on $[0,1]$, the Manski bounds for $\beta$ are $\left[\frac{p_D}{2}, 1 - \frac{p_D}{2}\right]$. The Supplementary Material shows that strong duality is satisfied whenever $b \in \left(\frac{p_D}{2}, 1 - \frac{p_D}{2}\right)$. Thus for this example, $B$ can be any convex, compact set in the interior of the Manski bounds.

\begin{remark}
	\label{Remark: always observed variables have finite support}
	
	Assumption \ref{Assumption: setting} \ref{Assumption: setting, support of always observed variables} asks that $X$ be finitely supported, ensuring that $\textbf{P}^b$ is characterized by a finite number of moments and hence that the dual problem \eqref{Definition: dual problem} is finite dimensional. In settings where assumption \ref{Assumption: setting} \ref{Assumption: setting, support of always observed variables} fails and $X$ is not finitely valued, one can still conduct a conservative breakdown point analysis. Specifically, requiring $Q_X$ match a finite number of moments of $P_{0X}$ will estimate a value no larger than $\delta^{BP}$. If this value is large enough to assuage missing data concerns, the researcher is assured the breakdown point is weakly larger. 
	
	To illustrate, consider requiring that $Q_X$ match the first moment of $P_{0X}$. Define  
	\begin{align*}
		&\tilde{h}(z, b) = \tilde{h}(y, x, b) = 
		\begin{pmatrix}
			g(y, x, b) \\
			x \\
			1
		\end{pmatrix},
		&&\tilde{c}(b) = 
		\begin{pmatrix}
			\frac{-p_D}{1-p_D} E[g(Y,X,b) \mid D = 1] \\
			E[X \mid D = 0] \\
			1
		\end{pmatrix}
	\end{align*}
	and consider the value of the problem 
	\begin{equation*}
		\tilde{V}(b) \equiv \sup_{\lambda \in \mathbb{R}^{d_g + d_x}} \lambda^\intercal \tilde{c}(b) - E\left[f^*\left(\lambda^\intercal \tilde{h}(Y, X, b)\right) \mid D = 1 \right].
	\end{equation*}
	This is the dual problem corresponding to the primal minimization problem $\inf_{Q \in \tilde{\textbf{P}}^b} d_f(Q \Vert P_1)$, where $\tilde{\textbf{P}}^b$ is the set of distributions that zero the moment conditions and match the first moment of $X$:
	\begin{equation*}
		\tilde{\textbf{P}}^b \equiv \left\{Q \; : \; Q \ll P_1, \; E_Q[X] = E_{P_{0X}}[X], \; p_D E_{P_1}[g(Y, X, b)] + (1-p_D) E_{Q}[g(Y, X, b)] = 0\right\}
	\end{equation*}
	The minimization problem $\inf_{Q \in \tilde{\textbf{P}}^b} d_f(Q \Vert P_1)$ is less constrained than the problem defining $\nu(b)$ in \eqref{Definition: primal problem}, and so has a lower value function. Minimizing $\tilde{V}(b)$ over $b \in B \cap \textbf{B}_0$ will therefore attain a lower value than $\delta^{BP}$. If $\inf_{b \in B \cap \textbf{B}_0} \tilde{V}(b)$ is large enough to assuage selection concerns, the reader is assured that $\delta^{BP}$ could only be larger. Moreover, the statistical properties of the estimator studied in section \ref{Section: estimators and asymptotics} are essentially unchanged when replacing $h$ and $c$ with $\tilde{h}$ and $\tilde{c}$ respectively. 
	
	As \cite{borwein1993failure} shows by counterexample, infinite dimensional analogues of \eqref{Definition: primal problem} and \eqref{Definition: dual problem} can fail to satisfy strong duality. \cite{borwein1993failure} also shows that duality can be restored by relaxing or penalizing the primal problem, and taking appropriate limits. This suggests another promising approach to relaxing assumption \ref{Assumption: setting} \ref{Assumption: setting, support of always observed variables}, but would considerably complicate estimation and so is left for future research.
\end{remark}

%% file: Section_Asymptotics.tex
\section{Estimation}
\label{Section: estimators and asymptotics}

Assumptions \ref{Assumption: setting} and \ref{Assumption: strong duality} are maintained throughout the remainder of the paper. Accordingly, the notation $\nu$ will be used for the value function of the dual problem as well. 

\subsection{The estimator}
\label{Section: estimators and asymptotics, the estimator}

The sample analogue of the dual problem provides an estimator of the value function, and suggests a simple plug-in estimator of the breakdown point. The asymptotic properties of these estimators are easier to study if the objective of the dual problem is expressed with a single unconditional expectation, which comes at the cost of additional notation. 

First define the matrix $J(D) = \begin{bmatrix} -D I_{d_g} & 0 \\ 0 & (1-D) I_K \end{bmatrix}$ where $I_{d_g}$ and $I_K$ are identity matrices. Notice that $E\left[\frac{J(D) h(DY, X, b)}{(1-p_D)}\right] = c(b)$ and
\begin{equation}
	\nu(b) = \sup_{\lambda \in \mathbb{R}^{d_g + K}} E\left[\frac{\lambda^\intercal J(D) h(DY, X, b)}{1-p_D} - \frac{D f^*\left(\lambda^\intercal h(DY, X, b)\right)}{p_D} \right]. \label{Definition: dual problem unconditional expectation}
\end{equation}
Define 
\begin{equation}
	\varphi(D, DY, X, b, \lambda, p) \equiv \frac{\lambda^\intercal J(D) h(DY, X, b)}{1-p} - \frac{D}{p} f^*(\lambda^\intercal h(DY, X, b)), \label{Definition: dual objective integrand}
\end{equation}
and observe that the dual problem is $\sup_{\lambda \in \mathbb{R}^{d_g + K}} E[\varphi(D, DY, X, b, \lambda, p_D)]$. The estimator of the value function 
is defined pointwise by
\begin{equation}
	\hat{\nu}_n(b) \equiv \sup_{\lambda \in \mathbb{R}^{d_g + K}} \frac{1}{n} \sum_{i=1}^n \varphi(D_i, D_i Y_i, X_i, b, \lambda, \hat{p}_{D,n}), \label{Definition: estimator of value function}
\end{equation}
where $\hat{p}_{D,n} \equiv \frac{1}{n}\sum_{i=1}^n D_i$ estimates $p_D$. Finally, $\hat{\delta}_n^{BP} \equiv \inf_{b \in B \cap \textbf{B}_0} \hat{\nu}_n(b)$ estimates the breakdown point.

\subsection{Asymptotic normality}
\label{Section: estimators and asymptotics, asymptotic normality}

The following assumption suffices for $\hat{\delta}_n^{BP}$ to be $\sqrt{n}$-consistent and asymptotically normal. First observe that the estimands $\theta_0(b) = (\nu(b), \lambda(b), p_D)$ solve the moment conditions $E[\phi(D, DY, X, b, \theta_0(b))] = 0$, where
\begin{equation}
	\phi(D, DY, X, b, \theta) = \phi(D, DY, X, b, v, \lambda, p) = 
	\begin{pmatrix}
		\varphi(D, DY, X, b, \lambda, p) - v \\
		\nabla_\lambda \varphi(D, DY, X, b, \lambda, p) \\
		D - p
	\end{pmatrix}, \label{Definition: Z-estimator integrand}
\end{equation}
Let $\text{Gr}(\theta_0) \equiv \left\{(b, \theta_0(b)) \; : \; b \in B\right\}$ denote the graph of $\theta_0$. For $\eta > 0$, the closed $\eta$-expansion about this graph is $\text{Gr}(\theta_0)^\eta \equiv \left\{(b, \theta) \in B \times \mathbb{R}^{d_g + K + 2} \; : \; \inf_{(b', \theta') \in \text{Gr}(\theta_0)} \lVert (b,\theta) - (b', \theta') \rVert \leq \eta\right\}$. 

\begin{restatable}[Estimation]{assumption}{assumptionEstimation}
	\label{Assumption: estimation}%
	\singlespacing
	
	Suppose that 
	\begin{enumerate}[label=(\roman*)]
		\item $\textbf{B}_0$ is closed, \label{Assumption: estimation, closed null hypothesis}
		
		\item $\min_{b \in B \cap \textbf{B}_0} \nu(b)$ has a unique solution, \label{Assumption: estimation, unique solution}
		
		\item the matrix $E[h(Y, X, b)h(Y, X, b)^\intercal \mid D = 1]$ is nonsingular for each $b \in B$, \label{Assumption: estimation, nonsingular second moments}
		
		\item $g(y, x, b)$ is continuously differentiable with respect to $b$ for each $(y,x)$, and \label{Assumption: estimation, continuously differentiable moment functions}
		
		\item there exists a convex, compact set $\Theta^B$ containing $\text{Gr}(\theta_0)^\eta$ for some $\eta > 0$ satisfying
		\begin{align*}
			&E\left[\sup_{(b, \theta) \in \Theta^B} \lVert \phi(D, DY, X, b, \theta) \rVert^2\right] < \infty &&\text{ and } &&E\left[\left(\sup_{(b,\theta) \in \Theta^B} \lVert \nabla_{(b,\theta)} \phi(D, DY, X, b, \theta) \rVert\right)^2\right] < \infty.
		\end{align*}
		\label{Assumption: estimation, moment conditions}
	\end{enumerate}
\end{restatable}

As previewed in section \ref{Section: measuring selection and breakdown analysis, preview of results}, $\hat{\delta}_n^{BP}$ is viewed as a two-step estimator where $\hat{\nu}_n$ estimates $\nu$ in the first step, and $\hat{\delta}_n^{BP} = \inf_{b \in B \cap \textbf{B}_0} \hat{\nu}_n(b)$ is a plug-in estimator for $\delta^{BP} = \inf_{b \in B \cap \textbf{B}_0} \nu(b)$. Conditions \ref{Assumption: estimation, nonsingular second moments}, \ref{Assumption: estimation, continuously differentiable moment functions}, and \ref{Assumption: estimation, moment conditions} imply $\sqrt{n}(\hat{\nu}_n - \nu)$ converges weakly in the space of bounded functions on $B$, to a limiting process that is almost surely continuous.
This is shown by linearizing $0 = \frac{1}{n}\sum_{i=1}^n \phi(D_i, D_i Y_i, X_i, b, \hat{\theta}_n(b))$ uniformly over $b \in B$. Conditions \ref{Assumption: estimation, closed null hypothesis} and \ref{Assumption: estimation, unique solution} ensure minimization over $B \cap \textbf{B}_0$ is a (Hadamard) differentiable map on the set of continuous functions of $B$. The delta method then implies $\sqrt{n}(\hat{\delta}_n^{BP} - \delta^{BP})$ converges in distribution to a normal distribution. 

Assumption \ref{Assumption: estimation} \ref{Assumption: estimation, closed null hypothesis} and \ref{Assumption: estimation, continuously differentiable moment functions} are easily verified by inspection of $\textbf{B}_0$ and $g$ respectively. Conditions \ref{Assumption: estimation, nonsingular second moments} and \ref{Assumption: estimation, moment conditions} are similar to conditions required of generalized empirical likelihood estimators (see, e.g., \cite{antoine2021robust} assumption 1 (v) and assumption 3 (iv), (vii)). Assumption \ref{Assumption: estimation} \ref{Assumption: estimation, unique solution} deserves additional scrutiny. When $\textbf{B}_0$ is a convex set, condition \ref{Assumption: estimation, unique solution} holds when $\nu$ is a strictly convex function. The following lemma shows that this is the case when $g(y,x,b)$ describes a linear model with the outcome being the only missing data value. 
\begin{restatable}[Convex value function, linear models]{lemma}{lemmaConvexDualValueFunctionLinearModels}
	\label{Lemma: linear models imply a convex value function}
	\singlespacing
	
	Suppose assumptions \ref{Assumption: setting} and \ref{Assumption: strong duality} hold, the sample is $\{D_i, D_i Y_i, X_{i1}, X_{i2}\}_{i=1}^n$ where $Y_i \in \mathbb{R}$, $X_{i1} \in \mathbb{R}^{d_{x1}}$, and $X_{i2} \in \mathbb{R}^{d_{x2}}$, and the parameter $\beta$ is identified by 
	\begin{equation*}
		E[(Y - X_1^\intercal \beta) X_2] = 0
	\end{equation*}
	Then $\hat{\nu}_n$ and $\nu$ are convex. If in addition $E[X_2X_1^\intercal]$ has full column rank, then $\nu$ is strictly convex.
\end{restatable}

\noindent Lemma \ref{Lemma: linear models imply a convex value function} covers instrumental variable models directly, and ordinary least squares as a special case (by setting $X_2 = X_1$). It also covers parameters of the form $\beta = E[\tilde{g}(Y, X)]$, because the OLS regression of $\tilde{g}(Y,X)$ on a constant recovers $E[\tilde{g}(Y, X)]$. Simulation evidence presented in the Supplementary Material suggests data generating processes and models not covered by lemma \ref{Lemma: linear models imply a convex value function} also produce convex $\nu$. Remark \ref{Remark: allowing for multiple minimizers} below discusses an approach to relaxing assumption \ref{Assumption: estimation} \ref{Assumption: estimation, unique solution}, at the cost of additional complexity.

Theorem \ref{Theorem: asymptotic normality} below formally states the convergence in distribution result along with consistency of an estimator of the asymptotic variance. The variance depends on the Jacobian term $\Phi(b) \equiv E[\nabla_\theta \phi(D, DY, X, b, \theta_0(b))]$, which is estimated with
\begin{equation}
	\hat{\Phi}_n(b) \equiv \frac{1}{n}\sum_{i=1}^n \nabla_{\theta} \phi(D, DY, X, b, \hat{\theta}_n(b)), \label{Definition: estimator of asymptotic variance}
\end{equation}
where $\hat{\theta}_n(b) \equiv (\hat{\nu}_n(b), \hat{\lambda}_n(b), \hat{p}_{D,n})$ and $\hat{\lambda}_n(b) \equiv \argmax_{\lambda \in \mathbb{R}^{d_g + K}} \frac{1}{n}\sum_{i=1}^n \varphi(D_i, D_i Y_i, X_i, b, \lambda, \hat{p}_{D,n})$. 

\begin{restatable}[Asymptotic normality]{theorem}{theoremAsymptoticNormality}
	\label{Theorem: asymptotic normality}
	\singlespacing
	
	Suppose assumptions \ref{Assumption: setting}, \ref{Assumption: strong duality}, and \ref{Assumption: estimation} hold. Let $\hat{b}_n \equiv $ \\ $\argmin_{b \in B \cap \textbf{B}_0} \hat{\nu}_n(b)$ and 
	\begin{equation*}
		\hat{\sigma}_n^2 \equiv \frac{1}{n}\sum_{i=1}^n \left((\hat{\Phi}_n(\hat{b}_n)^{-1})^{(1)} \phi(D, DY, X, \hat{b}_n, \hat{\theta}_n(\hat{b}_n))\right)^2
	\end{equation*}
	where $(\hat{\Phi}_n(\hat{b}_n)^{-1})^{(1)} $ is the first row of the matrix $\hat{\Phi}_n(\hat{b}_n)^{-1}$. Then $\sqrt{n}(\hat{\delta}_n^{BP} - \delta^{BP})/\hat{\sigma}_n \overset{d}{\rightarrow} N(0,1)$.
\end{restatable}

\subsection{Inference}
\label{Section: estimators and asymptotics, inference}

A large breakdown point implies the incomplete distribution $P_0$ would have to differ greatly from $P_1$ to rationalize the null hypothesis. If $\delta^{BP}$ is larger than the plausible amount of selection in the setting, the null hypothesis is similarly implausible. Skeptical readers following this argument may worry the point estimate $\hat{\delta}_n^{BP}$ is larger than $\delta^{BP}$ due to sample noise -- but the force of the argument is only strengthened if $\hat{\delta}_n^{BP}$ falls below $\delta^{BP}$. 

To address these concerns, researchers should report lower confidence intervals along with point estimates of the breakdown point. Theorem \ref{Theorem: asymptotic normality} implies that under assumptions \ref{Assumption: setting}, \ref{Assumption: strong duality}, and \ref{Assumption: estimation},
\begin{equation}
	\widehat{CI}_{L,n} \equiv \hat{\delta}_n - \frac{\hat{\sigma}_n}{\sqrt{n}} c_{1-\alpha} \label{Definition: lower confidence interval}
\end{equation}
satisfies $\lim_{n \rightarrow \infty} P(\widehat{CI}_{L,n} \leq \delta^{BP}) = 1-\alpha$ when $c_{1-\alpha}$ is the $1-\alpha$ quantile of the standard normal distribution. 

\begin{remark}
	\label{Remark: allowing for multiple minimizers}
	
	Assumption \ref{Assumption: estimation} \ref{Assumption: estimation, unique solution} can be relaxed at the cost of additional complexity. Without assumption \ref{Assumption: estimation} \ref{Assumption: estimation, unique solution}, $\sqrt{n}(\hat{\nu}_n - \nu)$ still converges in $\ell^\infty(B)$ to $\mathbb{G}_\nu$, a tight Gaussian process on $B$, and minimization of a function over $B \cap \textbf{B}_0$ remains a (Hadamard) \textit{directionally} differentiable map on the set of continuous functions of $B$. The delta method continues to imply $\sqrt{n}(\hat{\delta}_n^{BP} - \delta^{BP})$ converges in distribution to $\inf_{b \in \textbf{m}(\nu)} \mathbb{G}_\nu(b)$, where $\textbf{m}(\nu)$ is the set of minimizers of $\nu$. 
	
	Given a bootstrap $\hat{\nu}_n^*$ such that $\sqrt{n}(\hat{\nu}_n^* - \hat{\nu}_n)$ converges weakly in probability conditional on $\{D_i, D_i Y_i, X_i\}_{i=1}^n$ to $\mathbb{G}_\nu$,  confidence intervals can still be constructed by utilizing the tools developed in \cite{fang2019inference}. One approach is to estimate the set $\textbf{m}(\nu)$ through ``near maximizers'' of $\hat{\nu}_n$ and use this estimated set to form an estimator of the map $h \mapsto \inf_{b \in \textbf{m}(\nu)} h(b)$. The confidence interval for $\delta^{BP}$ is formed by replacing $\hat{\sigma}_n c_{1-\alpha}$ in display \eqref{Definition: lower confidence interval} with the $1-\alpha$ quantile of this estimated function applied to the bootstrap sample; see \cite{fang2019inference} theorem 3.2 and appendix lemma S.4.8.. As most cases of interest appear to satisfy assumption \ref{Assumption: estimation} \ref{Assumption: estimation, unique solution}, this extension is left for future research.
\end{remark}

%% file: Section_Simulations.tex
\section{Simulations}
\label{Section: simulations}

This section presents simulation results on a variety of different data generating processes. This serves both to illustrate the wide scope of models which can make use of breakdown point analysis and to investigate the finite sample properties of the proposed estimators. In each case, selection is measured using squared Hellinger divergence.

\subsection{Expectation}
\label{Section: simulations, simple mean}

Recall example \ref{Example: Squared Hellinger Expectation}. The parameter of interest is the mean of a scalar random variable $Y$, $\beta = E[Y] = p_D E_{P_1}[Y] + (1-p_D)E_{P_0}[Y]$, and the sample is $\{D_i, D_iY_i\}_{i=1}^n$. The distribution of $Y \mid D =1$ is the uniform distribution on $[0,1]$. The probability of observing $Y$ is $p_D = P(D= 1) = 0.7$. To support the claim $H_1 \; : \; \beta > 0.4$, let $H_0 \; : \; \beta \leq 0.4$. Recall that the true breakdown point, $\delta^{BP}$, of this example is just over $0.2$.

The following table summarizes 1,000 simulations for several different sample sizes.\footnote{Here $\text{CI Length} \equiv \hat{\delta}_n^{BP} - \widehat{CI}_{L,n}$.}

\input{Table_SqHellinger_unif_sims_summary.tex}

\noindent The simulations show little bias. Coverage is slightly above the targeted 95 percent significance level in smaller samples.

\subsection{Linear model} 
\label{Section: simulations, linear models}

Linear models are the among the most common tools used by empirical researchers. This subsection uses simulations to investigate linear regression with exogenous regressors.

Consider the model
\begin{equation}
	Y_1 = \beta_0 + \beta_1 X_1 + \beta_2 Y_2 + \beta_3 X_2 + \varepsilon = W^\intercal \beta + \varepsilon, \label{Simulation: MLR}
\end{equation}
where $W = \begin{pmatrix} 1 & X_1 & Y_2 & X_2 \end{pmatrix}^\intercal$ are the exogenous regressors: $E[W\varepsilon] = 0$. Here $Y_1$ is a continuously distributed dependent variable, $X_1 = \{0,1\}$ is the regressor of interest, $Y_2$ is a continuously distributed regressor, and $X_2 \in \{0, 1, 2\}$ is a discrete regressor. The conclusion to be investigated is that the coefficient on $X_1$ is positive:
\begin{align}
	&H_0 \; : \; \beta_1 \leq 0, &&H_1 \; : \; \beta_1 > 0 \label{Display: linear simulations breakdown point conclusion}
\end{align}

The data generating process specification takes inspiration from Mincerian wage equations. For worker $i$, let $Y_{1i}$ be $i$'s log-income, $X_{1i}$ an indicator for $i$ being a college graduate, $Y_{2i}$ be $i$'s work experience, and $X_{2i}$ the number of parents with college degrees ($0$, $1$, or $2$). Specifically, let $X_2$ be multinomial, $X_1 \sim \text{Binomial}\left(\frac{X_2 + 1}{4}\right)$, and $Y_2 \sim \text{Beta}(3-X_1, 3)$.\footnote{The distribution of $X_2$ is $P(X_2 = 0) = 0.4$, $P(X_2 = 1) = 0.25$, and $P(X_2 = 2) = 0.35$.} Let $\tilde{\varepsilon} \sim U[-1,1]$ (independent of all other variables), and $\varepsilon = (X_1+1)\tilde{\varepsilon}$. The coefficients are specified as $\beta_0 = \beta_1 = \beta_2 = 1$ and $\beta_3 = 0.5$. Finally, $Y_1$ is generated according to equation \eqref{Simulation: MLR}. Notice the support of $(Y_1, Y_2, X_1, X_2)$ is compact, ensuring the moment conditions in assumption \ref{Assumption: estimation} \ref{Assumption: estimation, moment conditions} are satisfied. The Supplementary Material shows simulation evidence that this model and data generating process produces a convex $\nu(\cdot)$, suggesting that assumption \ref{Assumption: estimation} \ref{Assumption: estimation, unique solution} holds.

For the missing data process, let $D = \mathbbm{1}\{\varepsilon X_1 + 10 X_1 + 5 (X_2 - 1) > \eta\}$, where $\eta \sim N(-5, 15^2)$. The population value of the breakdown point is approximated as the point estimate obtained from a sample with one million observations. This sample reveals $P(D = 1)$ is about $0.71$, and suffers from selection. Specifically, ignoring the incomplete observations is equivalent to solving $\frac{1}{n}\sum_{i=1}^n \frac{D_i}{\hat{p}_{D,n}} (Y_i - W_i^\intercal \hat{\beta}_n^{MCAR})W_i = 0$ for $\hat{\beta}_n^{MCAR}$, which results in $\hat{\beta}_n^{MCAR} = (1.08, 1.34, 1.02, 0.39)$. The squared Hellinger distance between $P_{0X}$ and $P_{1X}$ is about $0.08$. This large sample suggests the breakdown point of the conclusion $\beta_1 > 0$ is about 0.163, which is treated as the truth when evaluating the 1,000 simulations per sample size summarized in the following table:

\input{Table_SqHellinger_linear_sims_summary.tex}

\noindent The simulations again show little bias, with coverage slightly above the targeted 95 percent significance level in smaller samples.

\subsection{Logit model}
\label{Section: simulations, logistic regression}

The logit model is a popular choice for estimating the conditional probability of an event. Let $Z = (Z_1, Z_{-1}) \in \{0,1\} \times \mathbb{R}^d$ and suppose that $P(Z_1 = 1 \mid Z_{-1}) = \Lambda(Z_{-1}^\intercal \beta)$, where $\Lambda(t) \equiv \frac{\exp(t)}{1 + \exp(t)}$. The log-likelihood is concave, so estimating this model through maximum likelihood is equivalent to solving the first order condition
\begin{equation*}
	E[(Z_1 - \Lambda(Z_{-1}^\intercal \beta)) Z_{-1}] = 0.
\end{equation*}
The model can be viewed as nonlinear GMM, with moment function $g(z,b) = (z_1 - \Lambda(z_{-1}^\intercal b))z_{-1}$. The conclusion to be investigated is that $P(Z_1 = 1 \mid Z_{-1} = \bar{z}) = \Lambda(\bar{z}^\intercal \beta)$ is at least 0.5 for a fixed $\bar{z}$ of interest. The corresponding null and alternative hypotheses are
\begin{align}
	&H_0 : \Lambda(\bar{z}^\intercal \beta) \leq 0.5, &&H_1 : \Lambda(\bar{z}^\intercal \beta) > 0.5. \label{Display: logistic simulations breakdown point conclusion}
\end{align}

The data generating process is one where the dependent variable is always observed, and the regressors are sometimes missing. Specifically, $Y = Z_{-1} \in \mathbb{R}^3$ is constructed by drawing $\tilde{Y} \sim N(0, \Omega)$ and setting $Y^{(j)} = 2 \times (\Phi(\tilde{Y}^{(j)}) - 0.5)$ for each $j = 1, 2, 3$; the result is that each $Y^{(j)}$ has uniform marginal distribution on $[-1, 1]$, and together $(Y^{(1)}, Y^{(2)}, Y^{(3)})$ have a nontrivial joint distribution.\footnote{The matrix $\Omega$ is described by $\text{Var}(Y^{(j)}) = 1$ for each $j=1,2,3$, $\text{Cov}(Y^{(1)}, Y^{(2)}) = 0.5$, $\text{Cov}(Y^{(1)}, Y^{(3)}) = -0.1$, and $\text{Cov}(Y^{(2)}, Y^{(3)}) = 0.3$} The outcome is always observed: $X = Z_1$. The true underlying coefficients are $\beta = (1, -1, 0.1)$. Once again, the compact support of $(X, Y)$ ensures the moment conditions in assumption \ref{Assumption: estimation} \ref{Assumption: estimation, moment conditions} are satisfied. Simulation evidence presented in the Supplementary Material suggests that this model and data generating process produces a convex value function. Since $H_0$ is equivalent to $\bar{z}^\intercal \beta \leq \ln(0.5) - \ln(1 - 0.5) = 0$ and therefore defines a convex $\textbf{B}_0$, this suggests that assumption \ref{Assumption: estimation} \ref{Assumption: estimation, unique solution} holds. 

The missing data process is conditionally binomial with $P(D = 1 \mid X = x, Y = y) = \max\{0.8 - X, Y^{(3)}/2 + 0.5 \}$; that is, the probability of an observation being complete is at least $0.8$ when $X = 0$ and grows weakly with $Y^{(3)}$. The resulting samples suffer from selection. A sample with one million observations suggests that $P(D = 1)$ is about $0.65$. Ignoring the incomplete observations is equivalent to solving $\frac{1}{n}\sum_{i=1}^n \frac{D_i}{\hat{p}_{D,n}} g(D_i Y_i, X_i, \hat{\beta}_n^{MCAR}) = 0$, which results in $\hat{\beta}_n^{MCAR} = (1, -1, 0.79)$. The estimated squared Hellinger distance between $P_{0X}$ and $P_{1X}$ is $0.076$. The covariate value of interest is $\bar{y} = (-0.35, -0.25, 0.5)$. The true value for $\Lambda(\bar{y}^\intercal \beta)$ is $0.488$, while the estimate using the complete observations of the large sample above is $\Lambda(\bar{y}^\intercal \hat{\beta}_n^{MCAR}) = 0.573$. The point estimate for the breakdown point of the conclusion described by \eqref{Display: logistic simulations breakdown point conclusion} using this large sample is $0.108$. This is treated as the truth when evaluating the 1,000 simulations per sample size summarized in the following table:
\input{Table_SqHellinger_logistic_sims_summary.tex}

\noindent These simulations show essentially zero bias and correct coverage at relatively small sample sizes.

%% file: Table_SqHellinger_unif_sims_summary.tex
\begin{table}[H]
	\def\arraystretch{1.2}
	\caption{Simulations, expectation}
	\begin{center}
		\begin{tabular}{|c || c|c|c|c |} \hline 
		$n$ &   Bias &  St. Dev. &  Coverage &  Ave. CI Length \\ \hline \hline 
		1,000 &  0.005 &     0.056 &      98.5 &           0.090 \\ \hline
		3,000 &  0.002 &     0.032 &      96.3 &           0.051 \\ \hline
		5,000 &  0.001 &     0.025 &      95.8 &           0.039 \\ \hline 
		10,000 &  0.001 &     0.017 &      95.8 &           0.028 \\ \hline 
		\end{tabular}
	\end{center}
\end{table}

%% file: Table_SqHellinger_linear_sims_summary.tex
\begin{table}[H]
	\def\arraystretch{1.2}
	\caption{Simulations, linear model}
	\begin{center}
		\begin{tabular}{|c || c|c|c|c |} \hline 
		$n$ &   Bias &  St. Dev. &  Coverage &  Ave. CI Length \\ \hline \hline 
		1,000 &  0.016 &     0.048 &      98.9 &           0.076 \\ \hline
		3,000 &  0.007 &     0.026 &      95.8 &           0.041 \\ \hline
		5,000 &  0.004 &     0.019 &      95.4 &           0.031 \\ \hline
		10,000 &  0.003 &     0.013 &      94.5 &           0.022 \\ \hline 
		\end{tabular}
	\end{center}
\end{table}

%% file: Table_SqHellinger_logistic_sims_summary.tex
\begin{table}[H]
	\def\arraystretch{1.2}
	\caption{Simulations, logit model}
	\begin{center}
		\begin{tabular}{| c || c|c|c|c |} \hline 
		$n$ &   Bias &  St. Dev. &  Coverage &  Ave. CI Length \\ \hline \hline 
		1,000 &  0.003 &     0.018 &      94.5 &           0.029 \\ \hline 
		3,000 & -0.000 &     0.010 &      96.1 &           0.017 \\ \hline 
		5,000 &  0.001 &     0.008 &      94.8 &           0.013 \\ \hline 
		10,000 & -0.000 &     0.005 &      95.9 &           0.009 \\ \hline 
		\end{tabular}
	\end{center}
\end{table}

%% file: Section_Application.tex
\section{Application: attrition in randomized controlled trials}
\label{Section: application}

This section reports estimates of the breakdown point of conclusions drawn from a number of randomized controlled trials (RCTs) conducted in developing countries. There are several advantages to demonstrating breakdown point analysis on real world data in this way. First, these studies are known to suffer from missing data that is unlikely to be missing at random. Second, missingness in these studies is often a result of study subject migration. This illustrates an important point discussed in section \ref{Section: measuring selection and breakdown analysis, squared hellinger}: extra scrutiny should be given to conclusions involving variables that would predict migration, and hence missingness. Finally, these RCTs are evaluated using similar methodologies. The breakdown point estimates below thus show a range of values that might be expected due to variation in real world data, rather than significant variation in methodology. This provides useful context for researchers using breakdown point analysis to investigate the robustness of conclusions drawn from similar studies.

Missing data due to attrition is a prominent concern for studies conducted in developing countries. \cite{thomas2012cutting} notes that the primary cause of this attrition is researchers being unable to find respondents who moved after the baseline survey. The authors study the Indonesia Family Life Survey, which has a notably low attrition rate despite high mobility of the target population, and provide evidence that migrants differ from non-migrants along dimensions unlikely to be observed at a survey's baseline. Attrition due to study subject migration is also noted in \cite{molina2025attrition}, which studies randomized controlled trials conducted in developing countries. The authors show through examples that attempting to correct for attrition through inverse propensity weighting with baseline data does not make a notable difference to estimates -- but including individuals found only after intensive (and often costly) tracking does. Both papers suggest that missing data in these contexts is likely due to migration, and not missing at random. 

The breakdown point estimates below pertain to results found in \cite{barham2024experimental}, \cite{bandiera2020women}, and \cite{giacobino2024schoolgirls}, which all study randomized controlled trials conducted in developing countries. \cite{barham2024experimental} studies a conditional cash transfer (CCT) implemented by the Nicaraguan government to address poverty by improving health and education. Study subjects were randomized into early or late treatment groups in the baseline year, 2000, and the authors study the differential effects of receiving the treatment early. The breakdown point analysis below focuses on conclusions regarding the cohort of boys who were aged 9-12 in the year 2000. Those who received the CCT early -- when they were at higher risk of dropping out -- showed higher labor market participation and earnings in 2010. \cite{bandiera2020women} studies the impact of a program in Uganda designed to increase women's empowerment through training in vocational and life skills. The breakdown point analysis focuses on five conclusions regarding outcomes measured at midline: the treatment increased an index of entrepreneurial ability, increased the probability of being engaged in any income-generating activity, increased the probability of being self-employed, increased the probability of being employed for a wage, and increased expenditures on goods in the last month. Finally, \cite{giacobino2024schoolgirls} studies a scholarship for adolescent girls in Niger to attend middle school. The intervention was designed to deter child marriage. The breakdown point analysis below focuses on three conclusions: the scholarship reduced the probability of dropping out, reduced the probability of being married by endline, and increased life satisfaction as measured by a standardized 10-point Likert scale. 

Table \ref{Table: Application, MCAR estimates} reports the intent-to-treat (ITT) estimates when incomplete observations are dropped, referred to as missing completely at random (MCAR) estimates. Column (1) reports the total sample size, including subjects that attrited and could not be included in the estimates. Column (2) reports the number of complete observations on which the subsequent estimates are based. Column (3) reports the average of the outcome among untreated subjects. Columns (4) and (5) report coefficient estimates on an indicator for treatment status from a regression of the outcome on a constant, the indicator for treatment, and additional regressors. Column (4) replicates the results from the original papers, including the authors' choice of additional regressors and standard errors. To facilitate comparisons across studies, estimates in column (5) use indicators for the subject's region at baseline as the additional regressors and HC3 standard errors. Consistent with independent randomization of treatment assignment, changing the additional regressors does not meaningfully alter the estimates.

\input{Application_MCAR_table_formatted.tex}

Table \ref{Table: Application, BDP estimates} reports breakdown point analyses. In each case, the conclusion is that the ITT parameter takes the sign implied by the point estimate in table \ref{Table: Application, MCAR estimates}. Squared Hellinger is used to measure selection. Every subject's treatment status and region at baseline is observed, and used as the variables in $X$. Also reported is an estimate of $H^2(P_{0X}, P_{1X})$, formed by taking sample analogues of $P(X=x)$ for each possible value in $\mathcal{X}$ and plugging these into the definition of squared Hellinger. This provides an estimated lower bound on the amount of selection in the given setting, as described in remark \ref{Remark: squared hellinger lower bound}.

\input{Application_BDP_table_formatted.tex}

Tables \ref{Table: Application, MCAR estimates} and \ref{Table: Application, BDP estimates} show a number of patterns worth emphasizing. Compared to the other two studies, the larger sample of \cite{bandiera2020women} resulted in lower standard errors in table \ref{Table: Application, MCAR estimates} and lower confidence intervals that are closer to the breakdown point estimates in table \ref{Table: Application, BDP estimates}. However, the larger share of incomplete observations in this study results in generally smaller breakdown points. The magnitude of MCAR estimates and the amount of missing data are clearly determinants of the size of the breakdown point, but superficially similar results can have quite different breakdown points. For example, consider the conclusion that the CCT from \cite{barham2024experimental} differentially raised the probability of working somewhere other than the recipient's family farm, and the claim that the scholarship studied in \cite{giacobino2024schoolgirls} reduced the probability the subject is being married at endline. The corresponding MCAR estimates have a similar magnitude ($0.06$ compared to $-0.07$) and have similar standard errors ($0.02$ or $0.03$, depending on the specification). The two studies have a similar share of incomplete data. However, the breakdown point of the latter result appears notably larger than that of the former result. Notice also that the estimated effects of the treatment from \cite{bandiera2020women} on self-employment and wage employment under MCAR differ considerably ($0.06$ compared to $0.01$), but the claims that the corresponding ITT estimates are positive have similar breakdown points.

Several results appear fragile, while others are quite robust. Consider the claims that the treatment studied in \cite{bandiera2020women} increased expenditures or the probability of being self- or wage- employed. The breakdown point estimates of these claims are quite close to the estimate of $H^2(P_{0X}, P_{1X})$, implying that it would take only a small amount of selection on these outcomes to rationalize such claims being false.  In contrast, the results of \cite{giacobino2024schoolgirls} appear quite robust. The claim that the scholarship reduced the probability of dropping out could not be rationalized as false in the sample. 
The other claims from \cite{giacobino2024schoolgirls} have point estimates for the breakdown point that are close to 0.2. As discussed in remark \ref{Remark: 0.2 is a big number}, these are large values for a breakdown point implying the results are quite robust.

%% file: Application_MCAR_table_formatted.tex
\begin{table}[H]
	\def\arraystretch{1.1}
	\caption{Missing completely at random estimates}
	\label{Table: Application, MCAR estimates}
	\begin{center}
		\begin{tabular}{|l | c || c | c | c | c | c |}
			\hline
			\multirow{3}{*}{Paper} & \multirow{3}{*}{Outcome} & \multicolumn{2}{c|}{Observations} & Untreated & \multicolumn{2}{c|}{MCAR ITT estimates}   \\
			& & Total $(n)$ & Complete & mean & Replication & Region ind. \\
			& & (1) & (2) & (3) & (4) & (5) \\
			\hline \hline
			\multirow[c]{9}{*}{\begin{tabular}{l} Barham \\ et al. (2024)\end{tabular}} &  &  &  &  &  &  \\
			& Off-farm & 1,138 & 1,006 & 0.83 & 0.06 & 0.06 \\
			& employment &  &  &  & (0.02) & (0.03) \\
			&  &  &  &  &  &  \\
			& Rank of earnings & 1,138 & 1,006 & 497.15 & 41.78 & 42.87 \\
			& per mo. worked  &  &  &  & (19.50) & (21.58) \\
			&  &  &  &  &  &  \\
			& Read and write & 1,138 & 1,007 & 0.87 & 0.05 & 0.07 \\
			&  &  &  &  & (0.02) & (0.02) \\
			\cline{1-7}
			\multirow[c]{15}{*}{\begin{tabular}{l} Bandiera \\ et al. (2020)\end{tabular}} &  &  &  &  &  &  \\
			& Entrepreneurial & 5,966 & 4,765 & 71.77 & 5.63 & 5.46 \\
			& ability index &  &  &  & (0.98) & (0.71) \\
			&  &  &  &  &  &  \\
			& Income-generating & 5,966 & 4,831 & 0.10 & 0.07 & 0.07 \\
			& activity &  &  &  & (0.02) & (0.01) \\
			&  &  &  &  &  &  \\
			& Self-employed & 5,966 & 4,831 & 0.06 & 0.06 & 0.06 \\
			&  &  &  &  & (0.01) & (0.01) \\
			&  &  &  &  &  &  \\
			& Wage employed & 5,966 & 4,831 & 0.04 & 0.01 & 0.01 \\
			&  &  &  &  & (0.01) & (0.01) \\
			&  &  &  &  &  &  \\
			& Expenditures & 5,966 & 4,752 & 11.92 & 4.68 & 4.52 \\
			& in UGX, 1,000s  &  &  &  & (0.95) & (0.73) \\
			\cline{1-7}
			\multirow[c]{9}{*}{\begin{tabular}{l} Giacobino \\ et al. (2024) \end{tabular}} &  &  &  &  &  &  \\
			& Dropped out & 1,501 & 1,344 & 0.40 & -0.21 & -0.21 \\
			&  &  &  &  & (0.05) & (0.02) \\
			&  &  &  &  &  &  \\
			& Married & 1,501 & 1,344 & 0.14 & -0.07 & -0.07 \\
			&  &  &  &  & (0.03) & (0.02) \\
			&  &  &  &  &  &  \\
			& Life satisfaction & 1,501 & 1,344 & 0.00 & 0.25 & 0.25 \\
			&  &  &  &  & (0.11) & (0.05) \\
			\cline{1-7}
			
		\end{tabular}
	\end{center}
	
	\begin{flushleft}
		\vspace{-0.3 cm}
		{\scriptsize \textit{Notes:} Column (4) replicates each paper's results, including the authors' choice of additional regressors and standard errors. Column (5) uses region indicators as the additional regressors and reports HC3 standard errors in parentheses. For outcomes from Barham et al. (2024), column (3) reports the mean from subjects receiving the CCT late and estimates in column (5) make use of sampling weights from Molina-Mill\'an \& Macours (2025) rather than inverse propensity weights from Barham et al. (2024). For Bandiera et al. (2020), column (3) reports average outcomes measured at baseline.}
	\end{flushleft}
\end{table}

%% file: Application_BDP_table_formatted.tex
\begin{table}[H]
	\def\arraystretch{1.3}
	\caption{Breakdown point analyses}
	\label{Table: Application, BDP estimates}
	\begin{center}
		\begin{tabular}{|l | c || c | c | c | c |}
			\hline
			\multirow{2}{*}{Paper} & \multirow{2}{*}{Outcome}  & \multirow{2}{*}{$\hat{p}_{D,n}$} & \multirow{2}{*}{$\hat{\delta}_n^{BP}$} & \multirow{2}{*}{$\widehat{CI}_{L,n}$} & Estimate of \\
			& &  &  &  & $H^2(P_{0X}, P_{1X})$  \\
			\hline \hline
			\multirow[c]{3}{*}{Barham et al. (2024)} & Off-farm employment & 0.88 & 0.08 & 0.03 & 0.05 \\
			& Rank of earnings per mo. worked & 0.88 & 0.08 & 0.03 & 0.05 \\
			& Read and write & 0.88 & 0.07 & 0.04 & 0.05 \\
			\hline
			\multirow[c]{5}{*}{Bandiera et al. (2020)} & Entrepreneurial ability index & 0.81 & 0.08 & 0.06 & 0.03 \\
			& Income-generating activity & 0.82 & 0.05 & 0.04 & 0.03 \\
			& Self-employed & 0.82 & 0.04 & 0.03 & 0.03 \\
			& Wage employed & 0.82 & 0.03 & 0.03 & 0.03 \\
			& Expenditures in UGX, 1,000s & 0.81 & 0.04 & 0.04 & 0.03 \\
			\hline
			\multirow[c]{3}{*}{Giacobino et al. (2024)} & Dropped out & 0.90 & $\infty$ & $\infty$ & 0.05 \\
			& Married & 0.90 & 0.17 & 0.06 & 0.05 \\
			& Life satisfaction & 0.90 & 0.26 & 0.09 & 0.05 \\
			\hline			
		\end{tabular}
	\end{center}
\end{table}

%% file: Section_Conclusion.tex
\section{Conclusion}
\label{Section: conclusion}

This paper proposes breakdown point analysis as a tractable approach to assessing the sensitivity of a researcher's conclusion to the common assumption that the data are missing at random. When defined with squared Hellinger, the breakdown point $\delta^{BP}$ has a natural interpretation: if the result were false, the variables under study ($Z$) would have to predict an observation being selected into the sample ($D$) at least well enough that $H^2(P_0, P_1) = 1 - E[\sqrt{\text{Var}(D \mid Z)}]/\sqrt{\text{Var}(D)} \geq \delta^{BP}$. Estimators based on the sample analogue of the dual problem are shown $\sqrt{n}$-consistent and asymptotically normal, which facilitates the construction of lower confidence intervals. Researchers working with incomplete datasets should report the breakdown point estimate and lower confidence interval along with standard results, making transparent to their audience how robust the conclusion is to relaxing the assumption that the data are missing at random.

%% file: Appendix_Notation.tex
\section{Appendix: Notation}
\label{Appendix: notation}

This appendix summarizes notation and facts used throughout the appendices.

\subsection{Graphs}
\label{Appendix: notation, graphs}

Let $X \subseteq \mathbb{R}^{d_x}$ and $Y\subseteq \mathbb{R}^{d_y}$. For a function $f : X \rightarrow Y$, the \textit{graph} of the function refers to the set  $\text{Gr}(f) = \left\{(x,f(x)) \; : \; x \in X\right\} \subseteq X \times Y$. Define the closed $\delta$-expansion of the graph of $f$:
\begin{equation*}
	\text{Gr}(f)^\delta \equiv \left\{(x,y) \in X \times Y \; : \; \inf_{(x',y') \in \text{Gr}(f)} \lVert (x,y) - (x', y') \rVert \leq \delta\right\}
\end{equation*} 
Note that $\text{Gr}(f)^\delta$ is closed, and bounded if $\text{Gr}(f)$ is bounded. \\

Given $Z \subseteq \mathbb{R}^{d_z}$ and $g : X \rightarrow Z$, one can view $(f,g)$ as a function from $X$ to $(Y, Z)$:
\begin{align*}
	&(f,g) : X \rightarrow (Y, Z), &&(f,g)(x) = (f(x), g(x))
\end{align*}
Define the graph of this function, $\text{Gr}(f,g) = \left\{(x, f(x), g(x)) \; : \; x \in X\right\} \subseteq X \times Y \times Z$, and the closed $\delta$-expansion about this graph:
\begin{equation*}
	\text{Gr}(f,g)^\delta = \left\{(x,y, z) \in X \times Y  \times Z \; : \; \inf_{(x',y', z') \in \text{Gr}(f,g)} \lVert (x, y, z) - (x', y', z') \rVert \leq \delta\right\}
\end{equation*}

Several easily constructed subsets of $\text{Gr}(f,g)^\delta$ imply useful inequalities. For example,
\begin{align*}
	\inf_{(x', y', z') \in \text{Gr}(f, g)} \lVert (x, y, g(x)) - (x', y', z') \rVert &\leq \inf_{(x', y', z') \in \text{Gr}(f, g)} \lVert (x, y, g(x)) - (x', y', g(x)) \rVert \\
	&= \inf_{(x, y) \in \text{Gr}(f)^\delta} \lVert (x, y) - (x', y') \rVert 
\end{align*}
implies $\left\{(x, y, g(x)) \; : \; (x, y) \in \text{Gr}(f)^\delta\right\} \subseteq \text{Gr}(f, g)^\delta$. It follows that for a function $h : X \times Y \times Z \rightarrow \mathbb{R}$,
\begin{align*}
	\sup_{(x,y) \in \text{Gr}(f)^\delta} h(x, y, g(x)) \leq \sup_{(x,y,z) \in \text{Gr}(f,g)^\delta} h(x, y, z).
\end{align*}
Similarly, 
\begin{align*}
	\lVert (x, y, z) - (x', y', z') \rVert &\leq \lVert (x, y, z) - (x', y', z) \rVert + \lVert (x', y', z) - (x', y', z') \rVert \\
	&= \lVert (x, y) - (x', y') \rVert + \lVert z - z' \rVert
\end{align*}
implies that 
\begin{align*}
	&\inf_{(x', y') \in \text{Gr}(f)^{\delta/2}, \; z' \in \text{Gr}(g)^{\delta/2}} \lVert (x, y, z) - (x', y', z') \rVert \\
	&\hspace{2 cm} \leq \inf_{(x', y') \in \text{Gr}(f)^{\delta/2}} \lVert (x, y) - (x', y') \rVert + \inf_{z' \in \text{Gr}(g)^{\delta/2}} \lVert z - z' \rVert.
\end{align*}
It follows that $\left\{(x, y, z) \; : \; (x, y) \in \text{Gr}(f)^{\delta/2}, \; (x, z) \in \text{Gr}(g)^{\delta/2}\right\} \subseteq \text{Gr}(f, g)^\delta$, and hence for a function $h : X \times Y \times Z \rightarrow \mathbb{R}$,
\begin{align*}
	\sup_{(x,y) \in \text{Gr}(f)^{\delta/2}, \; z \in \text{Gr}(g)^{\delta/2}} h(x, y, ) \leq \sup_{(x,y,z) \in \text{Gr}(f,g)^\delta} h(x, y, z).
\end{align*}

Finally, note that any constant $\bar{y} \in Y$ can be viewed as a trivial function of $X$. The graph of this function is the set $\text{Gr}(\bar{y}) = \{(x,\bar{y}) \; : \; x \in X\}$ and $\text{Gr}(\bar{y})^\delta$ is the set $\left\{(x,y) \; : \; x \in X, \lVert y - \bar{y} \rVert \leq \delta\right\}$.

\subsection{Spaces of bounded functions}
\label{Appendix: notation, space of bounded functions}

For any set $T$, $\ell^\infty(T) \equiv \left\{f : T \rightarrow \mathbb{R} \; \text{ such that } \; \sup_{t \in T} \lvert f(t) \rvert < \infty\right\}$ denotes the set of real-valued bounded functions on $T$. $\ell^\infty(T)$ is equipped with the sup-norm: for $f \in \ell^\infty(T)$, $\lVert f \rVert_\infty \equiv \lVert f \rVert_{T} \equiv \sup_{t \in T} \lvert f(t) \rvert$. The space of bounded functions taking values in $\mathbb{R}^K$ for some $K \in \mathbb{N}$ is the product space $\ell^\infty(T)^K = \underbrace{\ell^\infty(T) \times \ldots \times \ell^\infty(T)}_{K \text{ times }}$, but can also be viewed as a process on $\ell^\infty(T \times \{1, \ldots, K\})$. The latter notation makes it clear that standard empirical process results, typically stated for scalar-valued processes, apply. \\

If $(T, d)$ is a compact metric space, the extreme value theorem implies that continuous functions on $T$ are also bounded and hence form of a subspace of $\ell^\infty(T)$. This subspace is denoted
\begin{equation*}
	\mathcal{C}(T, d) = \{f : T\rightarrow \mathbb{R} \; \text{ such that } \; f \text{ is continuous}\}
\end{equation*}
the notation $\mathcal{C}(T)$ will be used to mean $\mathcal{C}(T, d)$ when the metric $d$ is clear from context.

Some results will refer to subsets of bounded functions whose graphs fall into a particular set. Specifically, let $E^t \subseteq \mathbb{R}^{d_E}$ for each $t \in T$, $E^T \equiv \{(t,e) \; : \; t \in T, e \in E^t\}$, and $\ell^\infty(T, E^T)^{d_E}$ be the subset of $\ell^\infty(T)^{d_E}$ whose graph is a subset of $E^T$:
\begin{equation*}
	\ell^\infty(T, E^T)^{d_E} = \left\{g : T \rightarrow \mathbb{R}^{d_E} \; \text{ such that } \; g(t) \in E^t, \; \sup_{t \in T} \lVert g(t) \rVert < \infty\right\} \subset \ell^\infty(T)^{d_E}
\end{equation*}
For an example of how this will be used, let $\bar{x} > 0$ and note that the function $f(t,e) = \ln(t + e)$ is uniformly continuous on the set $\{(t,e) \; : \; t + e \geq \bar{x}\}$. Defining $E^t = \{e \in \mathbb{R} \; : \; e \geq \bar{x} - t\}$ and $E^T$ as above, we have that $f(t,e)$ is uniformly continuous on this set. This implies that $\tilde{f} : \ell^\infty(T, E^T) \rightarrow \ell^\infty(T)$ given by $\tilde{f}(g)(t) = f(t, g(t)) = \ln(t + g(t))$ is continuous (see lemma \ref{Technical Lemma: continuity of maps between bounded functions}).

\subsection{Matrices}
\label{Appendix: notation, matrices}

For a matrix $A \in \mathbb{R}^{J \times K}$, let $\lVert A \rVert_o = \sup_{x \; : \; \lVert x \rVert_2 = 1} \lVert A x \rVert_2$ be the operator norm of $A$, and $\lVert A \rVert_{\max} = \max_{ij} \lvert a_{ij} \rvert$, where $a_{ij} \in \mathbb{R}$ is the entry in the $i$-th row and $j$-th column of $A$. Let $\sigma_1(A) \geq \ldots \geq \sigma_K(A) \geq 0$ be the ordered singular values of $A$. For a square $K \times K$ real matrix $A$, let $\alpha_k(A)$ be the $k$-th largest eigenvalue of $A$; $\alpha_1(A) \geq \ldots \geq \alpha_k(A) \geq \ldots \geq \alpha_K(A)$. \\

Recall that all norms on finite dimensional real vector spaces are strongly equivalent, meaning that if $\lVert \cdot \rVert_1$ and $\lVert \cdot \rVert_2$ are any norms on $\mathbb{R}^{J \times K}$, there exist constants $c, C > 0$ such that $c \lVert A \rVert_1 \leq \lVert A \rVert_2 \leq C \lVert A \rVert_1$ for any matrix $A \in \mathbb{R}^{J \times K}$. If $A : T \rightarrow \mathbb{R}^{J \times K}$ for some set $T$, it follows that $E[\sup_t \lVert A(t) \rVert] < \infty$ for any norm if and only if $E[\sup_t \lVert A(t) \rVert_{\max}] < \infty$. Notice that strong equivalence with $\lVert \cdot \rVert_{\max}$ implies that, for any submatrix $\tilde{A}(t)$ of $A(t)$, $E[\sup_t \lVert A(t) \rVert] < \infty$ implies $E[\sup_t \lVert \tilde{A}(t) \rVert] < \infty$. \\

Recall that the singular values of a matrix $A \in \mathbb{R}^{J \times K}$ are related to the eigenvalues of the $K \times K$ square matrix $A^\intercal A$ by $\sigma_k(A) = \sqrt{\alpha_k(A^\intercal A)}$. The operator norm of a matrix is equal to its largest singular value, $\lVert A \rVert_o = \sigma_1(A)$, and for invertible matrix $A$ and any $k = 1, \ldots, K$, $\frac{1}{\sigma_k(A)}$ is a singular value of $A^{-1}$. These imply $\lVert A^{-1} \rVert_o = \frac{1}{\sigma_K(A)}$. Finally, for a vector $x \in \mathbb{R}^K$, $\lVert x x^\intercal \rVert_o = \lVert x^\intercal x \rVert_o = \lVert x \rVert_2$. 

\subsection{Additional calculations}
\label{Appendix: notation, calculations}

A number of expressions are useful for verifying conditions in proofs and for programming estimators. These are collected here for convenience.

Recall that $\theta_0(b) = (\nu(b), \lambda(b), p_D)$, where $\nu(b)$ is the population value of the value function, $\lambda(b)$ is the corresponding Lagrange multiplier, and $p_D = P(D = 1)$. The notation $\theta = (v, \lambda, p) \in \mathbb{R}^{d_g + K + 2}$ refers to a vector in Euclidean space. 
\begin{align}
	\varphi(D, DY, X, b, \lambda, p) &\equiv \frac{\lambda^\intercal J(D) h(DY, X, b)}{1-p} - \frac{D}{p} f^*(\lambda^\intercal h(DY, X, b)) \label{Appendix display: varphi, the dual objective integrand} \\
	\phi(D, DY, X, b, \theta) &= \phi(D, DY, X, b, v, \lambda, p) \equiv 
	\begin{pmatrix}
		\varphi(D, DY, X, b, \lambda, p) - v \\
		\nabla_\lambda \varphi(D, DY, X, b, \lambda, p) \\
		D - p
	\end{pmatrix} \label{Appendix display: phi, the Z-estimator integrand} \\
	\Phi(b) &= E[\nabla \phi(D, DY, X, b, \theta_0(b))] \label{Appendix display: Phi, the jacobian matrix} 
\end{align}
\begin{align}
	\nabla_\theta \phi(d, dy, x, b, \theta) &= \nabla_{v, \lambda, p} 
	\begin{pmatrix}
		\varphi(d, dy, x, b, \lambda, p) - v \\
		\nabla_\lambda \varphi(d, dy, b, \lambda, p) \\
		d - p
	\end{pmatrix} \notag\\
	&= 
	\begin{bmatrix}
		-1 & \nabla_\lambda \varphi(d, dy, x, b, \lambda, p)^\intercal & \nabla_p \varphi(d, dy, x, b, \lambda, p) \\
		0 & \nabla_\lambda^2 \varphi(d, dy, x, b, \lambda, p) & \nabla_p \nabla_\lambda \varphi(d, dy, x, b, \lambda, p) \\
		0 & 0 & -1
	\end{bmatrix} \label{Appendix display: jacobian matrix components} 
\end{align}
\begin{align}
	\nabla_\lambda \varphi(d, dy, x, b, \lambda, p) &= \frac{J(d) h(dy, x, b)}{1-p} - \frac{d}{p} (f^*)'(\lambda^\intercal h(dy, x, b))h(dy, x, b) \label{Appendix display: derivatives of varphi} \\
	\nabla_\lambda^2 \varphi(d, dy, x, b, \lambda, p) &= -\frac{d}{p} (f^*)''(\lambda^\intercal h(dy, x, b)) h(dy, x, b) h(dy, x, b)^\intercal \notag \\
	\nabla_p \varphi(d, dy, x, b, \lambda, p) &= \frac{\lambda^\intercal J(d) h(dy, x, b)}{(1-p)^2} + \frac{d}{p^2} f^*(\lambda^\intercal h(dy, x, b)) \notag \\
	\nabla_p \nabla_\lambda \varphi(d, dy, x, b, \lambda, p) &= \frac{J(d) h(dy, x, b)}{(1-p)^2} + \frac{d}{p^2} (f^*)'(\lambda^\intercal h(dy, x, b)) h(dy, x, b) \notag 
\end{align}

%% file: Appendix_BDP_Partial_ID.tex
\section{Appendix: Breakdown point analysis in a partial identification framework}

\label{Appendix: nominal identified sets}

The exercise proposed in section 2.3 can also be understood with a partial identification framework. This approach to exposition is used in \cite{kline2013sensitivity}, \cite{masten2020inference}, and \cite{diegert2025assessing}.

Under the assumption $d(P_0 \Vert P_1) \leq \delta$ and $P_0 \ll P_1$, the identified set for $\beta_P$ is a function of $\delta$: 
\begin{equation}
	\textbf{B}_{ID}(\delta) = \Big\{b \in \boldsymbol{B} \; : \; \exists Q, \;\; p_D \mathbb{E}_{P_1}[g(Z,b)] + (1-p_D) \mathbb{E}_Q[g(Z,b)] = 0, \text{ and } \; d(Q \; \Vert \; P_1) \leq \delta\Big\}
	\label{} \label{Definition: nominal identified set}
\end{equation}
Notice $\textbf{B}_{ID}(\delta)$ is always growing with $\delta$, in the sense that $\delta < \delta' \implies \textbf{B}_{ID}(\delta) \subseteq \textbf{B}_{ID}(\delta')$. 

The researcher is primarily interested in testing $H_0 \; : \; \beta \in \textbf{B}_0$ against $H_1 \; : \; \beta \in \textbf{B}_1 = \textbf{B} \setminus \textbf{B}_0 $. Naturally, if $\textbf{B}_{ID}(\delta)$ has trivial intersection with $\textbf{B}_0$ she is confident in rejecting $H_0$. This leads to the question ``what is the largest value of $\delta$ such that $\textbf{B}_{ID}(\delta)$ has empty intersection with $\textbf{B}_0$?'' Formally, define the \textit{breakdown point} as
\begin{equation}
	\bar{\delta}^{BP} = \sup\left\{\delta \in \mathbb{R}_+ \; : \; \textbf{B}_{ID}(\delta) \cap \textbf{B}_0 = \varnothing\right\} \label{Definition: breakdown point defined with nominal identified sets}
\end{equation}
if $\textbf{B}_{ID}(0) \cap \textbf{B}_0 = \varnothing$, otherwise define $\bar{\delta}^{BP} \equiv 0$. Lemma \ref{Lemma: characterization of breakdown point} shows that the definition of the breakdown point given in \eqref{Definition: breakdown point defined with nominal identified sets} is equivalent to that given by \eqref{Definition: breakdown point}. 
\begin{restatable}[Characterization of breakdown point]{lemma}{lemmaCharacterizationOfBreakdownPoint}
	\label{Lemma: characterization of breakdown point}
	\begin{equation*}
		\inf_{b \in \boldsymbol{B}_0} \nu(b) = \bar{\delta}^{BP}
	\end{equation*}
	
\end{restatable}
\begin{proof}
	Define the ``robust region'' as the set of $\delta \in \mathbb{R}_+$ where the identified set has trivial intersection with the null hypothesis: 
	\begin{equation*}
		RR = \{\delta \in \mathbb{R}_+ \; : \; \textbf{B}_{ID}(\delta) \cap \textbf{B}_0 = \varnothing\}
	\end{equation*}
	and let $RR^c = \mathbb{R}_+ \setminus RR = \{\delta \in \mathbb{R}_+ \; : \; \textbf{B}_{ID}(\delta) \cap \textbf{B}_0 \neq \varnothing\}$ be its complement in $\mathbb{R}_+$. Notice that 
	\begin{equation*}
		\bar{\delta}^{BP} = 
		\begin{cases} 
			\sup RR &\text{ if } RR \neq \varnothing \\
			0 &\text{ otherwise} 
		\end{cases}
	\end{equation*}
	
	The proof consists of two steps: 
	\begin{enumerate}
		\item Showing that 
		\begin{equation}
			\bar{\delta}^{BP} = \inf RR^c \label{Proof display: nominal identified sets breakdown point characterization, breakdown point is infimum over complement of robust region}
		\end{equation} 
		where the infimum over the empty set is defined to be $\infty$.
		
		\item Arguing that 
		\begin{align*}
			&\inf_{b \in \boldsymbol{B}_0} \nu(b) \leq \inf RR^c, &&\text{ and } &&\inf_{b \in \boldsymbol{B}_0} \nu(b) \geq \inf RR^c,
		\end{align*}
	\end{enumerate}

	Step 1.  is a consequence of $\textbf{B}_{ID}(\delta)$ being a growing set (in the sense that $\delta \leq \delta' \implies \textbf{B}_{ID}(\delta) \subseteq \textbf{B}_{ID}(\delta')$). Define $\bar{\delta}^* = \inf RR^c = \inf \{\delta \in \mathbb{R}_+ \; : \; \textbf{B}_{ID}(\delta) \cap \textbf{B}_0 \neq \varnothing\}$.
	There are three possibilities:
	\begin{enumerate}[label=(\roman*)]
		\item $\delta^{BP} = 0$. Then $RR^c$ contains $(0,\infty)$, hence $0 \leq \bar{\delta}^* = \inf RR^c  \leq \inf (0,\infty) = 0$.
		\item $\delta^{BP} \in (0,\infty)$. Notice that $\delta \leq \delta' \implies \textbf{B}_{ID}(\delta) \subseteq \textbf{B}_{ID}(\delta')$ implies that $\delta \leq \delta' \implies (\textbf{B}_{ID}(\delta) \cap \textbf{B}_0) \subseteq (\textbf{B}_{ID}(\delta') \cap \textbf{B}_0)$, from which it follows that 
		\begin{align*}
			&\delta \leq \delta' \text{ and } \delta' \in RR &&\implies &&\delta \in RR \\
			&\delta \leq \delta' \text{ and } \delta \in RR^c &&\implies &&\delta' \in RR^c
		\end{align*}
		since $\bar{\delta}^{BP} \in (0,\infty)$, we have $RR$ contains $[0, \bar{\delta}^{BP})$. Similarly, $RR^c$ contains $(\bar{\delta}^*, \infty)$, and since $RR \cap RR^c = \varnothing$, we have $\bar{\delta}^{BP} \leq \bar{\delta}^*$. For $n \in \mathbb{N}$, let $\delta_n \coloneqq \bar{\delta}^* - \frac{1}{n} \geq 0$, and notice that $\textbf{B}_{ID}(\delta_n) \cap B = \varnothing$, equivalently, $\delta_n \in RR$. Therefore
		\begin{equation*}
			\bar{\delta}^* - \frac{1}{n} \leq \bar{\delta}^{BP} \leq \bar{\delta}^*
		\end{equation*}
		let $n \rightarrow \infty$ to see that $\delta^{BP} = \delta^*$. 
		\item $\delta^{BP} = \infty$. Then the argument above implies $RR$ contains $[0,\infty)$, so $RR^c = \varnothing$ and $\delta^* = \infty$.
	\end{enumerate}
	Therefore \eqref{Proof display: nominal identified sets breakdown point characterization, breakdown point is infimum over complement of robust region} holds.

	For step 2., first notice that
	\begin{equation}
		\inf_{b \in \boldsymbol{B}_0} \nu(b) = \inf_{b \in \boldsymbol{B}_0} \inf_{Q \in \textbf{P}^b} d(Q \; \Vert \; P_1) = \inf \bigcup_{b \in \boldsymbol{B}_0}\left\{d(Q \; \Vert \; P_1) \; : \; Q \in \textbf{P}^b\right\} \label{Proof display: nominal identified sets breakdown point characterization inf}
	\end{equation}
	If $\delta$ is such that $\textbf{B}_{ID}(\delta) \cap \textbf{B}_0 \neq \varnothing$, then there exists $b \in \textbf{B}_0$ and $Q\in \textbf{P}^b$ such that $d(Q \; \Vert \; P_1) \leq \delta$. This implies
	\begin{align*}
		\inf RR^c = \inf \left\{\delta \in \mathbb{R}_+ \; : \; \textbf{B}_{ID}(\delta) \cap \textbf{B}_0 \neq \varnothing\right\} \geq \inf \bigcup_{b \in \textbf{B}_0} \left\{d(Q \; \Vert \; P_1) \; : \; Q \in \textbf{P}^b\right\}
	\end{align*}
	Conversely, for each real number $a$ satisfying $a = d(Q \; \Vert \; P_1)$ for some $Q \in \textbf{P}^b$, $b \in \textbf{B}_0$, we have that $a \in \left\{\delta \; : \; \textbf{B}_{ID}(\delta) \cap \textbf{B}_0 \neq \varnothing\right\}$. This implies 
	\begin{align*}
		\inf \bigcup_{b \in \textbf{B}_0} \left\{d(Q \; \Vert \; P_1) \; : \; Q \in \textbf{P}_b\right\}  \geq \inf \left\{\delta \; : \; \textbf{B}_{ID}(\delta) \cap \textbf{B}_0 \neq \varnothing\right\} = \inf RR^c
	\end{align*}
	Putting \eqref{Proof display: nominal identified sets breakdown point characterization, breakdown point is infimum over complement of robust region}, \eqref{Proof display: nominal identified sets breakdown point characterization inf}, and these two inequalities together we obtain
	\begin{align*}
		\inf_{b \in \textbf{B}_0} \nu(b) &= \inf \bigcup_{b \in \textbf{B}_0} \left\{d(Q \; \Vert \; P_1) \; : \; Q \in \textbf{P}^b\right\} = \inf \left\{\delta \; : \; \textbf{B}_{ID}(\delta) \cap \textbf{B}_0 \neq \varnothing\right\} = \bar{\delta}^{BP}
	\end{align*}
	as was claimed. $\square$
\end{proof}

%% file: Appendix_MeasuringSelectionBDP.tex
\section{Appendix: Proofs of squared Hellinger results}

\noindent \textbf{Lemma 2.1} 
\textit{
Let $(Z, D) \in \mathbb{R}^{d_z} \times \{0,1\}$ be random variables with $p_D = P(D=1) \in (0,1)$. Let $Z \mid D = 1 \sim P_1$ and $Z \mid D = 0 \sim P_0$. Then
\begin{equation}
	H^2(P_0, P_1) = 1 - \frac{E\left[\sqrt{\text{Var}(D \mid Z)}\right]}{\sqrt{\text{Var}(D)}} \label{Display: squared Hellinger interpretation} \tag{1}
\end{equation}
where the expectation is taken with respect to $p_D P_1 + (1-p_D) P_0$, the marginal distribution of $Z$.
}
\begin{proof}
	The marginal, unconditional distribution of $Z$ is $P = p_D P_1 + (1-p_D) P_0$. This distribution dominates $P_1$ and $P_0$, which have densities 
	\begin{align*}
		&f_1(z) = \frac{P(D = 1 \mid Z = z)}{p_D}, &&f_0(z) = \frac{(1-P(D=1 \mid Z = z))}{1-p_D},
	\end{align*}
	with respect to $P$. This implies
	\begin{align*}
		H^2(P_0, P_1) &= \frac{1}{2}\int\left(\sqrt{f_0(z)} - \sqrt{f_1(z)}\right)^2 dP(z) = \frac{1}{2}\left[\int f_0(z) + f_1(z) - 2\sqrt{f_1(z)f_0(z)} dP(z)\right] \\
		&= 1 - \frac{\int \sqrt{P(D=1 \mid Z = z)(1-P(D=1 \mid Z = z))} dP(z)}{\sqrt{p_D (1-p_D)}} \\
		&= 1 - \frac{E_P\left[\sqrt{\text{Var}(D \mid Z)}\right]}{\sqrt{\text{Var}(D)}}.
	\end{align*}
\end{proof}

\begin{restatable}{lemma}{lemmaSquaredHellingerInequality}
	
	\singlespacing
	
	Let \label{Lemma: squared hellinger inequality} $(Y, X, D) \in \mathbb{R}^{d_y} \times \mathbb{R}^{d_x} \times \{0,1\}$ be random variables with $p_D = P(D = 1) \in (0,1)$. Let $(Y,X) \mid D = d \sim P_d$, and $X \mid D = d \sim P_{dX}$. Then 
	\begin{equation*}
		H^2(P_0, P_1) = 1 - \frac{E\left[\sqrt{\text{Var}(D \mid Y, X)}\right]}{\sqrt{\text{Var}(D)}} \geq 1 - \frac{E[\sqrt{\text{Var}(D \mid X)}]}{\sqrt{\text{Var}(D)}} = H^2(P_{0X}, P_{1X})
	\end{equation*}
\end{restatable}
\begin{proof}
	Observe that the densities of $P_d$ and $P_{dX}$ with respect to $p_D P_1 + (1-p_D) P_0$ are $f_d(Y,X) = E[\mathbbm{1}\{D = d\} \mid Y, X] / P(D = d)$ and 
	\begin{equation*}
		f_{dX}(X) = \frac{E[\mathbbm{1}\{D = d\} \mid X]}{P(D = d)} = E\left[\frac{E[\mathbbm{1}\{D = d\} \mid Y, X]}{P(D = d)} \mid X\right] = E[f_d(Y,X) \mid X] 
	\end{equation*}
	respectively.  Use these to see that 
	\begin{align*}
		H^2(P_1, P_0) &= \frac{1}{2} E\left[\left(\sqrt{f_1(Y,X)} - \sqrt{f_0(Y,X)}\right)^2\right] = 1 - E\left[\sqrt{f_1(Y,X)f_0(Y,X)}\right] \\
		&= 1 - E\left[E\left[\sqrt{f_1(Y,X)}\sqrt{f_0(Y,X)} \mid X\right]\right] \\
		&\geq 1 - E\left[\left(E[f_1(Y,X) \mid X]E[f_0(Y,X) \mid X]\right)^{1/2}\right] \\
		&= 1 - E\left[\sqrt{f_{1X}(X) f_{0X}(X)}\right] \\
		&= H^2(P_{1X}, P_{0X})
	\end{align*}
	where the inequality comes from the Cauchy Schwarz inequality: $E\left[\sqrt{f_1(Y,X)}\sqrt{f_0(Y,X)} \mid X\right] \leq \left(E[f_1(Y,X) \mid X]E[f_0(Y,X) \mid X]\right)^{1/2}$. By lemma 2.1, $H^2(P_0, P_1) = 1 - E\left[\sqrt{\text{Var}(D \mid Y, X)}\right] / \sqrt{\text{Var}(D)}$ and $H^2(P_{0X}, P_{1X}) = 1 - E[\sqrt{\text{Var}(D \mid X)}] / \sqrt{\text{Var}(D)}$, which completes the proof.
\end{proof}
\begin{remark}
	\label{Remark: squared hellinger stronger inequality}
	Lemma \ref{Lemma: squared hellinger inequality} is a special case of example 14 in chapter 5.4 of \cite{pollard2002user}.
\end{remark}

%% file: Appendix_Duality.tex
\section{Appendix: Proofs of duality results}

\label{Appendix: duality}

\begin{restatable}[Unique primal solution]{lemma}{lemmaUniquePrimalSolution}
	\label{Lemma: strictly convex f ensures unique primal solution}
	Suppose $f$ is strictly convex on its domain, $p_D \in (0,1)$, and the infimum in \eqref{Definition: primal problem} is finite. Then any solution attaining the infimum in \eqref{Definition: primal problem} is unique, $P_1$-almost surely.
\end{restatable}
\begin{proof}
	Let $Q^0, Q^1 \in \textbf{P}^b$ attain the finite infimum in \eqref{Definition: primal problem}, and let  $q^0$ and $q^1$ denote their densities with respect to $P_1$. We have that $\frac{-p_D}{(1-p_D)} E_{P_1}[g(Y, X, b)] = E_{Q^0}[g(Y, X, b)] = E_{Q^1}[g(Y, X, b)]$ and $Q_X^0 = Q_X^1 = P_{0X}$. For any $\alpha \in (0,1)$, the measure $Q^\alpha = \alpha Q^1 + (1-\alpha) Q^0 \in \textbf{P}^b$ is feasible in \eqref{Definition: primal problem}, and characterized by the $P_1$-density $\alpha q^1 + (1-\alpha)q^0$. 
	
	Suppose for contradiction that $Q^0$ and $Q^1$ differ on a set of positive $P_1$-measure. Strict convexity implies that for any $(y, x)$ in that set,
	\begin{equation*}
		f(\alpha q^1(y, x) + (1-\alpha) q^0(y, x)) < \alpha f(q^1(y, x)) + (1-\alpha) q^0(y, x)
	\end{equation*}
	Integrating with respect to $P_1$ reveals $d_f(\alpha Q^1 + (1-\alpha) Q^0 \Vert P_1) < \alpha d_f(Q^1 \Vert P_1) + (1-\alpha) d_f(Q^0 \Vert P_1)$, contradicting optimality of $Q^0, Q^1$.
\end{proof}

\begin{restatable}[Weak duality]{lemma}{lemmaWeakDuality}
	\label{Lemma: weak duality}
	Let $\nu(b)$ and $V(b)$ be as defined in \eqref{Definition: primal problem} and \eqref{Definition: dual problem}, respectively. If assumption \ref{Assumption: setting} holds, then $V(b) \leq \nu(b)$ for any $b \in \textbf{B}$.
	
\end{restatable}
\begin{proof}
	First note that if $\nu(b) = \infty$ the inequality holds trivially.
	
	Suppose $\nu(b) < \infty$. Then $\textbf{P}^b \neq \varnothing$, hence there exists at least one density $q(z) = \frac{dQ}{dP_1}(z)$ satisfying $\int h(z,b) q(z) dP_1(z) = c(b)$. Notice that $f^*(r) = \sup_{t \in \mathbb{R}} \{rt - f(t)\}$ implies $f(t) + f^*(r) \geq f(t) + rt - f(t) = rt$. Use this to see that for any $Q \in \textbf{P}^b$ with $P_1$-density $q$,
	\begin{align*}
		f(q(z)) + f^*(\lambda^\intercal h(z, b)) &\geq \lambda^\intercal h(z, b) q(z) \\
		\implies f(q(z)) &\geq \lambda^\intercal h(z, b) q(z) - f^*(\lambda^\intercal h(z,b))
	\end{align*}
	integrating over $z$ with respect to $P_1$ gives
	\begin{align*}
		\int f(q(z)) dP_1(z) &\geq \lambda^\intercal \underbrace{\int h(z, b) q(z) dP_1(z)}_{=c(b)} - \int f^*(\lambda^\intercal h(z,b)) dP_1(z) \\
		\implies d_f(Q \Vert P_1) &\geq \lambda^\intercal c(b) - E\left[f^*(\lambda^\intercal h(Z,b)) \mid D = 1\right]
	\end{align*}
	the left hand side of the last inequality doesn't depend on $\lambda \in \mathbb{R}^{d_g + K}$, while the right hand side doesn't depend on $Q \in \textbf{P}^b$. Hence,
	\begin{align*}
		\nu(b) = \inf_{Q \in \textbf{P}^b} d_f(Q \Vert P_1) \geq \sup_{\lambda \in \mathbb{R}^{d_g + K}} \left\{\lambda^\intercal c(b) - E\left[f^*(\lambda^\intercal h(Z, b)) \mid D = 1\right]\right\} = V(b).
	\end{align*}
\end{proof}

\noindent \textbf{Theorem 3.1} (Strong duality). \textit{Suppose assumptions \ref{Assumption: setting} and \ref{Assumption: strong duality} hold. Then for each $b \in B$, $\nu(b) = V(b)$, with dual attainment.}

\begin{proof}
	Let $\textbf{M}$ be the set of measurable functions mapping $z = (x,y) \mapsto \mathbb{R}$. Consider the relaxed problem 
	\begin{align*}
		\tilde{\nu}(b) &= \inf_{q \in \tilde{\textbf{P}}^b} \int f(q(y, x)) dP_1(y, x) \\
		\tilde{\textbf{P}}^b &= \Bigg\{q \in \textbf{M} \; : \; \int h(y, x, b) q(y, x) dP_1(y, x) = c(b) \Bigg\}
	\end{align*}
	for any $q \in \tilde{\textbf{P}}^b$, $K(\psi) = \int \psi(y, x) q(y, x) dP_1(y, x)$ is a (possibly signed) measure with total measure one. Notice this problem has the same objective as the primal problem \eqref{Definition: primal problem}, but a larger feasible set (the set of finite signed measures with total measure one).
	
	Now apply Theorem II.2 of \citet{csiszar1999mem}, with trivial $\mathcal{K} = \{c(b)\}$. The dual of the relaxed problem is \eqref{Definition: dual problem}. Assumption \ref{Assumption: strong duality} \ref{Assumption: strong duality, constraint qualification} is the ``constraint qualification'' of \citet{csiszar1999mem} Theorem II.2, implying strong duality holds for the relaxed problem, $\tilde{\nu}(b) = V(b)$, and the dual problem's value is attained at a maximum. Let $\lambda(b)$ solve the dual problem. Assumption \ref{Assumption: strong duality} \ref{Assumption: strong duality, interior dual solution} allows application of the second part of Theorem II.2, implying the solution to the relaxed problem is given by
	\begin{equation*}
		q^b(y, x) = (f')^{-1}(\lambda(b)^\intercal h(y, x, b)) = (f^*)'(\lambda(b)^\intercal h(y, x, b))
	\end{equation*}
	By assumption \ref{Assumption: setting} \ref{Assumption: setting, divergence} and Lemma \ref{Lemma: strictly convex f ensures unique primal solution}, this solution is unique $P_1$-almost surely. 
	
	Now we show that $q^b$ in fact solves the primal problem, \eqref{Definition: primal problem}. Notice $q^b$ is nonnegative, because $f'$ is only defined on the non-negative reals. Furthermore,
	\begin{align*}
		\int q^b(x,y) dP_1(x,y) &= \int \sum_{k=1}^K \mathbbm{1}\{x = x_k\} q^b(x,y) dP_1(x,y) \\
		&= \sum_{k=1}^K \int \mathbbm{1}\{x = x_k\} q^b(x,y) dP_1(x,y) \\
		&= \sum_{k=1}^K P(X = x_k \mid D = 0) \\
		&= 1
	\end{align*}
	where the third equality follows from $\int h(y, x, b) q^b(x, y) dP_1(x, y) = c(b)$. The measure $Q^b$ given by $Q^b(\psi) = \int \psi(y, x) q^b(y, x) dP_1(y, x)$ is therefore a probability distribution dominated by $P_1$. Therefore $Q^b \in \textbf{P}^b$ is feasible in the primal problem \eqref{Definition: primal problem}. Being feasible in the primal and solving the relaxed problem, $Q^b$ must also solve the primal problem. 
\end{proof}

%% file: Appendix_Estimation_TechnicalLemmas.tex
\section{Appendix: Proofs of estimation results}

\label{Appendix: estimation}

\subsection{Technical lemmas}

\label{Appendix: estimation, technical lemmas}

These results are self-contained, with notation not related to the present paper.

\begin{restatable}[Uniform continuity of maps between bounded functions]{lemma}{lemmaContinuityOfMapsBetweenBoundedFunctions}
	
	\label{Technical Lemma: continuity of maps between bounded functions}
	
	Let $T$ be a set, $E^t \subseteq \mathbb{R}^{d_E}$ for each $t \in T$, $E^T = \{(t,e) \; : \; t \in T, e \in E^t\}$, and $\ell^\infty(T, E^T)^{d_E}$ be the subset of $\ell^\infty(T)^{d_E}$ whose graph is a subset of $E^T$:
	\begin{equation*}
		\ell^\infty(T, E^T)^{d_E} = \left\{g : T \rightarrow \mathbb{R}^{d_E} \; \text{ such that } \; g(t) \in E^t, \; \sup_{t \in T} \lVert g(t) \rVert < \infty\right\} \subset \ell^\infty(T)^{d_E}
	\end{equation*}
	Let $f : E^T \rightarrow \mathbb{R}^{d_f}$ be such that $\sup_{t \in T} \lVert f(t, g(t)) \rVert < \infty$ for any $g \in \ell^\infty(T, E^T)^{d_E}$, and define
	\begin{align*}
		&\tilde{f} : \ell^\infty(T, E^T)^{d_E} \rightarrow \ell^\infty(T)^{d_f}, &&\tilde{f}(g)(t) = f(t, g(t)).
	\end{align*}
	If $\{f(t, \cdot)\}_{t \in T}$ is uniformly equicontinuous, then $\tilde{f}$ is uniformly continuous.
	
\end{restatable}
\begin{proof}
	Let $\varepsilon > 0$, and use uniform equicontinuity to choose $\delta > 0$ such that
	\begin{equation*}
		\lvert e_1 - e_2 \rvert < \delta \implies \lvert f(t, e_1) - f(t, e_2) \rvert < \varepsilon/2
	\end{equation*}
	for any $t \in T$. Notice that if $ g_1, g_2 \in \ell^\infty(T, E^T)^{d_E}$ with $\lVert g_1 - g_2 \rVert_T = \sup_{t \in T} \lvert g_1(t) - g_2(t) \rvert < \delta$, then 
	\begin{equation*}
		\lVert \tilde{f}(g_1) - \tilde{f}(g_2) \rVert_T = \sup_{t \in T} \lvert f(t, g_1(t) - f(t, g_2(t)) \rvert \leq \varepsilon/2 < \varepsilon
	\end{equation*}
	and hence $\lVert g_1 - g_2 \rVert_T < \delta \implies \lVert \tilde{f}(g_1) - \tilde{f}(g_2) \rVert_T < \varepsilon$.
\end{proof}

\begin{remark}
	Lemma \ref{Technical Lemma: continuity of maps between bounded functions} implies many simpler special cases. For example, suppose that for all $t, t' \in T$,  $f(t,e) = f(t',e)$ and $E^t = E \subseteq \mathbb{R}$. Then lemma \ref{Technical Lemma: continuity of maps between bounded functions} simplifies to: if $f : E \rightarrow \mathbb{R}$ is uniformly continuous, then $\tilde{f} : \ell^\infty(T) \rightarrow \ell^\infty(T)$ defined pointwise by $\tilde{f}(g)(t) = f(g(t))$ is continuous. 
\end{remark}

\begin{restatable}[Restricted infimum is uniformly continuous]{lemma}{lemmaRestrictedInfimumIsUniformlyContinuous}
	\label{Technical Lemma: restricted infimum is uniformly continuous}
	
	For any $A \subseteq T$ and any $f,g \in \ell^\infty(T)$,
	\begin{equation*}
		\left\lvert \inf_{t \in A} f(t) - \inf_{t \in A} g(t) \right\rvert \leq \sup_{t \in A} \lvert f(t) - g(t) \rvert  
	\end{equation*}
	as a result, $\iota : \ell^\infty(T) \rightarrow \mathbb{R}$ given by $\iota(h) = \inf_{t \in A} h(t)$ is uniformly continuous.
\end{restatable}
\begin{proof}
	Notice that 
	\begin{align*}
		\sup_{t \in A} f(t) - \sup_{t \in A} g(t) &\leq \sup_{t \in A} \{f(t) - g(t)\} \leq \sup_{t \in A} \lvert f(t) - g(t) \rvert, \text{ and } \\
		-\left[\sup_{t \in A} f(t) - \sup_{t \in A} g(t)\right] = \sup_{t \in A} g(t) - \sup_{t \in A} f(t) &\leq \sup_{t \in A} \{g(t) - f(t)\} \leq \sup_{t \in A} \lvert g(t) - f(t) \rvert = \sup_{t \in A} \lvert f(t) - g(t) \rvert,
	\end{align*}
	hence $-\sup_{t \in A} \lvert f(t) - g(t) \rvert \leq \sup_{t \in A} f(t) - \sup_{t \in A} g(t) \leq \sup_{t \in A} \lvert f(t) - g(t) \rvert$, or equivalently 
	\begin{equation*}
		\left\lvert \sup_{t \in A} f(t) - \sup_{t \in A} g(t) \right\rvert \leq \sup_{t \in A} \lvert f(t) - g(t) \rvert.
	\end{equation*}
	Use this to see the claimed inequality:
	\begin{align*}
		\left\lvert \inf_{t \in A} f(t) - \inf_{t \in A} g(t) \right\rvert &= \left\lvert -\sup_{t \in A} \{-f(t)\} - \left(-\sup_{t \in A} \{-g(t)\}\right)\right\rvert = \left\lvert \sup_{t \in A} \{-g(t)\} - \sup_{t \in A} \{-f(t)\} \right\rvert \\
		&\leq \sup_{t \in A} \lvert -g(t) - \{-f(t)\}\rvert = \sup_{t \in A} \lvert f(t) - g(t) \rvert.
	\end{align*}
	
	Regarding the continuity claim, let $\varepsilon > 0$ and set $\delta = \varepsilon$. Then 
	\begin{align*}
		\lvert \iota(f) - \iota(g) \rvert \leq  \sup_{t \in A} \lvert f(t) - g(t) \rvert \leq \sup_{t \in T} \lvert f(t) - g(t) \rvert = \lVert f - g \rVert_T,
	\end{align*}
	hence $\lVert f - g \rVert_T < \delta$ implies $\lvert \iota(f) - \iota(g) \rvert < \varepsilon$.
\end{proof}

\begin{restatable}[Restricted infimum is Hadamard directionally differentiable]{lemma}{lemmaRestrictedInfimumIsHadamardDirectionallyDifferentiable}
	\label{Technical Lemma: restricted infimum is hadamard directionally differentiable}
	
	Let $(T, d)$ be a metric space, $A$ a compact subset of $T$, and 
	\begin{align*}
		&\iota : \ell^\infty(T) \rightarrow \mathbb{R}, &&\iota(f) = \inf_{t \in A} f(t)
	\end{align*}
	Then $\iota$ is Hadamard directionally differentiable at any $f \in \mathcal{C}(T, d)$ tangentially to $\mathcal{C}(T, d)$. $\Psi_A(f) = \argmin_{t \in A} f(t)$ is nonempty, and the directional derivative is given by
	\begin{align*}
		&\iota_f' : \mathcal{C}(T, d) \rightarrow \mathbb{R}, &&\iota_f'(h) = \inf_{t \in \Psi_A(f)} h(t)
	\end{align*}
	If $\Psi_A(f)$ is the singleton $\{t_f\}$, then $\iota$ is fully Hadamard differentiable at $f$ tangentially to $\mathcal{C}(T, d)$ and $\iota_f'(h) =  h(t_f)$. 
\end{restatable}
\begin{proof}
	
	The result is essentially a corollary of \cite{fang2019inference} Lemma S.4.9, which shows that $\phi : \ell^\infty(A) \rightarrow \mathbb{R}$ given by $\phi(f) = \sup_{t \in A} f(t)$ is Hadamard directionally differentiable at any $f \in \mathcal{C}(A, d)$ tangentially to $\mathcal{C}(A, d)$, with directional derivative
	\begin{align*}
		&\phi_f' : \mathcal{C}(A, d) \rightarrow \mathbb{R}, &&\phi_f'(h) = \sup_{t \in \Psi_A(f)} h(t).
	\end{align*}
	See \cite{fang2019inference} definition 2.1 for definitions of Hadamard directionally differentiable and fully Hadamard differentiable.
	
	Let $f \in \mathcal{C}(T, d)$ and note that $\Psi_A(f) = \argmin_{t \in A} f(t)$ is nonempty by the extreme value theorem. Let $\{h_n\}_{n=1}^\infty \subseteq \ell^\infty(T)$ and $\{r_n\}_{n=1}^\infty \subseteq \mathbb{R}_+$ be such that $h_n \rightarrow h \in \mathcal{C}(T, d)$ and $r_n \downarrow 0$. For $g \in \ell^\infty(T)$, let $g_A : A \rightarrow \mathbb{R}$ be the restriction of $g$ to $A$, given by $g_A(t) = g(t)$. Observe that $g \in \mathcal{C}(T, d)$ implies $g_A \in \mathcal{C}(A, d)$. Now notice that
	\begin{align*}
		&\left\lvert \frac{\iota(f + r_n h_n) - \iota(f)}{r_n} - \iota_a'(h) \right\rvert \\
		&\hspace{1 cm} = \left\lvert  \frac{\inf_{t \in A}\{f(t) + r_n h_n(t)\} - \inf_{t \in A} f(t)}{r_n} - \inf_{t \in \Psi_A(f)} h(t)\right\rvert \\
		&\hspace{1 cm} = \left\lvert \frac{-\sup_{t \in A} \{-f(t) - r_n h_n(t)\} - \left(-\sup_{t \in A} \{-f(t)\}\right)}{r_n} - \left(-\sup_{t \in \Psi_A(f)} \{-h(t)\}\right)\right\rvert \\
		&\hspace{1 cm} = \left\lvert \frac{\sup_{t \in A} \{-f(t) + r_n \left(-h_n(t)\right)\} - \left(\sup_{t \in A} \{-f(t)\}\right)}{r_n} - \left(\sup_{t \in \Psi_A(f)} \{-h(t)\}\right)\right\rvert \\
		&\hspace{1 cm} = \left\lvert \frac{\phi(-f_A + r_n (-h_{n,A})) - \phi(-f_A)}{r_n} - \phi_f'(-h_A)\right\rvert,
	\end{align*}
	where the last equality follows from the definitions and the fact that $\Psi_A(f) = \argmin_{a \in A} f(a) = \argmax_{a \in A} \{-f_A(a)\}$. 
	
	$h_n \rightarrow h \in h \in \mathcal{C}(T, d)$ implies $-h_{n,A} \rightarrow -h_A \in \mathcal{C}(A, d)$. Thus, \cite{fang2019inference} Lemma S.4.9 and the definition of Hadamard directional differentiability implies
	\begin{equation*}
		\lim_{n\rightarrow \infty} \left\lvert \frac{\iota(f + t_n h_n) - \iota(f)}{t_n} - \iota_a'(h) \right\rvert = \lim_{n\rightarrow \infty}\left\lvert \frac{\phi(-f_A + t_n (-h_{n,A})) - \phi(-f_A)}{t_n} - \phi_f'(-h_A)\right\rvert = 0.
	\end{equation*}
	Finally, if $\Psi_A(f) = \{t_f\}$ then $\inf_{t \in \Psi_A(f)}(h) = h(t_f)$ is linear in $h$, and hence $\iota$ is fully Hadamard differentiable at $f$. 
\end{proof}

\begin{restatable}[Uniform consistency of estimated moments]{lemma}{lemmaUniformConsistencyOfEstimatedMoments}
	\label{Technical Lemma: uniform consistency of estimated moments}
	
	Let $\mathcal{X} \subseteq \mathbb{R}^{d_X}$, $T \subseteq \mathbb{R}^{d_T}$, $E^t \subseteq \mathbb{R}^{d_E}$, $E^T = \{(t, e) \; : \; t \in T, e \in E^t\}$, 
	\begin{align*}
		&\hat{\gamma}_n, \gamma : T \rightarrow \mathbb{R}^{d_E}, &&\text{ and } &&f : \mathcal{X} \times E^T \rightarrow \mathbb{R}^{K \times J}.
	\end{align*}
	Suppose that 
	\begin{enumerate}[label=(\roman*)]
		\item $\{X_i\}_{i=1}^n$ is i.i.d.,
		\item $\sup_{t \in T} \lVert \hat{\gamma}_n(t) - \gamma(t) \rVert \overset{p}{\rightarrow} 0$,
		\item $\text{Gr}(\gamma) = \{(t, \gamma(t)) \; : \; t \in T\}$ is bounded, and
		\item there exists a finite $\varepsilon > 0$ such that $(t, e) \mapsto f(x, t, e)$ is continuous on 
		\begin{equation*}
			\text{Gr}(\gamma)^\varepsilon \equiv \left\{(t,e) \in E^T \; : \; \inf_{(t', e') \in \text{Gr}(\gamma)} \lVert (t, e) - (t', e') \rVert \leq \varepsilon\right\}
		\end{equation*}
		for all $x \in \mathcal{X}$, and 
		\begin{equation*}
			E\left[\sup_{(t, e) \in \text{Gr}(\gamma)^\varepsilon} \lVert f(X, t, e) \rVert\right] < \infty.
		\end{equation*}
	\end{enumerate}
	Then
	\begin{equation*}
		\sup_{t \in T} \left\lVert \frac{1}{n}\sum_{i=1}^n f(X_i, t, \hat{\gamma}_n(t)) - E[f(X, t, \gamma(t))] \right\rVert \overset{p}{\rightarrow} 0.
	\end{equation*}
\end{restatable}
\begin{proof}
	
	The triangle inequality implies
	\begin{align*}
		&\sup_{t \in T} \left\lVert \frac{1}{n}\sum_{i=1}^n f(X_i, t, \hat{\gamma}_n(t)) - E[f(X, t, \gamma(t))] \right\rVert \\
		&\hspace{1 cm} \leq \sup_{t \in T} \left\lVert \frac{1}{n}\sum_{i=1}^n f(X_i, t, \hat{\gamma}_n(t)) - E[f(X, t, \hat{\gamma}_n(t))] \right\rVert + \sup_{t \in T} \left\lVert E[f(X, t, \hat{\gamma}_n(t))] - E[f(X, t, \gamma(t))] \right\rVert 
	\end{align*}
	
	Consider the second term first. The dominated convergence theorem, $(t, e) \mapsto f(x, t, e)$ being continuous, and $E\left[\sup_{(t, e) \in \text{Gr}(\gamma)^\varepsilon} \lVert f(X, t, e) \rVert\right] < \infty$ implies that 
	\begin{align*}
		&\psi : \text{Gr}(\gamma)^\varepsilon \rightarrow \mathbb{R}^{K \times J}, &&\psi(t, e) = E[f(X, t, e)]
	\end{align*}
	is continuous. $\text{Gr}(\gamma)^\varepsilon$ is a closed and bounded subset of $\mathbb{R}^{d_T} \times \mathbb{R}^{d_E}$, hence compact by the Heine-Borel theorem. Thus, $\psi$ is in fact uniformly continuous by the Heine-Cantor theorem. Lemma \ref{Technical Lemma: continuity of maps between bounded functions} then implies
	\begin{align*}
		&\Psi : \ell^\infty(T, \text{Gr}(\gamma)^\varepsilon) \rightarrow \ell^\infty(T)^{K \times J}, &&\Psi(g)(t) = \psi(t, g(t))
	\end{align*}
	is continuous. $\sup_{t \in T} \left\lVert E[f(X, t, \hat{\gamma}_n(t))] - E[f(X, t, \gamma(t))] \right\rVert \overset{p}{\rightarrow} 0$ follows from $\sup_{t \in T} \lVert \hat{\gamma}_n(t) - \gamma(t) \rVert \overset{p}{\rightarrow} 0$ and the continuous mapping theorem.
	
	Now consider the first term. Compactness of $\text{Gr}(\gamma)^\varepsilon$, continuity of $(t, e) \mapsto f(x, t, e)$ on $\text{Gr}(\gamma)^\varepsilon$, and $E\left[\sup_{(t, e) \in \text{Gr}(\gamma)^\varepsilon} \lVert f(X, t, e) \rVert\right] < \infty$ implies that $\left\{f(X, t, e) \; : \; (t,e) \in \text{Gr}(\gamma)^\varepsilon\right\}$ is Glivenko-Cantelli by \citet{van2007asymptotic} example 19.8. With probability approaching one, $\sup_{t \in T} \lVert \hat{\gamma}_n(t) - \gamma(t) \rVert < \varepsilon$ and when this holds,
	\begin{align*}
		&\sup_{t \in T} \left\lVert \frac{1}{n}\sum_{i=1}^n f(X_i, t, \hat{\gamma}_n(t)) - E[f(X, t, \hat{\gamma}_n(t))] \right\rVert \leq \sup_{(t, g) \in \text{Gr}(\gamma)^\varepsilon} \left\lVert \frac{1}{n}\sum_{i=1}^n f(X_i, t, e) - E[f(X, t, e)] \right\rVert \overset{p}{\rightarrow} 0.
	\end{align*}
	This concludes the proof.
\end{proof}

\begin{restatable}[Uniform consistency of matrix inverses]{lemma}{lemmaUniformConsistencyOfMatrixInverses}
	
	\label{Technical Lemma: uniform consistency of matrix inverses}
	
	Let $\hat{\Phi}_n, \Phi : T \rightarrow \mathbb{R}^{K \times K}$. 
	If 
	\begin{enumerate}[label=(\roman*)]
		\item $\Phi(t)^{-1}$ exists for all $t \in T$,
		\item $\sup_{t \in T} \lVert \Phi(t) \rVert_o < \infty$ and $\sup_{t \in T} \lVert \Phi(t)^{-1} \rVert_o < \infty$, and
		\item $\sup_{t \in T} \lVert \hat{\Phi}_n(t) - \Phi(t) \rVert_o \overset{p}{\rightarrow} 0$, 
	\end{enumerate}
	then with probability approaching one, the function mapping $T$ to $\hat{\Phi}_n(t)^{-1}$ is well defined and 
	\begin{equation*}
		\sup_{t \in T} \lVert\hat{\Phi}_n(t)^{-1} - \Phi(t)^{-1}\rVert_o \overset{p}{\rightarrow} 0.
	\end{equation*}
\end{restatable}
\begin{proof}
	It suffices to show that the singular values of $\hat{\Phi}_n(t)$ converge in probability to the singular values of $\Phi(t)$, uniformly over $t \in T$:
	\begin{equation}
		\sup_{t \in T} \max_k \lvert \sigma_k(\hat{\Phi}_n(t)) - \sigma_k(\Phi(t)) \rvert \overset{p}{\rightarrow} 0. \label{Proof display: uniform consistency of matrix inverses, uniform consistency of singular values}
	\end{equation}
	To see why, notice that $\infty > \sup_{t \in T} \lVert \Phi(t)^{-1} \rVert_o = \sup_{t \in T} \frac{1}{\sigma_K(\Phi(t))} = \frac{1}{\inf_{t \in T} \sigma_K(\Phi(t))}$ implies $\varepsilon \equiv \inf_{t \in T} \sigma_K(\Phi(t)) > 0$. Condition \eqref{Proof display: uniform consistency of matrix inverses, uniform consistency of singular values} implies that with probability approaching one, 
	\begin{equation*}
		\sup_{t \in T} \max_k \left\lvert \sigma_k(\hat{\Phi}_n(t)) - \sigma_k(\Phi(t)) \right\rvert < \varepsilon/2,
	\end{equation*}
	and on this event the function mapping $T$ to $\hat{\Phi}_n(t)^{-1}$ is well defined. Then notice that 
	\begin{align*}
		\left\lVert \hat{\Phi}_n(t)^{-1} - \Phi(t)^{-1} \right\rVert_o &= \left\lVert \hat{\Phi}_n(t)^{-1}(\Phi(t) - \hat{\Phi}_n(t))\Phi(t)^{-1} \right\rVert_o \\
		&\leq \left\lVert \hat{\Phi}_n(t)^{-1} \right\rVert_o \left\lVert \Phi(t) - \hat{\Phi}_n(t)\right\rVert_o \left\lVert \Phi(t)^{-1} \right\rVert_o,
	\end{align*}
	implying 
	\begin{equation}
		\sup_{t \in T} \left\lVert \hat{\Phi}_n(t)^{-1} - \Phi(t)^{-1} \right\rVert_o \leq \sup_{t \in T} \left\lVert \hat{\Phi}_n(t)^{-1} \right\rVert_o \sup_{t \in T} \left\lVert \Phi(t) - \hat{\Phi}_n(t)\right\rVert_o \sup_{t \in T} \left\lVert \Phi(t)^{-1} \right\rVert_o. \label{Proof display: uniform consistency of matrix inverses, main inequality}
	\end{equation}
	Recall that $\sup_{t \in T} \lVert \Phi(t)^{-1} \rVert_o < \infty$ and $\sup_{t \in T} \lVert \hat{\Phi}_n(t) - \Phi(t) \rVert_o \overset{p}{\rightarrow} 0$ are assumed, the latter implying $\sup_{t \in T} \lVert \hat{\Phi}_n(t) \rVert_o = O_p(1)$ by the continuous mapping theorem. Thus \eqref{Proof display: uniform consistency of matrix inverses, main inequality} implies \\ $\sup_{t \in T} \left\lVert \hat{\Phi}_n(t)^{-1} - \Phi(t)^{-1} \right\rVert_o \overset{p}{\rightarrow} 0$. \\
	 
	 The argument that \eqref{Proof display: uniform consistency of matrix inverses, uniform consistency of singular values} holds is broken into three steps:
	 \begin{enumerate}
	 	\item Show that $\hat{\Phi}_n(t)^\intercal \hat{\Phi}_n(t)$ is uniformly consistent for $\Phi(t)^\intercal \Phi(t)$. 
	 	
	 	Notice that 
	 	\begin{align*}
	 		&\sup_{t \in T} \left\lVert \hat{\Phi}_n(t)^\intercal \hat{\Phi}_n(t) - \Phi(t)^\intercal \Phi(t) \right\rVert_o \\
	 		&\hspace{1 cm} \leq \sup_{t \in T} \left\lVert \hat{\Phi}_n(t)^\intercal \hat{\Phi}_n(t) - \hat{\Phi}_n(t)^\intercal \Phi(t) \right\rVert_o +  \sup_{t \in T} \left\lVert \hat{\Phi}_n(t)^\intercal \Phi(t) - \Phi(t)^\intercal \Phi(t) \right\rVert_o \\
	 		&\hspace{1 cm} \leq \sup_{t \in T}\left\lVert \hat{\Phi}_n(t)^\intercal \right\rVert_o \sup_{t \in T} \left\lVert \hat{\Phi}_n(t) - \Phi(t) \right\rVert_o + \sup_{t \in T} \left\lVert \hat{\Phi}_n(t)^\intercal - \Phi(t)^\intercal  \right\rVert_o \sup_{t \in T} \left\lVert \Phi(t) \right\rVert_o 
	 	\end{align*}
	 	Recall that for any square matrix $A \in \mathbb{R}^{K \times K}$, 
	 	\begin{equation*}
	 		\lVert A^\intercal \rVert_{\text{max}} = \lVert A \rVert_{\text{max}} \leq \lVert A \rVert_o \leq K \lVert A \rVert_{\text{max}} = K \lVert A^\intercal \rVert_{\text{max}}.
	 	\end{equation*}
	 	Use this to see that
	 	\begin{align*}
	 		&\sup_{t \in T} \lVert \hat{\Phi}_n(t)^\intercal \rVert_o \leq K \sup_{t \in T} \lVert \hat{\Phi}_n(t) \rVert_o &&\text{ and } &&\sup_{t \in T} \left\lVert \hat{\Phi}_n(t)^\intercal - \Phi(t)^\intercal  \right\rVert_o \leq K \sup_{t \in T} \left\lVert \hat{\Phi}_n(t) - \Phi(t) \right\rVert_o,
	 	\end{align*}
	 	and therefore
	 	\begin{align}
	 		&\sup_{t \in T} \left\lVert \hat{\Phi}_n(t)^\intercal \hat{\Phi}_n(t) - \Phi(t)^\intercal \Phi(t) \right\rVert_o \notag \\
	 		&\hspace{1 cm} \leq K\left(\sup_{t \in T} \lVert \hat{\Phi}_n(t) \rVert + \sup_{t \in T} \lVert \Phi(t) \rVert\right) \sup_{t \in T} \lVert \hat{\Phi}_n(t) - \Phi(t) \rVert \label{uniform consistency of matrix inverses, step one inequality}
	 	\end{align}
	 	$\sup_{t \in T} \lVert \Phi(t)^{-1} \rVert_o < \infty$ and $\sup_{t \in T} \lVert \hat{\Phi}_n(t) - \Phi(t) \rVert_o \overset{p}{\rightarrow} 0$ by assumption, implying $\sup_{t \in T} \lVert \hat{\Phi}_n(t) \rVert_o = O_p(1)$ by the continuous mapping theorem, and thus $\sup_{t \in T} \left\lVert \hat{\Phi}_n(t)^\intercal \hat{\Phi}_n(t) - \Phi(t)^\intercal \Phi(t) \right\rVert_o \overset{p}{\rightarrow} 0$ by \eqref{uniform consistency of matrix inverses, step one inequality}.
	 	
	 	\item Show the eigenvalues of $\hat{\Phi}_n(t)^\intercal \hat{\Phi}_n(t)$ converge to the eigenvalues of $\Phi(t)^\intercal \Phi(t)$ uniformly over $t \in T$.
	 	
	 	Apply Weyl's perturbation theorem, found in \citet{Bhatia1997} as corollary III.2.6: for Hermitian matrices $A$ and $B$,
	 	\begin{equation*}
	 		\max_k \lvert \alpha_k(A) - \alpha_k(b) \rvert \leq \lVert A - B \rVert_o
	 	\end{equation*}
	 	For real matrices Hermitian is equivalent to symmetric, so Weyl's perturbation theorem implies
	 	\begin{align*}
	 		&\sup_{t \in T} \max_k \lvert \alpha_k(\hat{\Phi}_n(t)^\intercal \hat{\Phi}_n(t)) - \alpha_k(\Phi(t)^\intercal \Phi(t)) \rvert \\
	 		&\hspace{1 cm} \leq \sup_{t \in T} \lvert \hat{\Phi}_n(t)^\intercal \hat{\Phi}_n(t) - \Phi(t)^\intercal \Phi(t) \rVert_o \overset{p}{\rightarrow} 0
	 	\end{align*}
	 	In other words, the eigenvalues of $\hat{\Phi}_n(t)^\intercal \hat{\Phi}_n(t)$ converge to the eigenvalues of $\Phi(t)^\intercal \Phi(t)$ uniformly over $t \in T$. These eigenvalues are the squared singular values of $\Phi_n(t)$.  
	 	
	 	\item Apply the continuous mapping theorem to conclude \eqref{Proof display: uniform consistency of matrix inverses, uniform consistency of singular values} holds.
	 	
	 	Let $\ell^\infty(T, [0,\infty))$ denote the subset of $\ell^\infty(T)$ consisting of functions $h$ taking nonnegative real values: $h : T \rightarrow [0, \infty)$. Lemma \ref{Technical Lemma: continuity of maps between bounded functions} shows that if $f : [0,\infty) \rightarrow \mathbb{R}$ is uniformly continuous, then $\tilde{f} : \ell^\infty(T, [0,\infty)) \rightarrow \ell^\infty(T)$ given pointwise by $\tilde{f}(h)(t) = f(h(t))$ is continuous. It is well known that the square root function $x \mapsto \sqrt{x}$ is uniformly continuous on $[0, \infty)$. Thus \eqref{Proof display: uniform consistency of matrix inverses, uniform consistency of singular values} follows by the continuous mapping theorem.
	 \end{enumerate}
\end{proof}

%% file: Appendix_Estimation_Consistency.tex
\subsection{Consistency}
\label{Appendix: estimation, consistency proofs}

\begin{restatable}[Unique dual solution]{lemma}{lemmaUniqueDualSolution}
	\label{Lemma: unique dual solution}
	
	Suppose assumptions \ref{Assumption: setting} and \ref{Assumption: strong duality} hold, $b \in B$, and \\ $E[h(Y, X, b)h(Y, X, b)^\intercal \mid D = 1]$ is nonsingular. Then $M(b,\lambda) \equiv E[\varphi(D, DY, X, b, \lambda, p_D)]$ is strictly concave in $\lambda$ and $\lambda(b) = \argmax_{\lambda \in \mathbb{R}^{d_g + K}} M(b,\lambda)$ is unique. 
\end{restatable}
\begin{proof}
	Let $\lambda \neq \tilde{\lambda}$ and $\alpha \in (0,1)$. Since $f$ is essentially smooth, $f^*$ is strictly convex and as a result,
	\begin{equation}
		f^*((\alpha \tilde{\lambda} + (1-\alpha) \lambda)^\intercal h(y, x, b)) < \alpha f^*(\tilde{\lambda}^\intercal h(y, x, b)) + (1-\alpha) f^*(\lambda^\intercal h(y, x, b)) \label{Proof display: unique dual solution}
	\end{equation}
	for any $(y, x)$ where $\tilde{\lambda}^\intercal h(y, x, b) \neq \lambda^\intercal h(y, x, b)$, equivalently, where $(\lambda - \tilde{\lambda})^\intercal h(y, x, b) \neq 0$. Since $\lambda - \tilde{\lambda} \neq 0$, nonsingularity of $E[h(Y, X, b) h(Y,X, b)^\intercal \mid D = 1]$ implies
	\begin{align*}
		0 < (\lambda - \tilde{\lambda})^\intercal E[h(Y, X, b)h(Y, X, b)^\intercal \mid D = 1](\lambda - \tilde{\lambda}) = E[\big((\lambda - \tilde{\lambda})^\intercal h(Y, X, b)\big)^2 \mid D = 1]
	\end{align*}
	implies $\left\{(y, x) \; : \; (\lambda - \tilde{\lambda})^\intercal h(y, x, b) \neq 0\right\}$ is a $P_1$-nonnegligible set. It follows that 
	\begin{align*}
		&E\left[\frac{D}{p_D}f^*((\alpha \tilde{\lambda} + (1-\alpha) \lambda)^\intercal h(DY, X, b))\right] \\
		&\hspace{2 cm} < \alpha E\left[\frac{D}{p_D}f^*(\tilde{\lambda}^\intercal h(DY, X, b))\right] + (1-\alpha) E\left[\frac{D}{p_D}f^*(\lambda^\intercal h(Y, X, b))\right]
	\end{align*}
	and hence
	\begin{align*}
		&M(b, \alpha \tilde{\lambda} + (1-\alpha) \lambda) = E[\varphi(D, DY, X, b, \alpha \tilde{\lambda} + (1-\alpha) \lambda, p_D)] \\
		&\hspace{1 cm} = E\left[\frac{(\alpha \tilde{\lambda} + (1-\alpha) \lambda)^\intercal J(D) h(DY, X, b)}{1-p_D} - \frac{D}{p_D} f^*((\alpha \tilde{\lambda} + (1-\alpha) \lambda)^\intercal h(DY, X, b)) \right] \\
		&\hspace{1 cm} > \alpha E\left[\frac{\tilde{\lambda}^\intercal J(D) h(DY, X, b)}{1-p_D}\right] + (1-\alpha)E\left[\frac{ \lambda^\intercal J(D) h(DY, X, b)}{1-p_D}\right] \\
		&\hspace{2 cm} - \alpha E\left[\frac{D}{p_D}f^*(\tilde{\lambda}^\intercal h(DY, X, b))\right] - (1-\alpha) E\left[\frac{D}{p_D}f^*(\lambda^\intercal h(Y, X, b))\right] \\
		&\hspace{1 cm} = \alpha M(b, \tilde{\lambda}) + (1-\alpha) M(b, \lambda)
	\end{align*}
	Therefore $M(b, \cdot)$ is strictly concave. $M(b, \cdot)$ attains a maximum by Theorem 3.1, and strict concavity implies this maximizer is unique.
\end{proof}

\begin{restatable}[Continuous dual solution and value function]{lemma}{lemmaContinuousDualSolutionAndValueFunction}
	\label{Lemma: continuous dual solution and value function}
	
	Suppose assumptions \ref{Assumption: setting}, \ref{Assumption: strong duality}, and \ref{Assumption: estimation} hold. Then $\lambda(b) = \argmax_{\lambda \in \mathbb{R}^{d_g + K}} M(b, \lambda)$, $\nu(b) = M(b, \lambda(b))$, and $\nabla_\lambda^2 M(b, \lambda(b))$ are all continuous. 
\end{restatable}
\begin{proof}
	
	Jensen's inequality and assumption \ref{Assumption: estimation} \ref{Assumption: estimation, moment conditions} imply that 
	\begin{equation*}
		E\left[\sup_{(b, \nu, \lambda, p) \in \text{Gr}(\theta_0)^\eta}\lVert \nabla_{(b,\nu, \lambda, p)} \phi(D, DY, X, b, \nu, \lambda, p)\rVert\right] < \infty
	\end{equation*}
	and, therefore, the following inequalities as well:
	\begin{align*}
		&E\left[\sup_{(b, \lambda) \in \text{Gr}(\lambda)^\eta} \lVert \nabla_\lambda \varphi(D, DY, X, b, \lambda, p_D) \rVert\right] < \infty, &&E\left[\sup_{(b, \lambda) \in \text{Gr}(\lambda)^\eta} \lVert \nabla_\lambda^2 \varphi(D, DY, X, b, \lambda, p_D) \rVert \right] < \infty
	\end{align*}
	where $\text{Gr}(\lambda) = \left\{(b, \lambda(b)) \; : \; b \in B\right\}$ and $\text{Gr}(\lambda)^\eta = \left\{(b,\lambda) \; : \; \inf_{(b',\lambda') \in \text{Gr}(\lambda)} \lVert (b, \lambda) - (b', \lambda')\rVert \leq \eta \right\}$. 
	It follows from the dominated convergence theorem that $M(b,\lambda) = E[\varphi(D, DY, X, b, \lambda, p_D)]$ is twice continuously differentiable with respect to $\lambda$ in a neighborhood of $\lambda(b)$ for every $b \in B$, with $\nabla_\lambda M(b, \lambda) = E\left[\nabla_\lambda \varphi(D, DY, X, b, \lambda, p_D)\right]$ and $\nabla_\lambda^2 M(b, \lambda) = E\left[\nabla_\lambda^2 \varphi(D, DY, X, b, \lambda, p_D)\right]$. 
	
	$\lambda(b)$ must therefore solve the first order condition 
	\begin{equation*}
		0 = \nabla_\lambda M(b, \lambda(b)) = E\left[\nabla_\lambda \varphi(D, DY, X, b, \lambda, p_D)\right].
	\end{equation*}
	Apply the implicit function theorem to this equation. The maps $(b, \lambda) \mapsto \nabla_\lambda M(b, \lambda)$ and $(b, \lambda) \mapsto \nabla_\lambda^2 M(b, \lambda)$ exist and are continuous on an open neighborhood of $(b,\lambda(b))$. Moreover, strict concavity of $M(b, \cdot)$ shown in lemma \ref{Lemma: unique dual solution} implies $\nabla_\lambda^2 M(b, \lambda(b))$ is negative definite and hence invertible. It follows from the implicit function theorem (found in \cite{zeidler1986nonlinear} as theorem 4.B) that $\lambda(b)$ is continuous in a neighborhood of $b$. Since this holds for every $b \in B$, the function $\lambda : B \rightarrow \mathbb{R}^{d_g + K}$ is continuous.
	
	Assumption \ref{Assumption: estimation} \ref{Assumption: estimation, moment conditions} and the dominated convergence theorem implies $M(b,\lambda) = E\left[\varphi(D, DY, X, b, \lambda, p_D) \right]$ and $(b, \lambda) \mapsto \nabla_\lambda^2 M(b, \lambda)$ are continuous. This implies $\nu(b) = M(b, \lambda(b))$ and $b \mapsto \nabla_\lambda^2 M(b, \lambda(b))$ are the composition of continuous functions and hence continuous. 
\end{proof}

\begin{restatable}[Uniform consistency of the dual objective]{lemma}{lemmaUniformConsistencyOfDualObjective}
	\label{Lemma: uniform consistency of the dual objective}
	
	Suppose assumptions \ref{Assumption: setting}, \ref{Assumption: strong duality}, and \ref{Assumption: estimation} hold, and let $\hat{M}_n(b,\lambda) \equiv \frac{1}{n}\sum_{i=1}^n \varphi(D_i, D_iY_i, X_i, b, \lambda, \hat{p}_{D,n})$. Then 
	\begin{equation*}
		\sup_{(b, \lambda) \in \text{Gr}(\lambda)^{\eta/2}} \lvert \hat{M}_n(b,\lambda) - M(b, \lambda) \rvert \overset{p}{\rightarrow} 0
	\end{equation*}
	where $\text{Gr}(\lambda)^{\eta/2} = \left\{(b, \lambda) \; : \; \inf_{(b', \lambda') \in \text{Gr}(\lambda)} \lVert (b, \lambda) - (b', \lambda') \rVert \leq \eta/2\right\}$ and $\text{Gr}(\lambda) = \left\{(b,\lambda(b)) \; : \; b \in B\right\}$.
\end{restatable}
\begin{proof}
	Note that 
	\begin{align*}
		&\sup_{(b, \lambda) \in \text{Gr}(\lambda_0)^{\eta/2}} \lvert \hat{\nu}_n(b,\lambda) - \nu(b, \lambda) \rvert \\
		&\hspace{1 cm} = \sup_{(b, \lambda) \in \text{Gr}(\lambda)^{\eta/2}} \left\lvert \frac{1}{n}\sum_{i=1}^n \varphi(D_i, D_i Y_i, X_i, b, \lambda, \hat{p}_{D,n}) - E[\varphi(D, DY, X, b, \lambda, p_D)\right\rvert
	\end{align*}
	and so the claim can be shown by applying technical lemma \ref{Technical Lemma: uniform consistency of estimated moments}, with $T \equiv \text{Gr}(\lambda)^{\eta/2}$ indexed by $t = (b, \lambda) $, and the constant map $\gamma(t) = p_D$ for all $t \in T$. Verify the conditions of lemma \ref{Technical Lemma: uniform consistency of estimated moments}:
	\begin{enumerate}[label=(\roman*)]
		\item $\{D_i, D_i Y_i, X_i\}_{i=1}^n$ is i.i.d. by assumption \ref{Assumption: setting}.
		\item $\sup_{(b, \lambda) \in \text{Gr}(\lambda)^{\eta/2}} \lvert \hat{p}_{D,n} - p_D \rvert = \lvert \hat{p}_{D,n} - p_D \rvert \overset{p}{\rightarrow} 0$ by the law of large numbers.
		\item $\text{Gr}(p_D) = \{(b, \lambda, p_D) \; : \; (b, \lambda) \in \text{Gr}(\lambda)^{\eta/2}\} = \{(b, \lambda(b), p_D) \; : \; b \in B\}$ is bounded because $\lambda$ is continuous (by lemma \ref{Lemma: continuous dual solution and value function}) and $B$ is compact by assumption \ref{Assumption: strong duality}.  
		\item $p_D \in (0,1)$ implies $\varepsilon \equiv \min\{\min\{p_D, 1- p_D\}, \eta \}/2 > 0$. Let \\ $\text{Gr}(p_D) \equiv \left\{(b, \lambda, p_D) \; : \; (b,\lambda) \in \text{Gr}(\lambda)^{\eta/2} \right\}$ and 
		\begin{align*}
			\text{Gr}(p_D)^\varepsilon &\equiv \left\{(b, \lambda, p) \; : \; \inf_{(b', \lambda', p') \in \text{Gr}(p_D)} \lVert (b, \lambda, p) - (b', \lambda', p') \rVert \leq \varepsilon\right\} \\
			&= \left\{(b, \lambda, p) \; : \; (b,\lambda) \in \text{Gr}(\lambda)^{\eta/2}, \; \lvert p - p_D \rvert \leq \varepsilon\right\}.
		\end{align*}
		Observe that 
		\begin{align*}
			(b, \lambda, p) \mapsto \varphi(d, dy, x, b, \lambda, p) = \frac{\lambda^\intercal J(d) h(dy, x, b)}{1-p} - \frac{d}{p} f^*(\lambda^\intercal h(dy, x, b)) 
		\end{align*}
		is continuous on $\text{Gr}(p_D)^\varepsilon$ for each $(d, dy, x)$. Moreover, $(b, \lambda, p) \in \text{Gr}(p_D)^\varepsilon$ implies
		\begin{align*}
			\inf_{(b',\lambda', p') \in \text{Gr}(p_D)} \lVert (b, \lambda, p) - (b', \lambda', p') \rVert &\leq  \inf_{(b,\lambda) \in \text{Gr}(\lambda)^{\eta/2}} \lVert (b, \lambda) - (b', \lambda') \rVert + \lvert p - p_D \rvert \\
			&\leq \eta/2 + \varepsilon \leq \eta
		\end{align*}
		and hence $\{(b, \nu(b), \lambda, p) \; : \; (b, \lambda, p) \in \text{Gr}(p_D)^\varepsilon\} \subseteq \text{Gr}(\theta_0)^\eta$. This implies
		\begin{align*}
			&E\left[\sup_{(b, \lambda, p) \in \text{Gr}(p_D)^\varepsilon} \lvert \varphi(D, DY, X, b, \lambda, p) \rvert\right] \leq E\left[\sup_{(b, v, \lambda, p) \in \text{Gr}(\theta_0)^\eta} \lvert \varphi(D, DY, X, b, \lambda, p) \rvert\right] \\
			&\hspace{1 cm} \leq E\left[\sup_{(b, v, \lambda, p) \in \Theta^B} \lvert \varphi(D, DY, X, b, \lambda, p) \rvert\right] < \infty.
		\end{align*}
	\end{enumerate}
	Thus, the result follows from lemma \ref{Technical Lemma: uniform consistency of estimated moments}. 
\end{proof}

\begin{restatable}[Uniform consistency of the first stage]{lemma}{lemmaUniformConsistencyFirstStage}
	\label{Lemma: uniform consistency of the first stage}
	
	Suppose assumptions \ref{Assumption: setting}, \ref{Assumption: strong duality}, and \ref{Assumption: estimation} hold, and let $\hat{\lambda}_n(b) = \argmax_{\lambda \in \mathbb{R}^{d_g + K}} \hat{M}_n(b,\lambda)$. Then 
	\begin{equation*}
		\sup_{b \in B} \lVert (\hat{\nu}_n(b), \hat{\lambda}_n(b), \hat{p}_{D,n}) - (\nu(b), \lambda(b), p_D) \rVert \overset{p}{\rightarrow} 0
	\end{equation*}
\end{restatable}
\begin{proof}
	Let $\Lambda(b) \equiv \left\{\lambda \; : \; \lVert \lambda - \lambda(b) \rVert \leq \eta/2\right\}$ and $\bar{\lambda}_n(b) = \argmax_{\lambda \in \Lambda(b)} \hat{M}_n(b, \lambda)$. The proof consists of three steps:
	\begin{enumerate}
		\item Show $\sup_{b \in B} \lVert \bar{\lambda}_n(b) - \lambda(b) \rVert \overset{p}{\rightarrow} 0$. \label{Proof step: first stage consistency, step 1}
		
		The following argument shows that for any $\varepsilon > 0$ there exists $\xi > 0$ such that $\sup_{b \in B} M(b, \lambda(b)) - M(b, \bar{\lambda}_n(b)) \leq \xi$ implies $\sup_{b \in B} \lVert \bar{\lambda}_n(b) - \lambda(b) \rVert < \varepsilon$, and the probability of the former event converges to one. Let $\varepsilon > 0$, and recall that $M(b,\lambda)$ and $\lambda(b)$ are continuous by lemma \ref{Lemma: continuous dual solution and value function}. This implies $M(b, \lambda(b)) - M(b, \lambda)$ is continuous in $(b,\lambda)$ and 
		\begin{align*}
			\bar{\Lambda}^{B, \varepsilon} \equiv \left\{(b, \lambda) \in \text{Gr}(\lambda)^{\eta/2} \; : \; \lVert \lambda - \lambda(b) \rVert \geq \varepsilon/2\right\}
		\end{align*}
		is compact. It follows by the extreme value theorem that $\sup_{(b,\lambda) \in \bar{\Lambda}^{B, \varepsilon}} M(b, \lambda(b)) - M(b, \lambda)$ is attained, say by $(b^s, \lambda^s)$. Lemma \ref{Lemma: unique dual solution} shows $\lambda(b)$ is the unique maximizer of $M(b,\cdot)$ over $\Lambda(b)$, which is a subset of $\left\{\lambda \; : \; (b,\lambda) \in \text{Gr}(\lambda)^{\eta/2}\right\}$, and therefore $\xi \equiv M(b^s, \lambda(b^s)) - M(b^s, \lambda^s) > 0$. Observe that $M(b, \lambda(b)) - M(b, \bar{\lambda}_n(b)) < \xi$ implies $\lVert \bar{\lambda}_n(b) - \lambda(b) \rVert < \varepsilon/2$, and thus
		\begin{align}
			&\sup_{b \in B} M(b, \lambda(b)) - M(b, \bar{\lambda}_n(b)) < \xi &&\implies &&\sup_{b \in B} \lVert \bar{\lambda}_n(b) - \lambda(b) \rVert \leq \varepsilon/2 < \varepsilon \label{Proof display: first stage consistency, well separated}
		\end{align}
		Now notice that 
		\begin{align}
			&\sup_{b \in B} M(b, \lambda(b)) - M(b, \bar{\lambda}_n(b)) \notag \\
			&\hspace{2 cm} \leq \sup_{b \in B} \left\{M(b, \lambda(b)) - \hat{M}_n(b, \lambda(b))\right\} + \underbrace{\sup_{b \in B}\left\{\hat{M}_n(b, \lambda(b)) - \hat{M}_n(b, \bar{\lambda}_n(b))\right\}}_{\leq 0 \text{ by defn of }\bar{\lambda}_n(b)} \notag \\
			&\hspace{4 cm} + \sup_{b \in B}\left\{\hat{M}_n(b, \bar{\lambda}_n(b)) - M(b, \bar{\lambda}_n(b))\right\} \notag \\
			&\hspace{2 cm} \leq \sup_{b \in B} \left\lvert \hat{M}_n(b, \lambda(b)) - M(b, \lambda(b)) \right\rvert + \sup_{b \in B} \left\lvert \hat{M}_n(b, \bar{\lambda}_n(b)) - M(b, \bar{\lambda}_n(b))\right\rvert \notag \\
			&\hspace{2 cm} \leq 2\sup_{(b, \lambda) \in \text{Gr}(\lambda)^{\eta/2}} \left\lvert \hat{M}_n(b, \lambda) - M(b, \lambda) \right\rvert. \label{Proof display: first stage consistency, step 1 inequality}
		\end{align}
		Lemma \ref{Lemma: uniform consistency of the dual objective} implies that $\sup_{(b, \lambda) \in \text{Gr}(\lambda)^{\eta/2}} \left\lvert \hat{M}_n(b, \lambda) - M(b, \lambda) \right\rvert < \xi/2$ holds with probability approaching one. When it does, \eqref{Proof display: first stage consistency, well separated} and \eqref{Proof display: first stage consistency, step 1 inequality} imply $\sup_{b \in B} \lVert \bar{\lambda}_n(b) - \lambda(b) \rVert < \varepsilon$. Therefore $\sup_{b \in B} \lVert \bar{\lambda}_n(b) - \lambda(b) \rVert \overset{p}{\rightarrow} 0$. \\
		
		\item Show $\sup_{b \in B} \lvert \hat{M}_n(b, \bar{\lambda}_n(b)) - M(b, \lambda(b)) \rvert \overset{p}{\rightarrow} 0$. \label{Proof step: first stage consistency, step 2}
		
		The claim follows from lemma \ref{Lemma: uniform consistency of the dual objective}, because 
		\begin{align*}
			\sup_{b \in B} \lvert \hat{M}_n(b, \bar{\lambda}_n(b)) - M(b, \lambda(b)) \rvert &= \sup_{b \in B} \left\lvert \sup_{\lambda \in \Lambda^b} \hat{M}_n(b,\bar{\lambda}_n(b)) - \sup_{\lambda \in \Lambda^b} M(b,\bar{\lambda}_n(b)) \right\rvert \\ 
			&\leq \sup_{b \in B} \sup_{\lambda \in \Lambda^b} \left\lvert \hat{M}_n(b,\lambda) - M(b, \lambda) \right\rvert \\
			&\leq \sup_{(b, \lambda) \in \text{Gr}(\lambda)^{\eta/2}} \left\lvert \hat{M}_n(b,\lambda) - M(b, \lambda) \right\rvert \overset{p}{\rightarrow} 0.
		\end{align*}
		
		\item Show that with probability approaching one, $\sup_{b \in B} \lVert \hat{\lambda}_n(b) - \bar{\lambda}_n(b) \rVert = 0$. \label{Proof step: first stage consistency, step 3}
		
		This follows from an argument similar to the proof of Theorem 2.7 in \cite{newey1994large}. With probability approaching one, $\sup_{b \in B} \lVert \bar{\lambda}_n(b) - \lambda(b) \rVert < \eta/2$ and on this event, $\bar{\lambda}_n(b) \in \text{int}(\Lambda(b)) = \{\lambda \; : \; \lVert \lambda - \lambda(b) \rVert < \eta/2\}$ for every $b \in B$. Since 
		\begin{align*}
			\hat{M}_n(b, \lambda) &= \frac{1}{n}\sum_{i=1}^n \varphi(D_i, D_i Y_i, X_i, b, \lambda, \hat{p}_{D,n}) \\
			&= \frac{1}{n}\sum_{i=1}^n \frac{\lambda^\intercal J(D_i) h(D_i Y_i, X_i, b)}{1-\hat{p}_{D,n}} - \frac{D_i}{\hat{p}_{D,n}} f^*(\lambda^\intercal h(D_i Y_i, X_i, b))
		\end{align*}
		is concave in $\lambda$, no $\lambda$ outside of $\text{int}(\Lambda(b))$ could make the objective larger than $\bar{\lambda}_n(b)$. Thus when $\sup_{b \in B} \lVert \bar{\lambda}_n(b) - \lambda(b) \rVert < \eta/2$ holds, $\hat{\lambda}_n(b) = \bar{\lambda}_n(b)$ for every $b \in B$ or equivalently, $\sup_{b \in B} \lVert \hat{\lambda}_n(b) - \bar{\lambda}_n(b) \rVert = 0$.
	\end{enumerate}
\end{proof}

\begin{restatable}[Consistency of $\delta^{BP}$]{theorem}{theoremConsistencyOfBDP}
	\label{Theorem: consistency of BDP}
	
	Suppose assumptions \ref{Assumption: setting}, \ref{Assumption: strong duality}, and \ref{Assumption: estimation} hold. Then $\hat{\delta}_n^{BP} \overset{p}{\rightarrow} \delta^{BP}$.
\end{restatable}
\begin{proof}
	Lemma \ref{Lemma: uniform consistency of the first stage} implies $\hat{\nu}_n$ converges in probability to $\nu$ in $\ell^\infty(B)$, and lemma \ref{Technical Lemma: restricted infimum is uniformly continuous} shows $\iota : \ell^\infty(B) \rightarrow \mathbb{R}$ given by $\iota(f) = \inf_{b \in B \cap \textbf{B}_0} f(b)$ is continuous. Since $\hat{\delta}_n^{BP} = \iota(\hat{\nu}_n)$ and $\delta^{BP} = \iota(\nu)$, the result follows from the continuous mapping theorem.
\end{proof}

%% file: Appendix_Estimation_Inference.tex
\subsection{Inference}
\label{Appendix: estimation, inference proofs}

\begin{restatable}[Bounds on Jacobian terms]{lemma}{lemmaBoundsOnJacobianTerms}
	\label{Lemma: bounds on jacobian terms}
	
	Suppose assumption \ref{Assumption: setting}, \ref{Assumption: strong duality}, and \ref{Assumption: estimation} hold. Then \\ $\sup_{b \in B} \lVert \Phi(b) \rVert_o < \infty$ and $\sup_{b \in B} \lVert \Phi(b)^{-1} \rVert_o < \infty$.
	
\end{restatable}
\begin{proof}
	
	Recall that $\Phi(b) = E[\nabla_\theta \phi(D, DY, X, b, \theta_0(b)) ]$. Jensen's inequality and convexity of norms implies
	\begin{align*}
		\sup_{b \in B} \lVert \Phi(b) \rVert_o &= \sup_{b \in B} \lVert E[\nabla_\theta \phi(D, DY, X, b, \theta) ] \rVert_o \leq E\left[\sup_{b \in B} \lVert \nabla_\theta \phi(D, DY, X, b, \theta) \rVert_o \right] \\
		&\leq E\left[\sup_{(b, \theta) \in \Theta^B} \lVert \nabla_\theta \phi(D, DY, X, b, \theta) \rVert_o\right],
	\end{align*}
	and $E\left[\sup_{(b, \theta) \in \Theta^B} \lVert \nabla_\theta \phi(D, DY, X, b, \theta) \rVert\right] < \infty$ is implied by assumption \ref{Assumption: estimation} \ref{Assumption: estimation, moment conditions} and Jensen's inequality. Therefore $\sup_{b \in B} \lVert \Phi(b) \rVert_o < \infty$.
	
	To establish $\sup_{b \in B} \lVert \Phi(b)^{-1} \rVert_o < \infty$, first use expression \eqref{Appendix display: jacobian matrix components} to see that 
	\begin{align*}
		\Phi(b) &= E\left[\nabla_\theta \phi(D, DY, X, b, \nu(b), \lambda(b), p_D)\right] \\
		&= \begin{bmatrix}
			-1 & 0 & E[\nabla_p \varphi(D, DY, X, b, \lambda(b), p_D)] \\
			0 & E[\nabla_\lambda^2 \varphi(D, DY, X, b, \lambda(b), p_D)] & E[\nabla_p \nabla_\lambda \varphi(D, DY, X, b, \lambda(b), p_D)] \\
			0 & 0 & -1
		\end{bmatrix} 
	\end{align*}
	where $E[\nabla_\lambda \varphi(D, DY, X, b, \lambda(b), p_D)]^\intercal = 0$ is the first order condition of the dual problem. 
	
	The middle matrix, $E[\nabla_\lambda^2 \varphi(D, DY, X, b, \lambda(b), p_D)]$, is invertible for each $b \in B$. To see this, first recall that this matrix equals $\nabla_\lambda^2 M(b, \lambda(b)$, which lemma \ref{Lemma: continuous dual solution and value function} shows is also continuous in $b$. The mapping from matrices to eigenvalues is continuous (see \citet{Bhatia1997} corollary III.2.6 or its application in the proof of lemma \ref{Technical Lemma: uniform consistency of matrix inverses}), so the extreme value theorem implies $\sup_{b \in B} \alpha_1(\nabla_\lambda^2 M(b, \lambda(b)))$ is attained by some $\bar{b} \in B$. Lemma \ref{Lemma: unique dual solution} argues that $M(\bar{b}, \lambda)$ is strictly concave in $\lambda$, hence $\nabla_\lambda^2 M(\bar{b},\lambda(\bar{b}))$ is negative definite and thus $\alpha_1(\nabla_\lambda^2 M(\bar{b}, \lambda(\bar{b}))) < 0$. To summarize,
	\begin{equation*}
		\sup_{b \in B} \alpha_1(\nabla_\lambda^2 M(b, \lambda(b))) = \alpha_1(\nabla_\lambda^2 M(\bar{b}, \lambda(\bar{b}))) < 0
	\end{equation*}
	which implies $\nabla M(b, \lambda(b)) = E[\nabla_\lambda^2 \varphi(D, DY, X, b, \lambda(b), p_D)]$ is invertible for each $b \in B$. With this invertibility claim, it is straightforward to verify that for each $b\in B$, $\Phi(b)^{-1}$ exists and is given by
	\begin{align*}
		&\Phi(b)^{-1} =\\
		&\begin{bmatrix}
			-1 & 0 & -E[\nabla_p \varphi(D, DY, X, b, \lambda(b), p_D)] \\
			0 & E[\nabla_\lambda^2 \varphi(D, DY, X, b, \lambda(b), p_D)]^{-1} & E[\nabla_\lambda^2 \varphi(D, DY, X, b, \lambda(b), p_D)]^{-1} E[\nabla_p \nabla_\lambda \varphi(D, DY, X, b, \lambda(b), p_D)] \\
			0 & 0 & -1
		\end{bmatrix}
	\end{align*}
	
	To see that $\sup_{b \in B} \lVert \Phi(b)^{-1} \rVert_o$ is finite, first recall that for conformable matrices,
	\begin{align*}
		\left\lVert 
		\begin{bmatrix}
			A_{11} & A_{12} \\
			A_{21} & A_{22}
		\end{bmatrix}
		\right\rVert_o
		&\leq
		\lVert A_{11} \rVert_o + \lVert A_{12} \rVert_o + \lVert A_{21} \rVert_o + \lVert A_{22} \rVert_o, \text{ and } \\
		\lVert AB \rVert_o &\leq \lVert A \rVert_o \lVert B \rVert_o.
	\end{align*}
	Apply these inequalities to find that
	\begin{align}
		&\sup_{b \in B} \lVert \Phi(b)^{-1} \rVert_o \notag \\
		&\hspace{1 cm} \leq 2 + \sup_{b \in B} \left\lvert E[\nabla_p \varphi(D, DY, X, b, \lambda(b), p_D)] \right\rvert + \sup_{b \in B} \lVert E[\nabla_\lambda^2 \varphi(D, DY, X, b, \lambda(b), p_D)]^{-1} \rVert_o \notag\\
		&\hspace{2 cm} + \sup_{b \in B} \lVert E[\nabla_\lambda^2 \varphi(D, DY, X, b, \lambda(b), p_D)]^{-1} \rVert_o \times \sup_{b \in B} \left\lVert E[\nabla_p \nabla_\lambda \varphi(D, DY, X, b, \lambda(b), p_D)]  \right\rVert \label{Proof display: bounds on jacobian terms, inverse matrix inequality}
	\end{align}
	$\left\lvert E[\nabla_p \varphi(D, DY, X, b, \lambda(b), p_D)] \right\rvert$ and $\left\lVert E[\nabla_p \nabla_\lambda \varphi(D, DY, X, b, \lambda(b), p_D)]\right\rVert$ are the operator norms of submatrices of $\Phi(b)$. Thus $\sup_{b \in B} \lVert \Phi(b) \rVert_o < \infty$, argued above, implies 
	\begin{align}
		&\sup_{b \in B} \left\lvert E[\nabla_p \varphi(D, DY, X, b, \lambda(b), p_D)] \right\rvert < \infty &&\text{ and } &&\sup_{b \in B} \left\lVert E[\nabla_p \nabla_\lambda \varphi(D, DY, X, b, \lambda(b), p_D)]\right\rVert < \infty. \label{Proof display: bounds on jacobian terms, inverse matrix inequality, first components}
	\end{align}
	Finally, since $E[\nabla_\lambda^2 \varphi(D, DY, X, b, \lambda(b), p_D)] = \nabla_\lambda^2 M(b, \lambda(b))$ is symmetric and negative definite, $\lVert E[\nabla_\lambda^2 \varphi(D, DY, X, b, \lambda(b), p_D)]^{-1} \rVert_o = \lVert \nabla_\lambda^2 M(b, \lambda(b))^{-1} \rVert_o = \frac{1}{\lvert \alpha_1(\nabla_\lambda^2 M(b, \lambda(b))) \rvert}$ and
	\begin{align}
		&\sup_{b \in B} \lVert E[\nabla_\lambda^2 \varphi(D, DY, X, b, \lambda(b), p_D)]^{-1} \rVert_o \notag \\
		&\hspace{2 cm} = \sup_{b \in B} \lVert \nabla_\lambda^2 M(b, \lambda(b))^{-1} \rVert_o = \sup_{b \in B} \frac{1}{\lvert \alpha_1(\nabla_\lambda^2 M(b, \lambda(b))) \rvert} \notag \\
		&\hspace{2 cm} = \frac{1}{\inf_{b \in B}\lvert \alpha_1(\nabla_\lambda^2 M(b, \lambda(b))) \rvert} = \frac{1}{\lvert \sup_{b \in B} \alpha_1(\nabla_\lambda^2 M(b, \lambda(b))) \rvert} < \infty, \label{Proof display: bounds on jacobian terms, inverse matrix inequality, second component}
	\end{align}
	where the final claim follows from $\sup_{b \in B} \alpha_1(\nabla_\lambda^2 M(b, \lambda(b))) < 0$ as argued above. Taken together, \eqref{Proof display: bounds on jacobian terms, inverse matrix inequality}, \eqref{Proof display: bounds on jacobian terms, inverse matrix inequality, first components}, and \eqref{Proof display: bounds on jacobian terms, inverse matrix inequality, second component} show that $\sup_{b \in B} \lVert \Phi(b)^{-1} \rVert_o < \infty$.
\end{proof}

\begin{restatable}[Donsker influence functions]{lemma}{lemmaDonskerInfluenceFunctions}
	\label{Lemma: Donsker influence functions}
	
	Suppose assumptions \ref{Assumption: setting}, \ref{Assumption: strong duality}, and \ref{Assumption: estimation} hold. Then the class of functions
	\begin{equation*}
		\left\{\phi(D, DY, X, b, \theta) \; : \; (b,\theta) \in \Theta^B\right\}
	\end{equation*}
	is Donsker.
\end{restatable}
\begin{proof}
	By verifying the conditions of \cite{van2007asymptotic} example 19.7. 
	
	$\Theta^B$ is a compact subset of a finite dimensional space, hence bounded. Let $(b_1, \theta_1), (b_2, \theta_2) \in \Theta^B$ and apply the mean value inequality (e.g., \cite{coleman2012calculus} Corollary 3.2) to find 
	\begin{align*}
		&\lVert \phi(d, dy, x, b_1, \theta_1) - \phi(d, dy, x, b_2, \theta_2) \rVert  \\
		&\hspace{1 cm} \leq \left[\sup_{t \in (0,1)} \left\lVert\nabla_{(b,\theta)} \phi(d, dy, x, t b_1 + (1-t) b_2, t\theta_1 + (1-t) \theta_2) \right\rVert_o\right] \lVert (b_1, \theta_1) - (b_2, \theta_2) \rVert \\
		&\hspace{1 cm} \leq \left[\sup_{(b,\theta) \in \Theta^B} \left\lVert\nabla_{(b,\theta)} \phi(d, dy, x, b, \theta) \right\rVert_o\right] \lVert (b_1, \theta_1) - (b_2, \theta_2) \rVert
	\end{align*} 
	Assumption \ref{Assumption: estimation} \ref{Assumption: estimation, moment conditions} includes $E\left[\left(\sup_{(b,\theta) \in \Theta^B} \left\lVert\nabla_{(b,\theta)} \phi(d, dy, x, b, \theta) \right\rVert_o\right)^2\right] < \infty$. Therefore the class $\left\{\phi(D, DY, X, b, \theta) \; : \; (b,\theta) \in \Theta^B\right\}$ is a special case of \cite{van2007asymptotic} example 19.7, and thus Donsker. 
\end{proof}

\begin{restatable}[Weak convergence of the first stage]{lemma}{lemmaWeakConvergenceOfTheFirstStage}
	\label{Lemma: weak convergence of the first stage}
	
	Suppose assumptions \ref{Assumption: setting}, \ref{Assumption: strong duality}, and \ref{Assumption: estimation} hold. Let $\mathcal{I} = \{1, \ldots, d_g + K +2\}$, and view $\hat{\theta}_n(b) \equiv (\hat{\nu}_n(b), \hat{\lambda}_n(b), \hat{p}_{D,n})$ and $\theta_0(b) \equiv (\nu(b), \lambda(b), p_D)$ as functions mapping $B \times \mathcal{I}$ to $\mathbb{R}$. Then 
	\begin{align*}
		&\sqrt{n}(\hat{\theta}_n - \theta_0) \overset{L}{\rightarrow} \mathbb{G}
	\end{align*}
	where $\mathbb{G}$ is a tight, mean zero Gaussian process in $\ell^\infty(B \times \mathcal{I})$. The covariance function of $\mathbb{G}$ is given by
	\begin{align*}
		&\text{Cov}(\mathbb{G}(b_1, i_1), \mathbb{G}(b_2, i_2)) \\
		&\hspace{1 cm} = E\left[(\Phi(b_1)^{-1})^{(i_1)} \phi(D, DY, X, b_1, \theta_0(b_1)) \left\{(\Phi(b_2)^{-1})^{(i_2)} \phi(D, DY, X, b_2, \theta_0(b_2))\right\}\right].
	\end{align*}
	where $(\Phi(b)^{-1})^{(i)}$ is the $i$-th row of the matrix $\Phi(b)^{-1} = E[\nabla_\theta \phi(D, DY, X, b, \theta_0(b))]^{-1}$. 
\end{restatable}
\begin{proof}
	
	For legibility, the proof is presented in six steps:
	\begin{enumerate}
		
		\item Mean value theorem. \label{Proof step: first stage weak convergence, step 1}
		
		For each $b \in B$, apply the mean value theorem to each coordinate of \\ $0 = \frac{1}{n}\sum_{i=1}^n \phi(D_i, D_i Y_i, X_i, b, \hat{\theta}_n(b))$ and stack the results to obtain
		\begin{align}
			0 &= \frac{1}{n}\sum_{i=1}^n \phi(D_i, D_iY_i, X_i, b, \theta_0(b)) \notag \\
			&\hspace{1 cm} + 
			\underbrace{
				\begin{bmatrix}
					\frac{1}{n}\sum_{i=1}^n \nabla_\theta \phi^{(1)}(D_i, D_i Y_i, X_i, b, \bar{\theta}_n^1(b)) \\
					\vdots \\
					\frac{1}{n}\sum_{i=1}^n \nabla_\theta \phi^{(d_g + K + 2)}(D_i, D_i Y_i, X_i, b, \bar{\theta}_n^{d_g + K + 2}(b))
				\end{bmatrix}
			}_{\equiv \bar{\Phi}_n(b)}
			(\hat{\theta}_n(b) - \theta_0(b)) \label{Proof display: first stage weak convergence, mean value theorem}
		\end{align}
		where $\nabla_\theta \phi^{(j)}(D_i, D_i Y_i, X_i, b, \theta)$ is the $j$-th coordinate of the vector $\nabla_\theta \phi(D_i, D_i Y_i, X_i, b, \theta)$, and $\bar{\theta}_n^j = \theta_0(b) + a_n^j(b)\times (\hat{\theta}_n(b) - \theta_0(b)) \in \mathbb{R}^{d_g + K +2}$ for some $a_n^j(b) \in (0,1)$.\footnote{See \cite{newey1994large} footnote 25.} Notice
		\begin{align*}
			\lVert \bar{\theta}_n^j(b) - \theta_0(b) \rVert &= \left\lVert \theta_0(b) + a_n^j(b)\times (\hat{\theta}_n(b) - \theta_0(b)) - \theta_0(b)\right\rVert \\
			&= a_n^j(b) \times \lVert \hat{\theta}_n(b) - \theta_0(b) \rVert \\
			&\leq \lVert \hat{\theta}_n(b) - \theta_0(b) \rVert
		\end{align*}
		and $\sup_{b \in B} \lVert \hat{\theta}_n(b) - \theta_0(b) \rVert \overset{p}{\rightarrow} 0$, shown in lemma \ref{Lemma: uniform consistency of the first stage}, implies $\sup_{b \in B} \lVert \bar{\theta}_n^j(b) - \theta_0(b) \rVert \overset{p}{\rightarrow} 0$.
		
		\item Show $\sup_{b \in B} \lVert \bar{\Phi}_n(b) - \Phi(b) \rVert \overset{p}{\rightarrow} 0$. \label{Proof step: first stage weak convergence, step 2}
		
		First, notice that 
		\begin{equation*}
			\bar{\Phi}_n(b) = \sum_{j=1}^{d_g + K + 2} e_{jj} \frac{1}{n}\sum_{i=1}^n \nabla_\theta \phi(D_i, D_i Y_i, X_i, b, \bar{\theta}_n^j(b)) 
		\end{equation*}
		where $e_{jj}$ is the square $(d_g + K + 2) \times (d_g + K + 2)$ matrix whose $(j, j)$-th entry is one and all other entries are zero.\footnote{
			When premultiplying a square matrix $A$, $e_{jj}$ ``selects'' the $j$-th row. For example,
			\begin{equation*}
				e_{22}A = 
				\begin{bmatrix}
					0 & 0 & 0 & \ldots & 0 \\
					0 & 1 & 0 & \ldots & 0 \\
					0 & 0 & 0 & \ldots & 0 \\
					\vdots & & & \ddots & \vdots \\
					0 & 0 & 0 & \ldots & 0
				\end{bmatrix}
				\begin{bmatrix}
					a_{11} & a_{12} & a_{13} & \ldots & a_{1J} \\
					a_{21} & a_{22} & a_{23} & \ldots & a_{2J} \\
					a_{31} & a_{32} & a_{33} & \ldots & a_{3J} \\
					\vdots & & & \ddots & \vdots \\
					a_{J1} & a_{K2} & a_{K3} & \ldots & a_{JJ} \\
				\end{bmatrix} 
				=
				\begin{bmatrix}
					0 & 0 & 0 & \ldots & 0 \\
					a_{21} & a_{22} & a_{23} & \ldots & a_{2J} \\
					0 & 0 & 0 & \ldots & 0 \\
					\vdots & & & \ddots & \vdots \\
					0 & 0 & 0 & \ldots & 0
				\end{bmatrix}
				\end{equation*}
			}
		
		Apply lemma \ref{Technical Lemma: uniform consistency of estimated moments} to $\frac{1}{n}\sum_{i=1}^n \nabla_\theta \phi(D_i, D_i Y_i, X_i, b, \bar{\theta}_n^j(b))$ for each $j  \in \{1, \ldots, d_g + K + 2\}$ to argue this is consistent for $E[\nabla_\theta \phi(D, DY, X, b, \theta_0(b))]$ uniformly over $b \in B$. 
		\begin{enumerate}[label=(\roman*)]
			\item $\{D_i, D_i Y_i, X_i\}_{i=1}^n$ is i.i.d. by assumption \ref{Assumption: setting}.
			\item $\sup_{b \in B} \lVert \bar{\theta}_n^j(b) - \theta_0(b) \rVert \overset{p}{\rightarrow} 0$ is shown in step \ref{Proof step: first stage weak convergence, step 1}. 
			\item $\theta_0(b) = (\nu(b), \lambda(b), p_D)$ is bounded, since $B$ is compact by assumption \ref{Assumption: strong duality} and $\nu(\cdot)$, $\lambda(\cdot)$ are continuous as shown by lemma \ref{Lemma: continuous dual solution and value function}. 
			\item $(b, \theta) \mapsto \nabla_\theta \phi(d, dy, x, b, \theta)$ is continuous at any $(b,\theta) \in \text{Gr}(\theta_0)^\eta$, by examination of equations \eqref{Appendix display: jacobian matrix components} and \eqref{Appendix display: derivatives of varphi}. Moreover, $E\left[\sup_{(b,\theta) \in \text{Gr}(\theta_0)^\eta} \lVert \nabla_\theta \phi(D, DY, X, b, \theta)\rVert_o \right]$ is finite; $\nabla_\theta \phi(D, DY, X, b, \theta)$ is a submatrix of $\nabla_{(b,\theta)} \phi(D, DY, X, b, \theta)$, while assumption \ref{Assumption: estimation} \ref{Assumption: estimation, moment conditions} and Jensen's inequality imply
			\begin{align*}
				E\left[\sup_{(b,\theta) \in \text{Gr}(\theta_0)^\eta} \lVert \nabla_{(b,\theta)} \phi(D, DY, X, b, \theta)\rVert_o\right] \leq E\left[\sup_{(b,\theta) \in \Theta^B} \lVert \nabla_{(b,\theta)} \phi(D, DY, X, b, \theta)\rVert_o\right] < \infty.
			\end{align*}
		\end{enumerate} 
		So by Lemma \ref{Technical Lemma: uniform consistency of estimated moments}, $\sup_{b \in B} \left\lVert \frac{1}{n}\sum_{i=1}^n \nabla_\theta \phi(D_i, D_i Y_i, X_i, b, \bar{\theta}_n^j(b)) - E[\nabla_\theta \phi(D, DY, X, b, \theta_0(b))] \right\rVert_o \overset{p}{\rightarrow} 0$ for each $j \in \{1, \ldots, d_g + K + 2\}$, from which it follows that 
		\begin{align*}
			&\sup_{b \in B} \lVert \bar{\Phi}_n(b) - \Phi(b) \rVert_o \\
			&\hspace{1 cm} = \sup_{b \in B} \left\lVert \sum_{j=1}^{d_g + K + 2} e_{jj} \left(\frac{1}{n}\sum_{i=1}^n \nabla_\theta \phi(D_i, D_i Y_i, X_i, b, \bar{\theta}_n^j(b)) - E[\nabla_\theta \phi(D, DY, X, b, \theta_0(b))] \right)\right\rVert_o \\
			&\hspace{1 cm} \leq \sum_{j=1}^{d_g + K + 2} \sup_{b \in B} \left\lVert \frac{1}{n}\sum_{i=1}^n \nabla_\theta \phi(D_i, D_i Y_i, X_i, b, \bar{\theta}_n^j(b)) - E[\nabla_\theta \phi(D, DY, X, b, \theta_0(b))] \right\rVert_o \\
			&\hspace{1 cm} \leq (d_g + K + 2) \max_j \sup_{b \in B} \left\lVert \frac{1}{n}\sum_{i=1}^n \nabla_\theta \phi(D_i, D_i Y_i, X_i, b, \bar{\theta}_n^j(b)) - E[\nabla_\theta \phi(D, DY, X, b, \theta_0(b))] \right\rVert_o \\
			&\hspace{1 cm} \overset{p}{\rightarrow} 0
		\end{align*}
		
		\item Uniform linearization. \label{Proof step: first stage weak convergence, step 3}
		
		Lemma \ref{Lemma: bounds on jacobian terms} shows $\sup_{b \in B} \lVert \Phi(b) \rVert_o < \infty$ and $\sup_{b \in B} \lVert \Phi(b)^{-1} \rVert_o < \infty$. Since with $\sup_{b \in B} \lVert \bar{\Phi}_n(b) - \Phi(b) \rVert_o \overset{p}{\rightarrow} 0$ is shown in step \ref{Proof step: first stage weak convergence, step 2}, lemma \ref{Technical Lemma: uniform consistency of matrix inverses} implies that with probability approaching one, $\bar{\Phi}_n(b)^{-1}$ is well defined as a function on $B$. When it is, rearrange expression \eqref{Proof display: first stage weak convergence, mean value theorem} to find 
		\begin{align*}
			\sqrt{n}(\hat{\theta}_n(b) - \theta_0(b)) &= \bar{\Phi}_n(b)^{-1} \frac{1}{\sqrt{n}} \sum_{i=1}^n \phi(D_i, D_i Y_i, X_i, b, \theta_0(b)) \\
			&= G_n(b) + R_n(b) \\
			\text{where } G_n(b) &= \Phi(b)^{-1} \frac{1}{\sqrt{n}} \sum_{i=1}^n \phi(D_i, D_i Y_i, X_i, b, \theta_0(b)), \\
			\text{and } R_n(b) &= \left[\bar{\Phi}_n(b)^{-1} - \Phi(b)^{-1}\right]\frac{1}{\sqrt{n}} \sum_{i=1}^n \phi(D_i, D_i Y_i, X_i, b, \theta_0(b)).
		\end{align*}
		
		\item Show $G_n \overset{L}{\rightarrow} \mathbb{G}$ in $\ell^\infty(B \times \mathcal{I})$. \label{Proof step: first stage weak convergence, step 4} 
		
		Define $\tilde{G}_n : B \rightarrow \mathbb{R}^{d_g + K + 2}$ pointwise as 
		\begin{equation*}
			\tilde{G}_n(b) = \frac{1}{\sqrt{n}} \sum_{i=1}^n \phi(D_i, D_i Y_i, X_i, b, \theta_0(b))
		\end{equation*}
		$\{\phi(D, DY, X, b, \theta_0(b)) \; : \; b \in B\}$ is a subset of the class considered in lemma \ref{Lemma: Donsker influence functions}, and is therefore Donsker (see \cite{vaart1997weak} theorem 2.10.1). Thus, $\tilde{G}_n \overset{L}{\rightarrow} \tilde{\mathbb{G}}$ in $\ell^\infty(B)^{d_g + K + 2}$, where $\tilde{\mathbb{G}}$ is a tight, mean-zero Gaussian process with covariance function
		\begin{equation*}
			\text{Cov}(\tilde{\mathbb{G}}(b_1), \tilde{\mathbb{G}}(b_2)) = E\left[\phi(D, DY, X, b_1, \theta_0(b_1)) \phi(D, DY, X, b, \theta_0(b_2))^\intercal\right]
		\end{equation*}
		Now define 
		\begin{align*}
			&L : \ell^\infty(B)^{d_g + K + 2} \rightarrow \ell^\infty(B \times \mathcal{I}), &&L(H)(b, i) = (\Phi(b)^{-1})^{(i)} H(b)
		\end{align*}
		and observe that $G_n = L(\tilde{G}_n)$. Note that $L$ is a linear operator on $H$. Lemma \ref{Lemma: bounds on jacobian terms} shows $\sup_{b \in B} \lVert \Phi(b)^{-1} \rVert_o < \infty$, which along with
		\begin{align*}
			\lVert LH \rVert_B = \sup_{b \in B} \lVert \Phi(b)^{-1} H(b) \rVert \leq \sup_{b \in B} \lVert \Phi(b)^{-1} \rVert_o \sup_{b \in B} \lVert H(b) \rVert = \left(\sup_{b \in B} \lVert \Phi(b)^{-1} \rVert_o\right) \lVert H \rVert_B
		\end{align*}
		shows that $L$ is bounded, hence continuous. The continuous mapping theorem then implies
		\begin{equation*}
			L(\tilde{G}_n) \overset{L}{\rightarrow} L(\tilde{\mathbb{G}})
		\end{equation*}
		where $L(\tilde{\mathbb{G}})$ is a tight, mean-zero Gaussian process on $\ell^\infty(B \times \mathcal{I})$. Letting $(\Phi(b))^{(i)}$ be the $i$-th row of the matrix $\Phi(b)^{-1}$, the covariance function of $L(\tilde{\mathbb{G}})$ is 
		\begin{align*}
			&\text{Cov}(\mathbb{G}(b_1, i_1), \mathbb{G}(b_2, i_2)) \\
			&\hspace{1 cm} = E\left[(\Phi(b_1)^{-1})^{(i_1)} \phi(D, DY, X, b_1, \theta_0(b_1)) \left\{(\Phi(b_2)^{-1})^{(i_2)} \phi(D, DY, X, b_2, \theta_0(b_2))\right\}\right].
		\end{align*}
		Notice that the marginals of $L(\tilde{\mathbb{G}})$ are equal in distribution to those of $\mathbb{G}$. By \cite{vaart1997weak} lemma 1.5.3, this implies the two distributions are the same and hence $G_n = L(\tilde{G}_n) \overset{L}{\rightarrow} \mathbb{G}$.
		
		\item Uniform linearization remainder control. \label{Proof step: first stage weak convergence, step 5}
		
		Since $\{\phi(D, DY, X, b, \theta_0(b)) \; : \; b \in B\}$ is Donsker and $E[\phi(D, DY, X, b, \theta_0(b))] = 0$ for all $b \in B$, $\sup_{b \in B} \left\Vert \frac{1}{\sqrt{n}} \sum_{i=1}^n \phi(D_i, D_i Y_i, X_i, b, \theta_0(b)) \right\rVert_o = O_p(1)$ by the continuous mapping theorem. Lemma \ref{Lemma: bounds on jacobian terms} shows $\lVert \Phi(b) \rVert_o < \infty$ and step \ref{Proof step: first stage weak convergence, step 2} that $\lVert \Phi(b)^{-1} \rVert_p < \infty$, and $\lVert \bar{\Phi}_n(b) - \Phi(b) \rVert \overset{p}{\rightarrow} 0$, so lemma \ref{Technical Lemma: uniform consistency of matrix inverses} implies $\sup_{b \in B} \left\lVert \bar{\Phi}_n(b)^{-1} - \Phi(b)^{-1} \right\rVert = o_p(1)$. Thus,
		\begin{align*}
			\sup_{b \in B} \lVert R_n(b) \rVert &= \sup_{b \in B} \left\lVert \left[\bar{\Phi}_n(b)^{-1} - \Phi(b)^{-1}\right]\frac{1}{\sqrt{n}} \sum_{i=1}^n \phi(D_i, D_i Y_i, X_i, b, \theta_0(b)) \right\rVert \\
			&\leq \sup_{b \in B} \left\lVert\bar{\Phi}_n(b)^{-1} - \Phi(b)^{-1} \right\rVert \sup_{b \in B} \left\lVert \frac{1}{\sqrt{n}} \sum_{i=1}^n \phi(D_i, D_i Y_i, X_i, b, \theta_0(b))  \right\rVert \\
			&\overset{p}{\rightarrow} 0.
		\end{align*}
		
		\item Conclusion. \label{Proof step: first stage weak convergence, step 6}
		
		As elements of $\ell^\infty(B \times \mathcal{I})$, $G_n \overset{L}{\rightarrow} \mathbb{G}$ and $R_n \overset{p}{\rightarrow} 0$, so 
		\begin{align*}
			&(G_n, R_n) \overset{L}{\rightarrow} (\mathbb{G}, 0) && \text{ in } \ell^\infty(B \times \mathcal{I})
		\end{align*}
		by \cite{van2007asymptotic} theorem 18.10. The continuous mapping theorem (\cite{van2007asymptotic} theorem 18.11) then implies
		\begin{equation*}
			\sqrt{n}(\hat{\theta}_n - \theta_0) = G_n + R_n \overset{L}{\rightarrow} \mathbb{G} + 0
		\end{equation*}
		which concludes the proof.
		
	\end{enumerate}
\end{proof}

\begin{restatable}[Support of $\mathbb{G}_\nu$]{lemma}{lemmaSupportOfValueFunctionAsymptoticDistribution}
	\label{Lemma: support of value function asymptotic distribution} 
	
	Suppose assumptions \ref{Assumption: setting}, \ref{Assumption: strong duality}, and \ref{Assumption: estimation} hold, let $\mathbb{G}$ be the random element of $\ell^\infty(B \times \mathcal{I})$ from lemma \ref{Lemma: weak convergence of the first stage}, and let $\mathbb{G}_\nu \in \ell^\infty(B)$ be the mean-zero Gaussian process on $B$ defined pointwise by $\mathbb{G}_\nu(b) = \mathbb{G}(b, 1)$. Then $\sqrt{n}(\hat{\nu}_n - \nu) \overset{L}{\rightarrow} \mathbb{G}_\nu$ in $\ell^\infty(B)$ and $P(\mathbb{G}_\nu \in \mathcal{C}(B)) = 1$, where $\mathcal{C}(B)$ is the set of continuous functions defined on $B$. 
\end{restatable}
\begin{proof}
	\singlespacing
	
	Lemma \ref{Lemma: weak convergence of the first stage} and the continuous mapping theorem implies $\sqrt{n}(\hat{\nu}_n - \nu) \overset{L}{\rightarrow} \mathbb{G}_\nu$. The Portmanteau theorem (\cite{vaart1997weak} theorem 1.3.4) shows that this is equivalent to 
	\begin{equation*}
		\limsup_{n \rightarrow \infty} P(\sqrt{n}(\hat{\nu}_n - \nu) \in F) \leq P(\mathbb{G}_\nu \in F)
	\end{equation*}
	for all closed sets $F \subseteq \ell^\infty(B)$. Since $\mathcal{C}(B)$ is closed and $\nu(\cdot)$ is continuous by lemma \ref{Lemma: continuous dual solution and value function}, it suffices to show that $\hat{\nu}_n$ is continuous with probability approaching one. 
	
	The argument is based on the Berge maximum theorem (\cite{Aliprantis2006} theorem 17.31). Recall $\hat{\lambda}_n(b) \equiv \argmax_{\lambda \in \mathbb{R}^{d_g + K}} \hat{M}_n(b, \lambda)$ and $\hat{\nu}_n(b) = \hat{M}_n(b, \hat{\lambda}_n(b))$. Let $\Lambda(b) \equiv \left\{\lambda \; : \; \lVert \lambda - \lambda(b) \rVert\leq \eta/2\right\}$. Lemma \ref{Lemma: uniform consistency of the first stage} implies $\sup_{b \in B} \left\lVert \hat{\lambda}_n(b) - \lambda(b) \right\rVert < \eta/2$ holds with probability approaching one, and when it does,
	\begin{equation*}
		\sup_{b \in B} \lvert \hat{\nu}_n(b) - \max_{\lambda \in \Lambda(b)} \hat{M}_n(b,\lambda) \rvert = \sup_{b \in B} \left\lvert \sup_{\lambda \in \mathbb{R}^{d_g + K}} \hat{M}_n(b, \lambda) - \max_{\lambda \in \Lambda(b)} \hat{M}_n(b,\lambda)\right\rvert = 0
	\end{equation*}
	It thus suffices to show that $b \mapsto \max_{\lambda \in \Lambda(b)} \hat{M}_n(b,\lambda)$ is continuous with probability approaching one. This will follow from the Berge maximum theorem, once it is shown that $\Lambda(\cdot)$ is a continuous correspondence and $\hat{M}_n$ is continuous on $\text{Gr}(\Lambda)$. Since 
	\begin{equation*}
		\Lambda(b) \subseteq\lambda(B)^{\eta/2} \equiv \left\{\lambda \; : \; \inf_{\lambda' \in \Lambda(B)}\lVert \lambda - \lambda' \rVert \leq \eta/2\right\} = \left\{\lambda \; : \; \inf_{b' \in B} \lVert \lambda - \lambda(b') \rVert \leq \eta/2\right\}
	\end{equation*}
	we can view $\Lambda : B \rightrightarrows \lambda(B)^{\eta/2}$, and thus
	\begin{equation*}
		\text{Gr}(\Lambda) = \left\{(b,\lambda) \in B \times \lambda(B)^{\eta/2} \; : \; \lambda \in \Lambda(b)\right\} = \left\{(b,\lambda) \; : \; \lVert \lambda - \lambda(b) \rVert \leq \eta /2\right\}.
	\end{equation*}
	\begin{enumerate}		
		\item Consider continuity of the objective first. 
		
		Assumption \ref{Assumption: estimation} \ref{Assumption: estimation, continuously differentiable moment functions} implies $h(y,x,b)$ is continuous in $b$, and assumption \ref{Assumption: setting} \ref{Assumption: setting, divergence} includes that $f^*(\cdot)$ is essentially smooth. It follows that
		\begin{equation*}
			\hat{M}_n(b,\lambda) = \frac{1}{n}\sum_{i=1}^n \frac{\lambda^\intercal J(D_i) h(D_iY_i, X_i, b)}{1-\hat{p}_{D,n}} - \frac{D_i}{\hat{p}_{D,n}} f^*(\lambda^\intercal h(D_iY_i, X_i, b))
		\end{equation*}
		is continuous at $(b,\lambda)$ if and only if $\lambda^\intercal h(D_iY_i, X_i, b) \in (\ell^*, u^*)$ for every $i$, 
		which holds if and only if $\hat{M}_n(b,\lambda) < \infty$. Notice that $\text{Gr}(\lambda)^{\eta/2} \equiv \left\{(b, \lambda) \; : \; \inf_{(b', \lambda') \in \text{Gr}(\lambda)} \lVert (b, \lambda) - (b', \lambda') \rVert \leq \eta/2\right\}$ contains $\text{Gr}(\Lambda)$ because $(b', \lambda') = (b,\lambda(b))$ is an element of $\text{Gr}(\lambda) = \left\{(b,\lambda(b)) \; : \; b \in B\right\}$. 
		Assumption \ref{Assumption: estimation} \ref{Assumption: estimation, moment conditions} implies $\sup_{(b,\lambda) \in \text{Gr}(\lambda)^{\eta/2}} \lvert M(b,\lambda)\rvert$ is finite, and lemma \ref{Lemma: uniform consistency of the dual objective} shows that $\hat{M}_n$ is uniformly consistent for $M$ on $\text{Gr}(\lambda)^{\eta/2}$, thus the continuous mapping theorem implies $\sup_{(b,\lambda) \in \text{Gr}(\lambda)^{\eta/2}} \lvert \hat{M}_n(b,\lambda) \rvert \overset{p}{\rightarrow} \sup_{(b,\lambda) \in \text{Gr}(\lambda)^{\eta/2}} \lvert M(b,\lambda)\rvert$ and therefore $\sup_{(b,\lambda) \in \text{Gr}(\lambda)^{\eta/2}} \lvert \hat{M}_n(b,\lambda) \rvert$ is finite with probability approaching one. When it is,
		\begin{equation*}
			\sup_{(b,\lambda) \in \text{Gr}(\Lambda)} \hat{M}_n(b,\lambda) \leq \sup_{(b,\lambda) \in \text{Gr}(\lambda)^{\eta/2}} \lvert \hat{M}_n(b,\lambda) \rvert < \infty.
		\end{equation*}
		and $\hat{M}_n$ is continuous on $\text{Gr}(\Lambda)$.
		
		\item Now consider continuity of $\Lambda : B \rightrightarrows \lambda(B)^{\eta/2}$. 

		Upper hemicontinuity will follow by application of the Closed Graph Theorem, \cite{Aliprantis2006} theorem 17.11. $B$ is compact by assumption \ref{Assumption: strong duality} and \ref{Lemma: continuous dual solution and value function} shows that $\lambda(\cdot)$ is continuous, therefore $\lambda(B) = \{\lambda(b) \; : \; b \in B\}$ is compact, and hence $\lambda(B)^{\eta/2}$ is compact. Suppose $\{(b_n, \lambda_n)\}_{n=1}^\infty \subseteq \text{Gr}(\Lambda)$ converges to $(b,\lambda)$. Then $b \in B$ because $B$ is closed. Since $\lambda(\cdot)$ is continuous, $\lVert \lambda(b_n) - \lambda(b) \rVert \rightarrow 0$, and therefore
		\begin{equation*}
			\lVert \lambda - \lambda(b) \rVert \leq \underbrace{\lVert \lambda - \lambda_n \rVert}_{\rightarrow 0} + \underbrace{\lVert \lambda_n - \lambda(b_n) \rVert}_{\leq \eta/2} + \underbrace{\lVert \lambda(b_n) - \lambda(b) \rVert}_{\rightarrow 0}
		\end{equation*}
		shows that $\lVert \lambda - \lambda(b) \rVert \leq \eta/2$, i.e. $\lambda \in \Lambda(b)$. Thus $(b,\lambda) \in \text{Gr}(\Lambda)$, so $\text{Gr}(\Lambda)$ is closed. \cite{Aliprantis2006} theorem 17.11 then implies $\Lambda : B \rightrightarrows \lambda(B)^{\eta/2}$ is upper hemicontinuous.
		
		Regarding lower semicontinuity, note that $B \subseteq \mathbb{R}^{d_b}$ and $\lambda(B)^{\eta/2} \subseteq \mathbb{R}^{d_g + K}$ are both metric spaces and hence first countable. Thus \cite{Aliprantis2006} theorem 17.21 implies $\Lambda$ is lower hemicontinuous at $b \in B$ if and only if for any sequence $\{b_n\} \subseteq B$ with $b_n \rightarrow b$ and any $\lambda \in \Lambda(b)$, there exists a subsequence $\{b_{n_k}\}_{k=1}^\infty$ and elements $\lambda_k \in \Lambda(b_{n_k})$ for each $k$ such that $\lambda_k \rightarrow \lambda$. For the subsequence we can take the sequence itself. Notice that $\lambda_n \equiv \lambda(b_n) + \lambda - \lambda(b)$ satisfies
		\begin{equation*}
			\lVert \lambda_n - \lambda(b_n) \rVert = \lVert \lambda(b_n) + \lambda - \lambda(b) - \lambda(b_n) \rVert = \lVert \lambda - \lambda(b) \rVert \leq \eta /2
		\end{equation*}
		and therefore $\lambda_n \in \lambda(b_n)$. Continuity of $\lambda(\cdot)$ and $b_n \rightarrow b$ implies $\lambda_n \rightarrow \lambda$, and thus $\Lambda$ is lower semicontinuous.
	\end{enumerate}
	To summarize, 
	\begin{align*}
		&\sup_{b \in B} \left\lVert \hat{\lambda}_n(b) - \lambda(b) \right\rVert < \eta/2 &&\text{ and } &&\sup_{(b,\lambda) \in \text{Gr}(\Lambda)} \hat{M}_n(b,\lambda) < \infty
	\end{align*}
	hold with probability approaching one. When both hold, $\hat{\nu}_n(b) = \max_{\lambda \in \Lambda(b)} \hat{M}_n(b,\lambda)$ is continuous by the Berge Maximum theorem, implying $\sqrt{n}(\hat{\nu}_n - \nu) \in \mathcal{C}(B)$. Thus 
	\begin{equation*}
		P(\sqrt{n}(\hat{\nu}_n - \nu) \in \mathcal{C}(B)) \geq P\left(\sup_{b \in B} \left\lVert \hat{\lambda}_n(b) - \lambda(b) \right\rVert < \eta/2 \text{ and } \sup_{(b,\lambda) \in \text{Gr}(\Lambda)} \hat{M}_n(b,\lambda) < \infty\right)\rightarrow 1.
	\end{equation*}
	As argued above, the Portmanteau theorem implies $P(\mathbb{G}_\nu \in \mathcal{C}(B)) = 1$. 
\end{proof}

\begin{restatable}[$\sqrt{n}$-consistency and convergence in distribution]{lemma}{lemmaWeakConvergenceOfBreakdownPoint}
	\label{Lemma: weak convergence of the breakdown point}
	\singlespacing
	
	Suppose assumptions \ref{Assumption: setting} and \ref{Assumption: strong duality} hold, and \ref{Assumption: estimation} \ref{Assumption: estimation, closed null hypothesis}, \ref{Assumption: estimation, nonsingular second moments}, \ref{Assumption: estimation, continuously differentiable moment functions}, \ref{Assumption: estimation, moment conditions} hold, but do not assume $\textbf{m}(\nu) = \argmin_{b \in B\cap \textbf{B}_0} \nu(b)$ is unique. Then 
	\begin{align*}
		\sqrt{n}(\hat{\delta}_n^{BP} - \delta^{BP}) \overset{d}{\rightarrow} \inf_{b \in \textbf{m}(\nu)} \mathbb{G}_\nu(b)
	\end{align*}
	where $\mathbb{G}_\nu$ is the weak limit of $\sqrt{n}(\hat{\nu}_n - \nu)$ in $\ell^\infty(B)$. 
\end{restatable}
\begin{proof}
	\singlespacing
	
	Let $\iota : \ell^\infty(B) \rightarrow \mathbb{R}$ be given by $\iota(f) = \inf_{b \in B \cap \textbf{B}_0} f(b)$. Then
	\begin{equation*}
		\sqrt{n}(\hat{\delta}_n^{BP} - \delta^{BP}) = \sqrt{n}(\iota(\hat{\nu}_n) - \iota(\nu))
	\end{equation*}
	suggests applying the Delta method, found in \cite{fang2019inference} as theorem 2.1. There are two assumptions to verify:
	\begin{enumerate}
		\item On the map $\iota$:
		\begin{enumerate}[label=(\roman*)]
			\item $\iota$ maps $(\ell^\infty(B), \lVert \cdot \rVert_B)$ to $(\mathbb{R}, \lvert \cdot \rvert)$, which are both Banach spaces. 
			\item Lemma \ref{Technical Lemma: restricted infimum is hadamard directionally differentiable} implies that $\iota$ is Hadamard directionally differentiable at any $f \in \mathcal{C}(B)$ tangentially to $\mathcal{C}(B)$, and lemma \ref{Lemma: continuous dual solution and value function} shows that $\nu \in \mathcal{C}(B)$.
		\end{enumerate}
		
		\item On the estimator $\hat{\nu}_n$:
		\begin{enumerate}[label=(\roman*)]
			\item As noted in lemma \ref{Lemma: support of value function asymptotic distribution}, $\sqrt{n}(\hat{\nu}_n - \nu) \overset{L}{\rightarrow} \mathbb{G}_\nu$ in $\ell^\infty(B)$.
			\item $\mathbb{G}_\nu$ is tight. 
			Lemma \ref{Lemma: support of value function asymptotic distribution} shows that $P(\mathbb{G}_\nu \in \mathcal{C}(B)) = 1$, i.e. the support of $\mathbb{G}_\nu$ is included in $\mathcal{C}(B)$. 
		\end{enumerate}
	\end{enumerate}
	\cite{fang2019inference} theorem 2.1 then implies
	\begin{align*}
		\sqrt{n}(\iota(\hat{\nu}_n) - \iota(\nu)) = \iota_\nu'(\sqrt{n}(\hat{\nu}_n - \nu)) + o_p(1)
	\end{align*}
	Lemma \ref{Technical Lemma: restricted infimum is hadamard directionally differentiable} shows the directional derivative of $\iota$ at $\nu$ is given by
	\begin{align*}
		&\iota_\nu' : \mathcal{C}(B) \rightarrow \mathbb{R}, &&\iota_\nu'(h) = \inf_{b \in \textbf{m}(\nu)} h(b)
	\end{align*}
	and therefore 
	\begin{align*}
		\sqrt{n}(\hat{\delta}_n^{BP} - \delta^{BP}) = \sqrt{n}(\iota(\hat{\nu}_n) - \iota(\nu)) = \inf_{b \in \textbf{m}(\nu)} \left\{\sqrt{n}(\hat{\nu}_n(b) - \nu(b))\right\} + o_p(1) \overset{L}{\rightarrow} \inf_{b \in \textbf{m}(\nu)} \mathbb{G}_\nu(b).
	\end{align*}
\end{proof}

\noindent \textbf{Theorem 4.2} (Asymptotic normality). \textit{Suppose assumptions \ref{Assumption: setting}, \ref{Assumption: strong duality}, and \ref{Assumption: estimation} hold. Let $\hat{b}_n \equiv $ \\ $\argmin_{b \in B \cap \textbf{B}_0} \hat{\nu}_n(b)$ and 
\begin{equation*}
	\hat{\sigma}_n^2 \equiv \frac{1}{n}\sum_{i=1}^n \left((\hat{\Phi}_n(\hat{b}_n)^{-1})^{(1)} \phi(D, DY, X, \hat{b}_n, \hat{\theta}_n(\hat{b}_n))\right)^2
\end{equation*}
where $(\hat{\Phi}_n(\hat{b}_n)^{-1})^{(1)} $ is the first row of the matrix $\hat{\Phi}_n(\hat{b}_n)^{-1}$. Then $\sqrt{n}(\hat{\delta}_n^{BP} - \delta^{BP})/\hat{\sigma}_n \overset{d}{\rightarrow} N(0,1)$.}
\begin{proof}
	\singlespacing
	
	Since $\textbf{m}(\nu)$ is a singleton, say $\textbf{m}(\nu)= \{b_\nu\}$, lemmas \ref{Lemma: weak convergence of the first stage} and \ref{Lemma: weak convergence of the breakdown point} imply $\sqrt{n}(\hat{\delta}_n^{BP} - \delta^{BP}) \overset{d}{\rightarrow} N(0, \sigma^2)$ where 
	\begin{align*}
		\sigma^2 &= E\left[\left((\Phi(b_\nu)^{-1})^{(1)} \phi(D, DY, X, b_\nu, \theta_0(b_\nu))\right)^2\right] \\
		&= e_1^\intercal \Phi(b_\nu)^{-1} E\left[\phi(D, DY, X, b_\nu, \theta_0(b_0)) \phi(D, DY, X, b_\nu, \theta_0(b_0))^\intercal\right] (\Phi(b_\nu)^{-1})^\intercal e_1
	\end{align*}
	and $e_1 = (1, 0, \ldots, 0) \in \mathbb{R}^{d_g + K + 2}$. Now notice that $\hat{\sigma}_n^2$ is the sample analogue:
	\begin{align*}
		\hat{\sigma}_n^2 &\equiv \frac{1}{n}\sum_{i=1}^n \left((\hat{\Phi}_n(\hat{b}_n)^{-1})^{(1)} \phi(D, DY, X, \hat{b}_n, \hat{\theta}_n(\hat{b}_n))\right)^2 \\
		&= e_1^\intercal \hat{\Phi}_n(\hat{b}_n)^{-1} \left[\frac{1}{n}\sum_{i=1}^n \phi(D_i, D_i Y_i, X_i, \hat{b}_n, \hat{\theta}_n(\hat{b}_n))\phi(D_i, D_i Y_i, X_i, \hat{b}_n, \hat{\theta}_n(\hat{b}_n))^\intercal \right] (\hat{\Phi}_n(\hat{b}_n)^{-1})^{\intercal} e_1
	\end{align*}
	
	It suffices to show $\hat{\Phi}_n(\hat{b}_n) \overset{p}{\rightarrow} \Phi(b_\nu)$ and 
	\begin{align}
		&\frac{1}{n}\sum_{i=1}^n \phi(D_i, D_i Y_i, X_i, \hat{b}_n, \hat{\theta}_n(\hat{b}_n))\phi(D_i, D_i Y_i, X_i, \hat{b}_n, \hat{\theta}_n(\hat{b}_n))^\intercal \notag \\
		&\hspace{3 cm} \overset{p}{\rightarrow} E\left[\phi(D, DY, X, b_\nu, \theta_0(b_0)) \phi(D, DY, X, b_\nu, \theta_0(b_0))^\intercal\right]. \label{Proof display: consistent variance estimator, consistency of inner variance matrix}
	\end{align}
	With these, the continuous mapping theorem will imply $\hat{\sigma}_n \overset{p}{\rightarrow} \sigma$, hence 
	$(\sqrt{n}(\hat{\delta}_n^{BP} - \delta^{BP}), \hat{\sigma}) \overset{d}{\rightarrow} (N(0, \sigma^2), \sigma)$, and another application of the continuous mapping theorem gives the conclusion $\frac{\sqrt{n}(\hat{\delta}_n^{BP} - \delta^{BP})}{\hat{\sigma}_n} \overset{d}{\rightarrow} N(0,1)$. 
	
	To show $\hat{\Phi}_n(\hat{b}_n) \overset{p}{\rightarrow} \Phi(b_\nu)$ and \eqref{Proof display: consistent variance estimator, consistency of inner variance matrix}, first notice that 
	\begin{align*}
		&\left\{\phi(D, DY, X, b, \theta)\phi(D, DY, X, b, \theta)^\intercal \; : \; (b,\theta) \in \text{Gr}(\theta_0)^\eta\right\} \\
		&\left\{\nabla_\theta \phi(D, DY, X, b, \theta) \; : \; (b,\theta) \in \text{Gr}(\theta_0)^\eta\right\}
	\end{align*}
	are special cases of \cite{van2007asymptotic} example 19.8 and hence Glivenko-Cantelli. Specifically, $\text{Gr}(\theta_0)^\eta$ is closed and bounded and hence compact. $(b, \theta) \mapsto \phi(D, DY, X, b, \theta)\phi(D, DY, X, b, \theta)^\intercal$ and $(b,\theta) \mapsto \nabla_\theta \phi(D, DY, X, b, \theta)$ are continuous by inspection of \eqref{Appendix display: phi, the Z-estimator integrand}, \eqref{Appendix display: jacobian matrix components}, and \eqref{Appendix display: derivatives of varphi}. Finally, $E\left[\sup_{(b, \theta) \in \text{Gr}(\theta_0)^\eta} \rVert \nabla_\theta \phi(D, DY, X, b, \theta) \rVert\right] < \infty$ and
	\begin{equation*}
	E\left[\sup_{(b, \theta) \in \text{Gr}(\theta_0)^\eta} \rVert \phi(D, DY, X, b, \theta)\phi(D, DY, X, b, \theta)^\intercal \rVert_o\right] = E\left[\sup_{(b, \theta) \in \text{Gr}(\theta_0)^\eta} \rVert \phi(D, DY, X, b, \theta)\rVert^2\right] < \infty
	\end{equation*}
	are implied by assumption \ref{Assumption: estimation} \ref{Assumption: estimation, moment conditions}.
	
	Next, observe that $(\hat{b}_n, \hat{\theta}_n(\hat{b}_n)) \overset{p}{\rightarrow} (b_\nu, \theta_0(b_\nu))$. First, $\hat{b}_n \overset{p}{\rightarrow} b_\nu$ follows from a standard extremum estimator argument. The function $\nu : B \rightarrow \mathbb{R}$ is continuous, uniquely minimized over the compact $B \cap \textbf{B}_0$ at $b_\nu$, and $\sup_{b \in B} \lvert \hat{\nu}_n(b) - \nu(b)\rvert \overset{p}{\rightarrow} 0$ by lemma \ref{Lemma: uniform consistency of the first stage}. Thus \cite{newey1994large} theorem 2.1 implies $\hat{b}_n = \argmin_{b \in B \cap \textbf{B}_0} \hat{\nu}_n(b)$ are consistent for $b_\nu$. Use the triangle inequality, $\hat{b}_n \overset{p}{\rightarrow} b_\nu$, continuity of $\theta_0(b) = (\nu(b), \lambda(b), p_D)$ (lemma \ref{Lemma: continuous dual solution and value function}), and $\sup_{b \in B} \lVert \hat{\theta}_n(b) - \theta_0(b) \rVert = o_p(1)$ (lemma \ref{Lemma: uniform consistency of the first stage}) to see that
	\begin{align*}
		\lVert \hat{\theta}_n(\hat{b}_n) - \theta_0(b_\nu) \rVert 
		&\leq \underbrace{\sup_{b \in B} \lVert \hat{\theta}_n(b) - \theta_0(b) \rVert}_{=o_p(1)} + \underbrace{\lVert \theta_0(\hat{b}_n) - \theta_0(b_\nu) \rVert}_{=o_p(1) \text{ by CMT}} = o_p(1)
	\end{align*}
	
	Note that $(b,\theta) \mapsto E[\nabla_\theta \phi(D, DY, X, b, \theta)]$ is continuous on $\text{Gr}(\theta_0)^\eta$ by the dominated convergence theorem and continuity of $(b, \theta) \mapsto \nabla_\theta \phi(D, DY, X, b, \theta)$ visible in equations \eqref{Appendix display: jacobian matrix components} and \eqref{Appendix display: derivatives of varphi}. $(\hat{b}_n, \hat{\theta}_n(\hat{b}_n)) \overset{p}{\rightarrow} (b_\nu, \theta_0(b_\nu))$, so $(\hat{b}_n, \hat{\theta}_n(\hat{b}_n)) \in \text{Gr}(\theta_0)^\eta$ holds with probability approaching one and when it does,
	\begin{align*}
		\lVert \hat{\Phi}_n(\hat{b}_n) - \Phi(b_\nu) \rVert &= \left\lVert \frac{1}{n}\sum_{i=1}^n \nabla_\theta \phi(D_i, D_iY_i, X_i, \hat{b}_n, \hat{\theta}_n(\hat{b}_n)) - E\left[\nabla_\theta \phi(D, DY, X, b_\nu, \theta_0(b_\nu))\right] \right\rVert\\
		&\leq \left\lVert \frac{1}{n}\sum_{i=1}^n \nabla_\theta \phi(D_i, D_iY_i, X_i, \hat{b}_n, \hat{\theta}_n(\hat{b}_n)) - E\left[\nabla_\theta \phi(D, DY, X, \hat{b}_n, \hat{\theta}_n(\hat{b}))\right] \right\rVert \\
		&\hspace{1cm} + \left\lVert E\left[\nabla_\theta \phi(D, DY, X, \hat{b}_n, \hat{\theta}_n(\hat{b}))\right] - E\left[\nabla_\theta \phi(D, DY, X, b_\nu, \theta_0(b_\nu))\right] \right\rVert \\
		&\leq \underbrace{\sup_{(b,\theta) \in \text{Gr}(\theta_0)^\eta} \left\lVert \frac{1}{n}\sum_{i=1}^n \nabla_\theta \phi(D_i, D_iY_i, X_i, b,\theta) - E\left[\nabla_\theta \phi(D, DY, X, b,\theta)\right] \right\rVert}_{=o_p(1) \text{ by Glivenko-Cantelli}} \\
		&\hspace{1 cm} + \underbrace{\left\lVert E\left[\nabla_\theta \phi(D, DY, X, \hat{b}_n, \hat{\theta}_n(\hat{b}))\right] - E\left[\nabla_\theta \phi(D, DY, X, b_\nu, \theta_0(b_\nu))\right] \right\rVert}_{=o_p(1) \text{ by CMT}} \\
		&= o_p(1).
	\end{align*}
	Essentially the same argument implies \eqref{Proof display: consistent variance estimator, consistency of inner variance matrix} holds, which completes the proof.
\end{proof}

%% file: Appendix_Examples.tex
\section{Appendix: Examples}

\label{Appendix: examples}

\subsection{Expectation}
\label{Appendix: examples, expectation}

This simple example is useful primarily to illustrate the ideas in a concrete setting. \\

Suppose the parameter of interest is $\beta = E[Y] \in \mathbb{R}$, and the sample is $\{D_i, D_i Y_i\}_{i=1}^n$. The conclusion to be supported is that $\beta > \bar{b}$, motivating the null and alternative hypotheses
\begin{align*}
	&H_0 \; : \; \beta \leq \bar{b}, &&H_1 \; : \; \beta > \bar{b}
\end{align*}
The model is characterized by $g(y, b) = y - b$. For the dual problem, set $h(y, b) = \begin{pmatrix} y - b & 1 \end{pmatrix}^\intercal$. The dual problem is 
\begin{equation}
	\sup_{\lambda \in \mathbb{R}^2} \lambda^\intercal c(b) - E_{P_1}[f^*(\lambda^\intercal h(Y, b))] \label{Example: Appendix, Expectation Dual Problem}
\end{equation}
where $c(b) = \begin{pmatrix} \frac{-p_D}{1-p_D}(E_{P_1}[Y] - b) & 1 \end{pmatrix}^\intercal$. 

\subsubsection*{Dual solution when $d_f$ is Kullback-Leibler and $P_1$ is $\mathcal{U}[0,1]$}

Suppose that $P_1$, the distribution of $Y \mid D = 1$, is  $\mathcal{U}[0,1]$. Let $\mu_1 = E[Y \mid D = 1] = 1/2$. Note that, since the support of $P_0$ is contained within $[0,1]$ as well, we have $\beta = E[Y] \in \left[p_D \mu_1, p_D \mu_1 + (1-p_D)\right]$. The endpoints are only attained if $P_0$ concentrates degenerately at $0$ or $1$ respectively, distributions which violate $P_0 \ll P_1$. \\

For tractability, let the measure of selection be Kullback-Leibler. For this divergence we let $f(t) = t\log(t) - t + 1$, which has convex conjugate $f^*(r) = \exp(r) - 1$. The dual problem has first order condition
\begin{align*}
	0 &= c(b) - E_{P_1}\left[(f^*)'(\lambda^\intercal h(Y, b))h(Y,b)\right] \\
	&= 
	\begin{pmatrix}
		\frac{-p_D}{1-p_D}\left(\frac{1}{2} - b\right) \\
		1 
	\end{pmatrix}
	- E\left[
	\exp\left(
	\lambda_1(Y - b) + \lambda_2
	\right)
	\begin{pmatrix}
		(Y - b) \\
		1
	\end{pmatrix}
	\right]. 
\end{align*}
From the second equation we have
\begin{equation}
	\lambda_2 = -\log\left(E[\exp(\lambda_1(Y - b))]\right). \label{Example: Appendix, Expectation KL LM for Constant}
\end{equation}

Suppose $b = \frac{1}{2}$. Then the first equation requires
\begin{equation}
	0 = E\left[\exp(\lambda_1(Y - b) + \lambda_2)\left(Y - \frac{1}{2}\right)\right]. \label{Example: Appendix, Expectation KL LM for Moment b = 1/2}
\end{equation}
Notice that if $\lambda_1 = 0$, then \eqref{Example: Appendix, Expectation KL LM for Constant} implies $\lambda_2 = 0$, and \eqref{Example: Appendix, Expectation KL LM for Moment b = 1/2} holds.

Now suppose $b \neq 1/2$. Consider the dual objective, and notice that 
\begin{equation*}
	E_{P_1}[f^*(\lambda^\intercal h(Y, b))] = \int_0^1 \exp(\lambda^\intercal h(y, b)) - 1 dy.
\end{equation*}
Since $b \neq 1/2$, it follows that $\frac{-p_D}{1-p_D}(1/2 - b) \neq 0$ and so $\lambda_1 \neq 0$. Thus, the integral above can be solved with $u$-substitution, setting $u = \lambda_1(y - b) + \lambda_2$:
\begin{align*}
	E_{P_1}[f^*(\lambda^\intercal h(Y, b))] &=  \int_0^1 \exp(\lambda^\intercal h(y, b)) - 1 dy = \frac{1}{\lambda_1} \int_{\lambda_1(-b) + \lambda_2}^{\lambda_1(1-b) + \lambda_2} \exp(u) du - 1\\
	&= \frac{\exp(\lambda^\intercal \textbf{b}_1) - \exp(\lambda^\intercal \textbf{b}_0)}{\lambda^\intercal e_1} - 1,
\end{align*}
where $\textbf{b}_1 = \begin{pmatrix} 1 - b & 1 \end{pmatrix}^\intercal$, $\textbf{b}_0 = \begin{pmatrix} -b & 1 \end{pmatrix}^\intercal$, and $e_1 = \begin{pmatrix} 1 & 0 \end{pmatrix}^\intercal$. Thus \eqref{Example: Appendix, Expectation Dual Problem} becomes 
\begin{align*}
	\sup_{\lambda\in \mathbb{R}^2} \lambda^\intercal 
	\begin{pmatrix}
		\frac{-p_D}{1-p_D}(1/2 - b) \\
		1
	\end{pmatrix} 
	- \frac{\exp(\lambda^\intercal \textbf{b}_1) - \exp(\lambda^\intercal \textbf{b}_0)}{\lambda^\intercal e_1} + 1,
\end{align*}
from which we can compute the first order conditions
\begin{align*}
	0 = \begin{pmatrix}
		\frac{-p_D}{1-p_D}(1/2 - b) \\
		1
	\end{pmatrix} - \frac{\exp(\lambda^\intercal \textbf{b}_1)\textbf{b}_1 - \exp(\lambda^\intercal \textbf{b}_0)\textbf{b}_0}{\lambda^\intercal e_1}  + \frac{\exp(\lambda^\intercal \textbf{b}_1) - \exp(\lambda^\intercal \textbf{b}_0)}{(\lambda^\intercal e_1)^2}e_1.
\end{align*}
Once again, the second equation can be solved for $\lambda_2$. The following form will be more useful:
\begin{align}
	0 &= 1 - \frac{\exp(\lambda_1(1 - b) + \lambda_2) - \exp(\lambda_1(-b) + \lambda_2)}{\lambda_1} \notag \\
	\implies \frac{\lambda_1}{\exp(\lambda_2)} &= \exp(\lambda_1(1-b)) - \exp(\lambda_1(-b)) \label{Example: Appendix, Expectation KL Uniform FOC}
\end{align}
The first equation is
\begin{align*}
	\frac{-p_D}{1-p_D}\left(\frac{1}{2} - b\right) &= \frac{\exp(\lambda_1(1 - b) + \lambda_2)(1-b) - \exp(\lambda_1(-b) + \lambda_2)(-b)}{\lambda_1} \\
	&\hspace{4 cm} - \frac{\exp(\lambda_1(1 - b) + \lambda_2) - \exp(\lambda_1(-b) + \lambda_2)}{\lambda_1^2} \\
	&= \frac{\exp(\lambda_2)}{\lambda_1}\Bigg[\exp(\lambda_1(1-b)) - b[\exp(\lambda_1(1-b)) - \exp(\lambda_1(-b))]\\
	&\hspace{4 cm} - \frac{\exp(\lambda_1(1-b)) - \exp(\lambda_1(-b))}{\lambda_1}\Bigg] \\
	&= \frac{\exp(\lambda_1(1-b))}{\exp(\lambda_1(1-b)) - \exp(\lambda_1(-b))} - b - \frac{1}{\lambda_1} = \frac{\exp(\lambda_1)}{\exp(\lambda_1) - 1} - b - \frac{1}{\lambda_1}
\end{align*}
where the second to last equality uses \eqref{Example: Appendix, Expectation KL Uniform FOC} above. Rearranging gives 
\begin{align*}
	\frac{\exp(\lambda_1)}{\exp(\lambda_1) - 1} - \frac{1}{\lambda_1} &= \frac{-p_D(1/2 - b) + (1-p_D)b}{1-p_D} = \frac{2b - p_D}{2(1-p_D)}
\end{align*} \\

Now notice that $\frac{\exp(\lambda_1)}{\exp(\lambda_1) - 1} - \frac{1}{\lambda_1}$ is well defined and continuous whenever $\lambda_1 \neq 0$, takes values between $0$ and $1$, with limits
\begin{align*}
	&\lim_{\lambda_1 \rightarrow \infty} \frac{\exp(\lambda_1)}{\exp(\lambda_1) - 1} - \frac{1}{\lambda_1} = 1, &&\lim_{\lambda_1 \rightarrow -\infty} \frac{\exp(\lambda_1)}{\exp(\lambda_1) - 1} - \frac{1}{\lambda_1} = 0
\end{align*}
Repeated applications of l'H\^opital's rule shows that 
\begin{align*}
	\lim_{\lambda_1 \rightarrow 0} \frac{\exp(\lambda_1)}{\exp(\lambda_1) - 1} - \frac{1}{\lambda_1} &= \lim_{\lambda_1 \rightarrow 0} \frac{\lambda_1\exp(\lambda_1) - \exp(\lambda_1) + 1}{\lambda_1(\exp(\lambda_1) - 1)} = \frac{1}{2}
\end{align*}
Therefore there exists a solution whenever $\frac{2b - p_D}{2(1-p_D)} \in \left(0,\frac{1}{2}\right) \cup \left(\frac{1}{2}, 1\right)$. Given this solution, \eqref{Example: Appendix, Expectation KL Uniform FOC} can be rearranged to obtain
\begin{equation*}
	\lambda_2 = \log\left(\frac{\lambda_1}{\exp(\lambda_1(1-b)) - \exp(\lambda_1(-b))}\right)
\end{equation*}

Now notice that 
\begin{align*}
	\frac{2b - p_D}{2(1-p_D)} > 0 &\implies b > \frac{p_D}{2}, \\
	\frac{2b - p_D}{2(1-p_D)} < 1 &\implies b < 1 - \frac{p_D}{2}
\end{align*}
and recall that $b = 1/2$ implies $\lambda_1 = \lambda_2 = 0$ solves the dual problem. Therefore the dual problem has a solution whenever $b \in \left(\frac{p_D}{2}, 1 - \frac{p_D}{2}\right)$. \\

$P_1$ has compact support, and $f^*(\lambda_1(y - b) + \lambda_2) = \exp(\lambda_1(y - b) + \lambda_2) - 1$ is continuous in $y$ for any $(\lambda_1, \lambda_2)$. Thus the extreme value theorem implies the solution is in the interior of $\left\{\lambda \in \mathbb{R}^2 \; : \; E[\lvert f^*(\lambda^\intercal h(Y,b))\rvert] < \infty\right\} = \left\{\lambda \in \mathbb{R}^2 \; : \; \int \lvert \exp(\lambda_1 (y - b) + \lambda_2) - 1 \rvert dy < \infty\right\}$. The implied solution to the primal, $q^b(y) = (f^*)'(\lambda^\intercal h(y, b)) = \exp(\lambda_1(y - b) + \lambda_2)$ satisfies $0 < q^b(y) < \infty$ on the support of $P_1$ and solves the moment conditions. Thus assumption \ref{Assumption: strong duality} is satisfied for any convex, compact $B \subset \left(\frac{p_D}{2}, 1 - \frac{p_D}{2}\right)$.

\subsection{Linear models}
\label{Appendix: examples, linear models}

\noindent \textbf{Lemma 4.1} (Convex value function, linear models). \textit{Suppose assumptions \ref{Assumption: setting} and \ref{Assumption: strong duality} hold, the sample is $\{D_i, D_i Y_i, X_{i1}, X_{i2}\}_{i=1}^n$ where $Y_i \in \mathbb{R}$, $X_{i1} \in \mathbb{R}^{d_{x1}}$, and $X_{i2} \in \mathbb{R}^{d_{x2}}$, and the parameter $\beta$ is identified by 
\begin{equation*}
	E[(Y - X_1^\intercal \beta) X_2] = 0
\end{equation*}
Then $\hat{\nu}_n$ and $\nu$ are convex. If in addition $E[X_2X_1^\intercal]$ has full column rank, then $\nu$ is strictly convex.}
\begin{proof}
	Let $b^0, b^1 \in B$, be distinct, $\alpha \in (0,1)$, and $b^\alpha = \alpha b^1 + (1-\alpha) b^0$. The proof of theorem 3.1 shows that the primal problem at $b^0$ and $b^1$ is attained by $Q^0$ and $Q^1$ with densities $q^0$, $q^1$. The moment conditions are $0 = E\left[X_2(Y - X_1^\intercal\beta)\right] = E\left[X_2 Y\right] - E\left[X_2 X_1^\intercal\right]\beta$, so $Q^0 \in \textbf{P}^{b^0}$ and $Q^ \in \textbf{P}^{b^1}$ implies
	\begin{align}
		E_{Q^1}\left[X_2Y\right] - E_{P_{0X}}\left[X_2X_1^\intercal\right]b^1 &= \frac{-p_D}{1-p_D}\left(E_{P_1}\left[X_2Y\right] - E_{P_1}\left[X_2X_1^\intercal \right]b^1\right), \label{Proof display: convex value function, moment equations Q1} \\
		E_{Q^0}\left[X_2Y\right] - E_{P_{0X}}\left[X_2X_1^\intercal\right]b^0 &= \frac{-p_D}{1-p_D}\left(E_{P_1}\left[X_2Y\right] - E_{P_1}\left[X_2X_1^\intercal \right]b^0\right) \label{Proof display: convex value function, moment equations Q0}
	\end{align}
	implying that 
	\begin{equation*}
		E_{\alpha Q^1 + (1-\alpha)Q^0} \left[X_2Y\right] - E_{P_{0X}}\left[X_2X_1^\intercal\right]b^\alpha = \frac{-p_D}{1-p_D}\left(E_{P_1}\left[X_2Y\right] - E_{P_1}\left[X_2X_1^\intercal \right]b^\alpha\right)
	\end{equation*}
	Similarly, $E_{Q^0}[\mathbbm{1}\{X = x_k\}] = E_{Q^1}[\mathbbm{1}\{X = x_k\}] = E_{P_{0X}}[\mathbbm{1}\{X = x_k\}]$ for all $k = 1,\ldots, K$. It follows that $Q^\alpha \equiv \alpha Q^1 + (1-\alpha) Q^0$ is feasible for $b^\alpha = \alpha b^1 + (1-\alpha) b^0$. This implies
	\begin{equation*}
		d_f(Q^\alpha \Vert P_1) \geq \inf_{Q \in \textbf{P}^{b^\alpha}} d_f(Q \Vert P_1) = \nu(b^\alpha)
	\end{equation*}
	$Q^\alpha$ has $P_1$-density $q^\alpha = \alpha q^1 + (1-\alpha) q^0$. Convexity of $f$ implies that for any $(y,x)$,
	\begin{equation*}
		\alpha f(q^1(y,x)) + (1-\alpha) f(q^0(y,x)) \geq f(\alpha q^1(y,x) + (1-\alpha) q^0(y,x)) = f(q^\alpha(y, x))
	\end{equation*}
	integrating with respect to $P_1$ shows that 
	\begin{equation*}
		\alpha d_f(Q^1 \Vert P_1) + (1-\alpha) d_f(Q^0 \Vert P_1) \geq d_f(Q^\alpha \Vert P_1) \geq \nu(b^\alpha)
	\end{equation*}
	Since the left hand side equals $\alpha \nu(b^1) + (1-\alpha) \nu(b^0)$, this shows $\nu$ is convex. Notice that no properties of $P_1$, $P_{0X}$ were specified in the argument above, so the same argument works to show $\hat{\nu}_n(b)$ is convex in $b$ by replacing $P_1$, $P_{0X}$ with their empirical counterparts. 
	
	Finally, to see that $\nu$ is strictly convex when $E[X_2 X_1^\intercal]$ has full column rank, use equations \eqref{Proof display: convex value function, moment equations Q1} and \eqref{Proof display: convex value function, moment equations Q0} to see that
	\begin{align*}
		(1-p_D)\left[E_{Q^{1,n}}\left[X_2Y\right] - E_{Q^{0,n}}\left[X_2Y\right]\right] &= \underbrace{\left[p_DE_{P_1}[X_2X_1^\intercal] + (1-p_D) E_{P_0}[X_2X_1^\intercal]\right]}_{=E[X_2 X_1^\intercal]}(b^1 - b_0) 
	\end{align*}
	Since $E[X_2X_1^\intercal]$ has full column rank and $b^1 - b^0 \neq 0$,
	\begin{align*}
		(1-p_D)\left[E_{Q^1}\left[X_2Y\right] - E_{Q^0}\left[X_2Y\right]\right] \neq 0
	\end{align*}
	and thus $Q^1$ differs from $Q^0$, implying $q^1$ differs from $q^0$ on a set of positive $P_1$ measure. For $(y,x)$ in that set, strict convexity of $f$ assumed in \eqref{Assumption: setting} \ref{Assumption: setting, divergence} implies 
	\begin{equation*}
		\alpha f(q^1(y,x)) + (1-\alpha) f(q^0(y,x)) > f(\alpha q^1(y,x) + (1-\alpha) q^0(y,x)) = f(q^\alpha(y,x))
	\end{equation*}
	integrating with respect to $P_1$ implies $\alpha d_f(Q^1 \Vert P_1) + (1-\alpha) d_f(Q^0 \Vert P_1) > d_f(Q^\alpha\Vert P_1)$, and thus $\alpha \nu(b^1) + (1-\alpha) \nu(b^0) > d_f(Q^\alpha\Vert P_1) \geq \nu(b^\alpha)$. 
\end{proof}

Simulations suggest that ordinary least squares more generally produces convex $\nu(b)$. Consider the data generating process described in section 5.2. Here the data is of the form \\ $\{D_i, D_iY_{i1}, D_i Y_{i2}, X_{i1}, X_{i2}\}_{i=1}^n$, and the model is given by
\begin{align*}
	&Y_{i1} = \beta_0 + \beta_1 X_1 + \beta_2 Y_2 + \beta_3 X_2 + \varepsilon, &&E\left[\begin{pmatrix} 1 \\ X_1 \\ Y_2 \\ X_2 \end{pmatrix}\varepsilon\right] = 0
\end{align*}
 The following figure investigates convexity of the $\nu(b)$ (where $d_f(Q \Vert P) = H^2(Q, P)$) numerically, by looking for convexity along random line segments. Specifically, let $b_1$ and $b_0$ be points in the sample space and compute $\hat{\nu}_n(\lambda b_1 + (1-\lambda) b_0)$ for many values of $\lambda$ between $0$ and $1$. The following figure shows the results of this exercise for 10 randomly selected $(b_0, b_1)$ pairs, and shows that no deviation from convexity was detected. 
\begin{figure}[H]
	\begin{center}
		\includegraphics[scale=0.6]{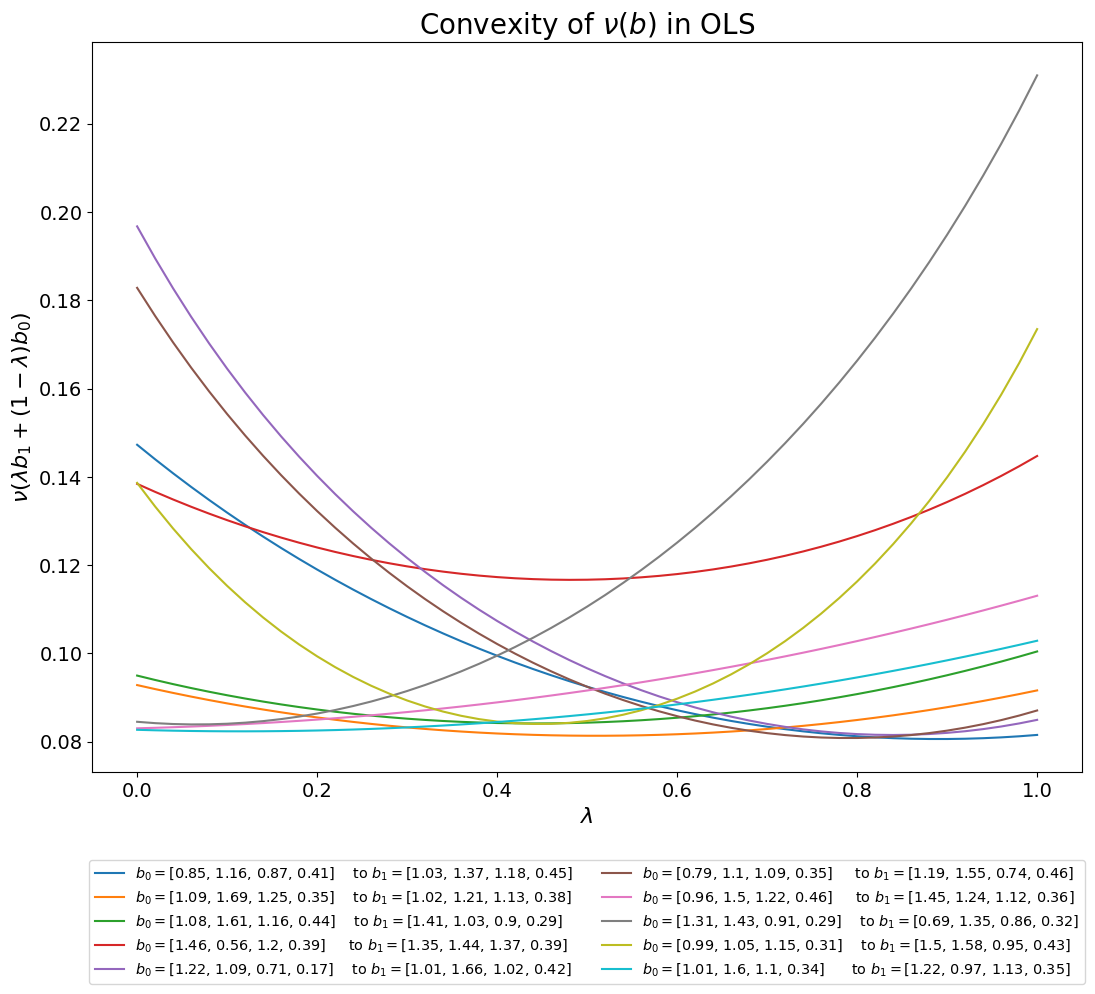}
	\end{center}
\end{figure}

\subsection{Binary choice models}
\label{Appendix: examples, binary choice models}

Let $V \in \{0,1\}$, $W \in \mathbb{R}^d$, and suppose interest is in $P(V = 1 \mid W = w)$. A common choice of model assumes 
\begin{equation*}
	P(V = 1 \mid W = w) = F(w^\intercal \beta) 
\end{equation*}
for a known CDF $F(\cdot)$. This model can be derived from a latent variable model of the form $V = \mathbbm{1}\{W'\beta \geq \xi\}$, where conditional on $W$, the unobserved ``latent variable'' $\xi$ has distribution $F(x)$.
\begin{equation*}
	P(V = 1 \mid W = w) = P(\xi \leq W^\intercal \beta \mid W = w) = F(w^\intercal \beta)
\end{equation*}
For example, the logit model uses $F(x) = \Lambda(x) = \frac{\exp(x)}{1 + \exp(x)}$, while the probit model uses $F(x) = \Phi(x) = \int_{\infty}^x \frac{1}{\sqrt{2\pi}} \exp(-t^2/2) dt$.

Given i.i.d. data of the form $\{V_i, W_i\}_{i=1}^n$, the model can be estimated through maximum likelihood. 
The likelihood of an observation $(V, W)$ is $F(W^\intercal b)^{V}(1-F(W^\intercal b)^{1-V}$, implying a population log-likelihood of 
\begin{align*}
	\ell(b) \equiv E\left[V \ln(F(W^\intercal b)) + (1 - V)\ln(1 - F(W^\intercal b))\right]
\end{align*}
Assuming $F(x)$ is differentiable with density $f(x)$ and that differentiation and expectation can be interchanged, the score is given by
\begin{align*}
	s(b) \equiv \nabla_b \ell(b) = E\left[\frac{f(W^\intercal b)}{F(W^\intercal b)\left(1 - F(W^\intercal b)\right)}\left(V - F(W^\intercal b)\right) W\right]
\end{align*}
Supposing $f(x)$ is differentiable with derivative $f'(x)$, the Hessian can be calculated and shown negative definite when $E[WW^\intercal]$ is full rank. This implies the log-likelihood is strictly concave, and hence the first order condition suffices for maximization. Therefore the model could also be viewed as a GMM model solving 
\begin{align*}
	0 = E\left[\frac{f(W^\intercal \beta)}{F(W^\intercal \beta)\left(1 - F(W^\intercal \beta)\right)}\left(V - F(W^\intercal \beta)\right) W\right]
\end{align*}

\subsubsection*{Logit model}
For the logit model, $F(x) = \Lambda(x) = \frac{\exp(x)}{1 + \exp(x)}$, we can compute that 
\begin{equation*}
	f(x) = \frac{\exp(x)}{(1 + \exp(x))^2} = F(x)(1 - F(x))
\end{equation*}
and thus the score simplifies to 
\begin{equation*}
	s(b) = E\left[\left(V - \Lambda(W^\intercal b)\right) W\right]
\end{equation*}
This makes it straightforward to compute the Hessian of the log-likelihood as
\begin{equation*}
	\nabla_b^2 \ell(b) = E\left[ -\Lambda(W^\intercal b)(1-\Lambda(W^\intercal b)) W W^\intercal\right]
\end{equation*}
Let $U \equiv \sqrt{\Lambda(W^\intercal b)(1-\Lambda(W^\intercal b))} W$ and observe that $\nabla_b^2 \ell(b) = -E[U U^\intercal]$ is negative definite if $E[WW^\intercal]$ is full rank. Thus, the logit model can be viewed as a GMM model, where $\beta$ solves
\begin{equation*}
	0 = E\left[\left(V - \Lambda(W^\intercal \beta)\right)W\right]
\end{equation*}
This model can be put into the form used in assumption \ref{Assumption: setting} with $Z = (Z_{(1)}, Z_{-1}) = (V,W)$, $g(z,b) = (z_{1} - \Lambda(z_{-1}^\intercal b))z_{-1}$, and $\nabla_b g(z,b) = -\Lambda(z_{-1}^\intercal b)(1 - \Lambda(z_{-1}^\intercal b))z_{-1} z_{-1}^\intercal$. 

Simulations suggest that the logit model may also produce a convex $\nu(b)$. Consider the data generating process described in section 5.3. The same numerical exercise described above results in a figure that again shows no deviation from convexity. 
\begin{figure}[H]
	\begin{center}
		\includegraphics[scale=0.6]{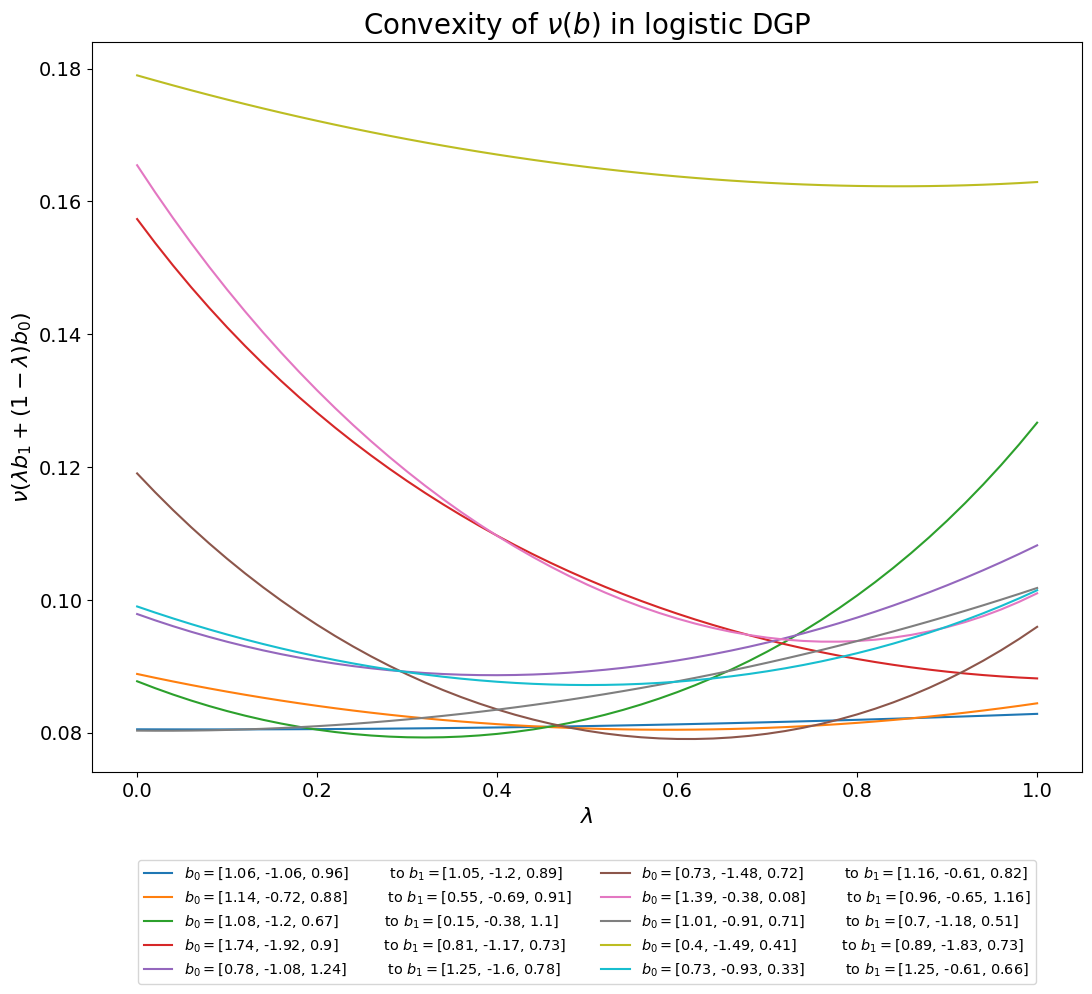}
	\end{center}
\end{figure}